\documentclass[aps, prd,twocolumn, amsmath, floats,floatfix,
superscriptaddress, nofootinbib,showpacs]{revtex4}

\usepackage{stackengine} 
\usepackage{doi}
\usepackage{array} \usepackage{ifpdf}
\ifpdf \usepackage[pdftex]{graphicx} \graphicspath{{figures/pdf/}}
\DeclareGraphicsExtensions{.pdf} \else \usepackage[dvips]{graphicx}
\usepackage{graphicx} \usepackage{pstricks} \usepackage{pst-node}
\usepackage{pst-blur} \graphicspath{{figures/eps/}}
\DeclareGraphicsExtensions{.eps} \definecolor{Pink}{rgb}{1.,0.75,0.8}
\definecolor{cyan2}{cmyk}{0.40,0,0,0} \fi
\newcommand{\Eref}[1]{Eq.~\eqref{#1}} 

\def\p0{\partial_0}

\def\apj{Astrophys. J.}  
\def\mnras{Mon. Not. R. Astron. Soc.}  \def\prd{Phys. Rev. D}
\def\prl{Phys. Rev. Lett.}  \def\cqg{Class. Quantum Grav.}

\usepackage{tikz}
\usepackage{amsmath} \usepackage{amssymb} \usepackage{listings}
\usepackage{mathtools} \usepackage{subfigure}

\begin{document}

\title{Numerical relativity simulations of thick accretion disks
  around tilted Kerr black holes}

\author{Vassilios Mewes} \affiliation{Departamento de Astronom\'{\i}a
  y Astrof\'{\i}sica, Universitat de Val\`encia, Doctor Moliner 50,
  46100, Burjassot (Val\`encia), Spain}
  
\author{Jos\'e A. Font} \affiliation{Departamento de Astronom\'{\i}a y
  Astrof\'{\i}sica, Universitat de Val\`encia, Doctor Moliner 50, 46100,
  Burjassot (Val\`encia), Spain} \affiliation{Observatori
  Astron\`omic, Universitat de Val\`encia, C/ Catedr\'atico Jos\'e
  Beltr\'an 2, 46980, Paterna (Val\`encia), Spain}
  
\author{Filippo \surname{Galeazzi}} \affiliation{ Institut f\"ur
  Theoretische Physik, Max-von-Laue-Stra\ss e 1, 60438 Frankfurt,
  Germany}
  
\author{Pedro J. Montero} \affiliation{Max-Planck-Institute f{\"u}r
  Astrophysik, Karl-Schwarzschild-Str. 1, 85748, Garching bei
  M{\"u}nchen, Germany}

\author{Nikolaos Stergioulas} \affiliation{Department of Physics,
  Aristotle University of Thessaloniki, Thessaloniki 54124, Greece}

\begin{abstract}
  In this work we present 3D numerical relativity simulations of thick accretion 
  disks around {\it tilted} Kerr black holes. We investigate the evolution of three 
  different initial disk models with a range of initial black hole spin magnitudes 
  and tilt angles. For all the disk-to-black hole mass ratios considered
  ($0.044-0.16$) we observe significant black hole precession and nutation during 
  the evolution. This indicates that for such mass ratios, neglecting the self-gravity 
  of the disks by evolving them in a fixed background black hole spacetime is not
  justified. We find that the two more massive models are unstable against the 
  Papaloizou-Pringle (PP) instability and that those PP-unstable models remain unstable for 
  all initial spins and tilt angles considered, showing that the development of the instability 
  is a very robust feature of such PP-unstable disks. Our lightest model, which 
  is the most astrophysically favorable outcome of mergers of binary compact objects, 
  is stable. The tilt between the black hole spin 
  and the disk is strongly modulated during the growth of the PP instability,
  causing a partial global realignment of black hole spin and disk angular 
  momentum in the most massive model with constant specific angular
  momentum $l$. For the model with non-constant $l$-profile we observe
  a long-lived $m=1$ non-axisymmetric structure which shows strong 
  oscillations of the tilt angle in the inner regions of the disk. This effect 
  might be connected to the development of Kozai-Lidov oscillations.
  Our simulations also confirm earlier findings that the development of the PP instability 
  causes the long-term emission of large amplitude gravitational waves, predominantly 
  for the $l=m=2$ multipole mode. The imprint of the BH precession on the 
  gravitational waves from tilted BH-torus systems remains an interesting open issue 
  that would require significantly longer simulations than those presented in this work.
\end{abstract}


\maketitle

\section{Introduction}
\label{sec:introduction}

Stellar mass black hole--torus systems are believed to be the end states of binary neutron star 
(NS-NS) or black hole neutron star (BH-NS) mergers, as well as of the rotational gravitational 
collapse of massive stars. BH-torus systems emit gravitational waves (GWs), which may eventually provide 
the direct means to study their actual formation and evolution and prove the current 
hypotheses that associate them to gamma-ray burst (GRB) engines~\cite{Woosley1993,Janka1999}. 
Such a possibility is out of reach to electromagnetic observations due to their intrinsic high density 
and temperature. For an overview of the event rate estimates of NS-NS and BH-NS mergers that 
are observable with initial and advanced LIGO see e.g.~\cite{Abadie:2010,Dominik2013,Dominik2014}.

Our theoretical understanding of the formation of BH-torus systems and their evolution relies 
strongly on numerical work. In recent years a significant number of numerical relativity
simulations have shown the feasibility of the formation of such systems from generic initial 
data (see e.g.~\cite{Rezzolla:2010,Kyutoku2011,
Hotokezaka2013, Hotokezaka2013c,Kastaun2015} for recent progress). 
In particular, the 3D simulations of~\cite{Rezzolla:2010} (see also references therein) have 
shown that unequal-mass NS-NS mergers lead to the self-consistent formation of {\it massive} 
accretion tori (or thick disks) around spinning black holes, thus meeting the necessary
requirements of the GRB's central engine hypothesis. However, if the energy released in a 
(short) GRB comes from the accretion torus, the BH-torus system has to survive for up to a 
few seconds~\cite{rees:94}. Any instability which might disrupt the system on shorter timescales, 
such as the runaway instability~\cite{Abramowicz1983} or the Papaloizou-Pringle instability 
(PPI hereafter)~\cite{Papaloizou1984}, could pose a severe problem for the prevailing GRB 
models.

Using perturbation theory, Papaloizou and Pringle found that tori with
constant specific angular momentum are unstable to non-axisymmetric
global modes. Such modes have a co-rotation radius within the torus, located 
in a narrow region where waves cannot propagate.  Waves can still tunnel 
through the co-rotation zone and interact with waves in the other region. The 
transmitted modes are amplified due to the feedback mechanism provided by 
the reflecting boundaries at the inner and outer edges of the torus. The
manifestation of the PPI, as~\cite{Hawley1991} first showed numerically through
hydrodynamical simulations in the fixed background metric of a Schwarzschild 
BH, is in the form of counter-rotating epicyclic vortices, or ``planets'", with $m$ 
planets emerging from the growth of a mode of order $m$.  Moreover, BH-torus
systems are characterized by the presence of a cusp-like inner edge
where mass transfer driven by the radial pressure gradient is possible. If due to 
accretion the cusp moves deeper inside the disk material, the mass transfer 
speeds up leading to a runaway process that destroys the disk on a dynamical 
timescale (see e.g.~\cite{Font2002} for test-fluid simulations in general
relativity of the occurrence of this instability and~\cite{Montero2010} for 
axisymmetric simulations where self-gravity was first taken into account). In 
most recent numerical relativity
simulations~\cite{Rezzolla:2010,Hotokezaka2013,Neilsen2014,Kastaun2015} 
the BH-torus systems 
under consideration did not manifest signs of dynamical instabilities on short 
dynamical timescales, because the non-constant angular momentum profiles 
of the massive disks that formed in the simulations seem to make them stable against
the development of the runaway instability. However, on longer timescales
non-axisymmetric instabilities can easily develop~\cite{Korobkin2011,Kiuchi2011}. 

Korobkin et al~\cite{Korobkin2011} have studied non-axisymmetric instabilities in 
self-gravitating disks around BHs using three-dimensional hydrodynamical simulations 
in full general relativity. Their models incorporate both moderately slender and 
slender disks with disk-to-BH mass ratios ranging from 0.11 to 0.24. While no sign 
of the runaway instability was observed for these particular models, unstable 
non-axisymmetric modes were indeed present. More recently~\cite{Korobkin2013} 
observed that by a suitable choice of model parameters, namely constant angular 
momentum tori exactly filling or slightly overflowing their Roche lobe, a rapid 
mass accretion episode with the characteristics of a runaway instability sets in. The 
astrophysical significance of such fine-tuned models is uncertain as, e.g.~they do 
not seem to be favored as the end-product of self-consistent numerical relativity 
simulations of binary neutron star
mergers~\cite{Rezzolla:2010,Hotokezaka2013,Kastaun2015,Sekiguchi2015}. 
Correspondingly, the 3D
general relativistic numerical simulations of BH-torus systems of~\cite{Kiuchi2011} 
showed that an $m=1$ non-axisymmetric instability grows for a wide range of
self-gravitating tori orbiting BHs. Their models included torus-to-BH mass ratios in the 
range $0.06-0.2$ and both constant and non-constant angular momentum profiles 
in the disks.~\cite{Kiuchi2011} found that the non-axisymmetric
structure persists for a timescale much longer than the dynamical one, becoming a 
strong emitter of large amplitude, quasi-periodic GWs, observable by 
forthcoming detectors. 

The vast majority of numerical relativity simulations to date that
have studied the formation of BH-torus systems have led to systems in
which the accretion torus is aligned with the equatorial plane of the
central Kerr BH. 
There are, however, several ways of arriving at {\it tilted} accretion
disks around Kerr BH. Possible scenarios include asymmetric supernova
explosions in binary systems~\cite{Fragos2010} or via BH-NS mergers 
in which the BH spin is misaligned with the orbital plane of the
binary. See also~\cite{Fragile2001,Maccarone2002} for arguments that
most BH-torus systems should be tilted.
Recently, full GRHD simulations of BH-NS mergers
with {\it tilted} initial BH spins have been 
performed by~\cite{Foucart2011,Foucart2013,Kawaguchi2015},
which can result in tilted accretion tori around the BH remnant. In
these simulations massive remnant disks only formed for high initial BH spins.
As shown in~\cite{Foucart2012}, the disk mass in BH-NS mergers 
increases with increasing initial BH spin and decreases with 
increasing initial BH mass. This is due to the size of the ISCO 
(innermost circular stable orbit) of the BH in the merger. The ISCO
grows with BH mass and decreases with the spin magnitude
of the BH. If the ISCO is large enough, the NS will be 'swallowed'
entirely by the BH before being tidally disrupted, leaving no
accretion torus behind. In order to estimate disk masses of 
resulting from BH-NS mergers, one needs thus an estimate for
the initial BH masses in these system. One such method, via
population synthesis considerations, favor larger 
BH masses~\cite{Belczynski2008,Belczynski2010,Fryer2012}, with a peak around
$8\, \mathrm{M}_{\odot}$ and a mass gap between the lightest expected
BHs and NS masses. This means that
these massive BH would need very large initial spins in order to be able
form massive remnant disks after the BH-NS merger. 

General relativity hydrodynamics and 
MHD simulations of such kind of {\it tilted} thick accretion disks around 
BHs have been performed by Fragile and collaborators within the test-fluid 
approximation (see e.g.~\cite{Fragile2005,Fragile2007a}). These authors 
have carried out an extensive program to study both the dynamics and 
observational signatures of thick accretion tori around tilted Kerr black 
holes. 

Motivated by these previous works we present in this paper an extended set of
numerical relativity simulations of tilted accretion disks around Kerr BHs in which 
the self-gravity of the disks is taken into account. 
In this first study of the effects of the disk self-gravity in tilted 
BH-torus systems, we have chosen low mass central black holes,
and small spins that are unlikely to produce massive tilted disks 
in astrophysical mergers for the reasons outlined above. As a first
study, we want to investigate the effects of the disk self-gravity 
around tilted BH with our existing initial data, where some of our 
initial models were known to be unstable
against the PPI in the untilted case~\cite{Korobkin2011}. 
We plan on studying these 
tilted BH-torus systems with
more astrophysically realistic data in future works.
Our simulations incorporate 
the effects of the self-consistent evolution of BH-torus systems with misaligned 
spins via the solution of the full 3+1 Einstein equations coupled to the hydrodynamics
equations. Due to the Lense-Thirring effect, the inner regions of the disks start 
precessing and might become twisted and warped affecting their dynamical behavior 
(for a definition of twist and warp see e.g.~\cite{Nelson2000}). Those inner regions 
might also be forced into the equatorial plane of the BH due to the so-called 
Bardeen-Petterson effect~\cite{Bardeen1975}. Our simulations allow us to investigate 
these issues and find out how the dynamics of the tilted disks affect the central BH, 
assessing in turn the importance of neglecting the disk's self-gravity on the overall 
dynamics of these systems.

This paper is organized as follows: In Section~\ref{sec:numericalmethods} we 
summarize the mathematical and numerical framework we employ in our numerical 
study, explaining in detail the methods we use for the spacetime and matter evolution. 
We also describe several diagnostics we use to monitor and analyze the simulations. 
In Section~\ref{sec:ID} we describe the setup of our initial data, both in the untilted 
and tilted cases. We also provide a table which contains the key parameters of the 
models considered in this study. Our results are presented and discussed in detail 
in Section~\ref{sec:results}. Finally, we present our conclusions in Section~\ref{sec:conclusions}.  
Throughout the paper, we use units in which ${c}={G}={M}_{\odot}=1$ where
${c}$, ${G}$ and ${M}_{\odot}$ are the speed of light, the gravitational constant and the 
solar mass, respectively. Latin indices run from 1 to 3 while Greek indices run from 
0 to 3.

\section{Mathematical and numerical framework}
\label{sec:numericalmethods}

\subsection{Spacetime evolution}

The simulations reported in this paper are performed using the
publicly available Einstein Toolkit (ET)
code~\cite{ET,Loffler2012}. The ET is a code for relativistic
astrophysics simulations, which uses the modular \verb:Cactus:
framework~\cite{Cactus} (consisting of general modules commonly called `thorns') 
and provides adaptive mesh refinement (AMR) via the \verb:Carpet: driver \cite{Carpet}.
The set of complex time-dependent partial differential equations 
is integrated using the method-of-lines where the time integration 
is performed using a  4th order Runge-Kutta algorithm.
The toolkit provides numerical solvers and the necessary infrastructure 
to evolve the Einstein equations,
\begin{equation}\label{eqs:einstein}
  G^{\mu \nu} = 8\, \pi \, T^{\mu \nu}\,,
\end{equation}
coupling the evolution of spacetime to the dynamic of the matter
content through the stress-energy tensor $T^{\mu \nu}$. 
The methods to evolve the Einstein equations are based on the 3+1 
ADM~\cite{Arnowitt2008} splitting of spacetime, in
which the metric becomes
\begin{equation}
  ds^2 = \left(-\alpha^2 + \beta_i \beta^i \right) dt^2 + 2 \beta_i \,dt \,dx^i + \gamma_{ij} \,dx^i \,dx^j\,,
\end{equation}
where $\alpha$ is the lapse function, $\beta^i$ the shift vector, and
$\gamma_{ij}$ is the spatial metric tensor. The extrinsic curvature tensor
$K_{ij}$ is constructed from the spatial metric $\gamma_{ij}$ as follows:
\begin{equation}
  K_{ij} \equiv -\frac{1}{2 \alpha} \left(\partial_t -
    \mathcal{L}_{\beta} \right) \, \gamma_{ij} \, ,
\end{equation}
where $\mathcal{L}_{\beta}$ stands for the Lie derivative with respect to the
shift vector.

The left hand side of the Einstein field equations is evolved 
using the \verb:McLachlan: code~\cite{Brown2009,Reisswig2011}, 
which solves a conformal-traceless ``$3+1$'' formulation of the Einstein
equations known as BSSN~\cite{Nakamura1987,Shibata1995,Baumgarte1998}.

The BSSN evolution equations are evolved using a fourth-order accurate,
centered finite-differencing operator, while the advection terms for the shift vector
are evolved with a fourth-order upwind stencil. We apply fifth-order Kreiss-Oliger 
Dissipation to all spacetime variables to achieve overall fourth-order for the 
spacetime evolution. Using the BSSN evolution system and the gauge conditions described
in Appendix~\ref{sec:BSSN}, the \verb:McLachlan: code can evolve stably a single puncture 
for several light crossing times, see e.g.~\cite{Loffler2012}. 

We apply Sommerfeld outgoing boundary conditions for all the BSSN 
evolution variables with an extrapolation to include the part of the boundary conditions 
that does not behave like a simple outgoing wave~\cite{Alcubierre2000}.

\subsection{Matter Evolution}

The evolution of the hydrodynamics is performed by 
\verb:GRHydro:~\cite{Baiotti2005,Loffler2012,Mosta2014}, that
solves the general relativistic hydrodynamics equations in flux-conservative
form, in the so-called Valencia formulation~\cite{Marti91,Banyuls1997,FontJoseA.2008}, 
using high-resolution shock-capturing (HRSC) methods. 
The GRHD evolutions equations are
described in Appendix~\ref{sec:grhydro}.

As a grid-based code \verb:GRHydro: cannot evolve regions without
matter content due to the breakdown of the EOS and evolution equations
in regions of zero rest-mass density. As is customary, a low density
atmosphere (typically several orders of magnitude below the maximum
density of the initial data) fills those grid points that should
belong to vacuum regions. In all our simulations we set the density of the 
atmosphere to $10^{-8}$ times
the initial maximum rest-mass density in the torus. Although fully evolved, the dynamical
effects of the atmosphere regions can be neglected.

We use the piecewise parabolic method (PPM)~\cite{Colella1984} to reconstruct the 
primitive variables at the cell interfaces and Marquina's flux formula~\cite{Donat1996,Aloy1999} 
to compute the numerical fluxes. Differently from the original implementation reported in 
Ref.~\cite{Baiotti07}, we reconstruct the quantities $W v^i$ instead of the three-velocities
$v^i$. This guarantees that the velocities reconstructed at the cell
boundaries remain subluminal even under extreme conditions like those
encountered near the apparent horizon (AH) of the BH.

\subsection{Diagnostics}

We measure the BH mass during the evolution using the \verb:AHFinderDirect: thorn
which implements the AH finder described in~\cite{Thornburg2003}. The BH spin 
magnitude and direction are measured using the \verb:QuasiLocalMeasures: 
thorn~\cite{Dreyer2003,Schnetter2006} (see also the review of quasi-local methods in
\cite{Szabados2009}). We measure the spin direction using the so-called flat-space 
rotational Killing vector method of~\cite{Campanelli2007}, which can be derived using 
Weinberg's pseudotensor in Gaussian coordinates and is equal to the Komar angular 
momentum integral when the latter is expressed in a foliation adapted to the axisymmetry 
of spacetime~\cite{Mewes2015}. To track the moving location of the puncture during the 
evolution, we use the \verb:PunctureTracker: thorn. Gravitational waves are
extracted using a multipole expansion of the Weyl scalar at different
radii. The thorns \verb:Multipole: and \verb:WeylScal4: compute the
respective quantities. We calculate the mass
accretion rate $\dot{M}$ as the instantaneous flow of matter
through the AH, using the \verb:Outflow: thorn. 
To quantify the growth of non-axisymmetric modes in the disk, we use
the thorn \verb:GRHydroAnalysis: which computes 3D Fourier integrals
of the density,
described in~\cite{Baiotti2007a}. The amplitude of the $m$-th mode is given by:

\begin{equation}
  D_m = \int \, \alpha \, \sqrt{\gamma} \,\rho \, e^{-i\,m\,\phi} \, d^3x \, .
  \label{eq:mode-calculation}
\end{equation}
Since the computational domain is very large (see
section~\ref{subsection:num_setup}
below), the total mass of the 
atmosphere can be a non-negligible fraction of the total 
rest-mass of the torus.
We therefore have to ignore cells corresponding to the atmosphere
when integrating the torus rest-mass density and in the calculations 
of the non-axisymmetric modes in the disk. 
Despite the non-negligible mass of the atmosphere,
this mass is not dynamically
important because it is spatially homogeneous (and therefore its
gravitational pull on the central BH-torus system is zero) and its coordinate 
velocity is set to zero. 

\subsection{Twist and tilt angles}
\label{subsection:twist_tilt_diagnostics}

In order to keep track of the response of the tilted accretion disk
and to check for Lense-Thirring precession and the occurrence of the
Bardeen-Patterson effect, we measure two angles, the twist
(precession) and tilt (inclination) of the disk~\cite{Nelson2000}. We
closely follow the description of the angles given in~\cite{Fragile2005},
with a modification in the calculation of the twist, and furthermore
adapt the way they are calculated.

\begin{figure}[t]
  \centering
  \includegraphics[scale=0.42]{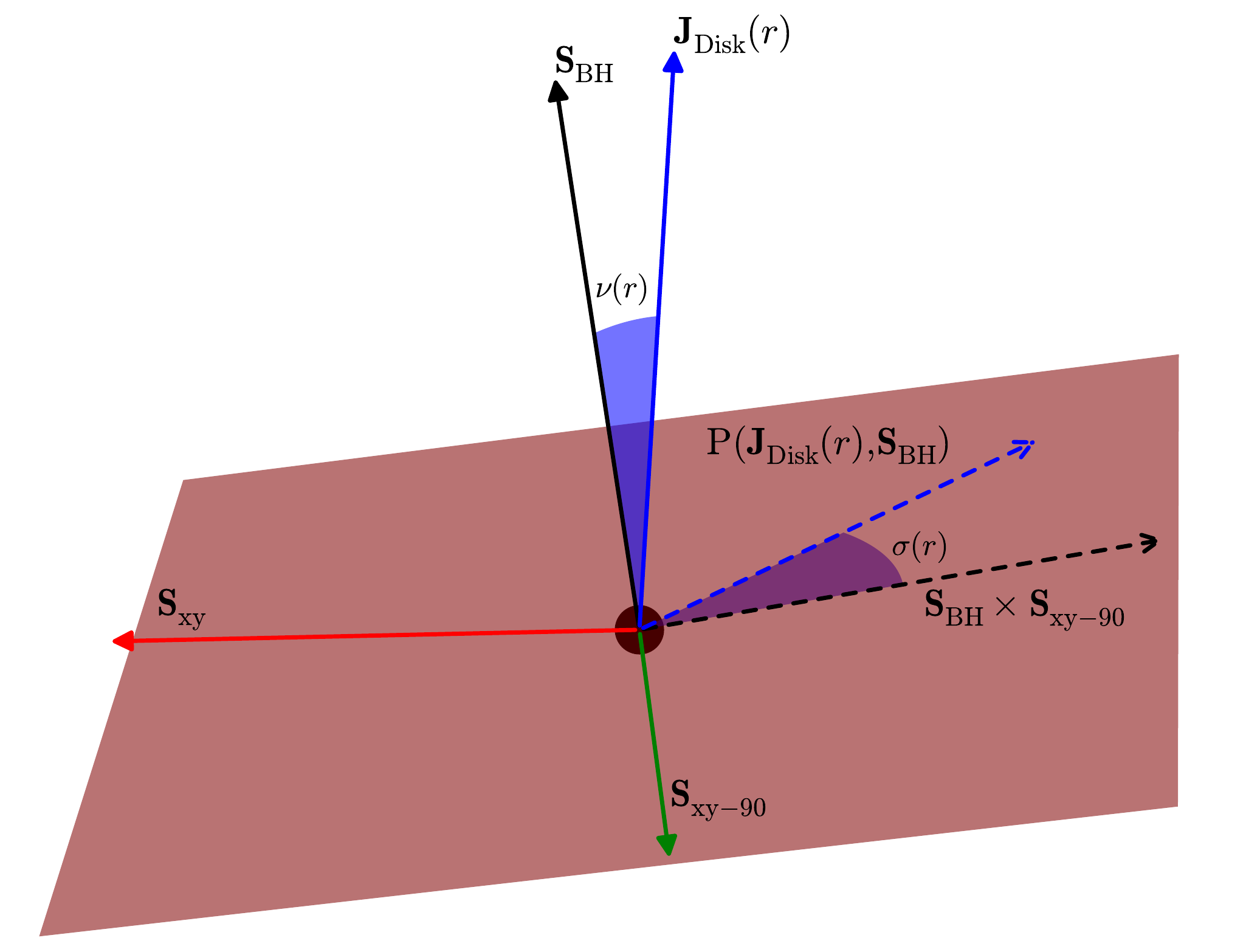}
  \caption{The figure visualizes the various vectors and projections
    we use to calculate the twist and tilt in the disk. Note that all
    vectors in the figure have been scaled to unity for better
    visualization.}
  \label{fig:tilt_twist_sketch}
\end{figure}

The main idea is to split the disk up into a series of annuli and
calculate the angular momentum vector of the matter,
$\mathbf{J}_{\rm{Disk}}(r)$, for each individual annulus. Using this
vector, we define two angles, the twist $\sigma(r)$
\begin{equation}\label{twist}
  \sigma(r) = \angle (\mathbf{S}_{\rm{BH}} \times
  \mathbf{S}_{\mathrm{xy}-90},
  \mathrm{P}(\mathbf{J}_{\mathrm{Disk}}(r),\mathbf{S}_{\mathrm{BH}}))\,,
\end{equation}
and the tilt $\nu(r)$
\begin{equation}\label{tilt}
  \nu(r) = \angle  (\mathbf{S}_{\rm{BH}},
  \mathbf{J}_{\rm{Disk}}(r))\,,
\end{equation}
where
\begin{equation}
  \mathrm{P}(\mathbf{a},\mathbf{b}) = 
  \mathbf{a} - \frac{\mathbf{a} \cdot
    \mathbf{b}}{|\mathbf{b}|^2}  \, \mathbf{b}\,,
\end{equation}
is the projection of vector $\mathbf{a}$ onto the plane with normal
$\mathbf{b}$.

The vector $\mathbf{S}_{\rm{BH}} \times \mathbf{S}_{\mathrm{xy}-90}$
is constructed in the following way: We project the BH spin
$\mathbf{S}_{\mathrm{BH}}$ onto the $xy$-plane,
$\mathbf{S}_{\mathrm{xy}}
=\mathrm{P}(\mathbf{S}_{\mathrm{BH}},\mathbf{z})$ and then rotate the
resulting vector $\mathbf{S}_{\mathrm{xy}}$ by $\pi/2$ about the
$z$-axis. The cross product of $\mathbf{S}_{\mathrm{BH}}$ and
$\mathbf{S}_{\mathrm{xy-90}}$ then lies in the equatorial plane of the
BH and the twist $\sigma(r)$ at $t=0$ is $0$ throughout the disk. A
sketch showing the construction of these vectors and twist and tilt
angles is provided in Fig.~\ref{fig:tilt_twist_sketch}.

The twist $\sigma(r)$ is then a measure of how much the angular
momentum vector of each annulus has precessed in the (dynamically
changing) equatorial plane of the BH, while the tilt $\nu(r)$ gives
the angle between the BH spin and the angular momentum vectors for
each annulus.  The disk is said to be \emph{twisted} if the twist
becomes a function of radius, $\sigma = \sigma(r)$, and said to be
\emph{warped} if the tilt becomes a function of radius, $\nu =
\nu(r)$.

The angular momentum vector for each annulus is calculated following
the procedure outlined in the section on spin  in
\cite{Weinberg1972} (pages 46f.). We calculate $\mathbf{J}_{\rm{Disk}}(r)$ from the
anti-symmetric angular momentum tensor
\begin{equation}\label{eq:ang_mom tensor}
  L^{\mu \nu} = \int d^3 x (x^{\mu}T^{\nu 0} -
  x^{\nu}T^{\mu 0})\,,
\end{equation}
in the following way:
\begin{equation}\label{eq:J_shell}
  J_{x} = L^{2 3}, \,\,J_{y} = J^{31} \:  \,\,\text{and} \,\, \: J_{z} = L^{12}\,.
\end{equation}

The relevant components of the stress-energy tensor $T^{\mu \nu}$ are
\begin{equation}
  T^{i 0} =  \rho \:h \:
  \left(\frac{W^{2}}{\alpha} \: v^{i}-\frac{W}{\alpha} \: \beta^{i} \right) + P \: \frac{\beta^{i}}{\alpha^{2}}\,,
\end{equation}
where Latin indices run from $1$ to $3$.

We note that our procedure differs slightly from that given
in~\cite{Fragile2005}, where the components of the disk shell
angular momentum vectors were computed using the intrinsic spin as follows:
\begin{equation}
  S_{\alpha} = \frac{1}{2} \, \epsilon_{\alpha \beta \gamma \delta}\,
  J^{\beta \gamma} \,U^{\delta}\,,
\end{equation}
where $\epsilon_{\alpha \beta \gamma \delta}$ is the 4-dimensional
Levi-Civita symbol and $ U^{\delta} =
p^{\delta}/(-p_{\beta}p^{\beta})^\frac{1}{2} $ is the total 4-velocity
of the system, with $p^{\alpha}$ being the 4-momentum.

Using the extrinsic spin, the contribution to the total angular
momentum stemming from choosing a reference point is accounted for in
the linear momentum of the center of mass of the system. We choose to
use the angular momentum tensor $L^{\mu \nu}$, calculated about the
center of the BH instead, using formula (\ref{eq:ang_mom tensor}) in
the calculation of $\sigma(r)$ and $\nu(r)$. We therefore pick a
reference frame centered on the BH, as it is moving around the grid and
we are interested in the total angular momentum of the disk shell
about the BH center.
We note that the calculation of the
twist and tilt angles is in fact gauge dependent. We have checked
the impact of the gauge choice on the twist and tilt angles by 
measuring the angles for the initial data with different lapse and shift
profiles to find that the angles are calculated correctly (as the initial 
tilt and twist profiles are parameters of our initial data), which also
serves as an important consistency check for the correct measure
of the initial angles.
In the actual numerical implementation, we do not use the inverse
cosine to calculate the angles, as the function becomes very
inaccurate when the vectors become close to being parallel or
anti-parallel, instead we use the \verb:atan2: function which is free of this
deficiency. The tilt $\nu(r)$ is then calculated using the following
formula:
\begin{equation}\label{angle}
  \nu(r) = \tan^{-1} \left[\:|\hat{\mathbf{S}}_{\rm{BH}} \times \hat{\mathbf{J}}_{\rm{Disk}}(r)| \:, \: \hat{\mathbf{S}}_{\rm{BH}} \cdot \hat{\mathbf{J}}_{\rm{Disk}}(r) \right]
\end{equation}
where the hat indicates unit vectors.

For the twist $\sigma(r)$, we need to be more careful. In the formula
above, we implicitly assume that $\nu(r) < \pi \,, \forall \,t$,
because Eq.~(\ref{angle}) will always return angles $\leq
\pi$. Although it does not generally make much sense to consider
angles $\geq \pi$ between two 3D vectors, because there is an up and
down direction, we are nevertheless interested in directional (signed)
angles $\geq \pi$ for the twist. The reason is that we want to be able
to track cumulative twists larger than $\pi$. In order to calculate
the directional angle $\sigma(r)$ from the reference vector
$\mathbf{S}_{\rm{BH}} \times \mathbf{S}_{\mathrm{xy}-90}$ to the
target vector
$\mathrm{P}(\mathbf{J}_{\mathrm{Disk}}(r),\mathbf{S}_{\mathrm{BH}})$
in a fixed sense of rotation, we need a third reference vector that
always lies above the plane spanned by the two original vectors. As
the plane in which both vectors live is constructed to be the
equatorial plane of the BH, we can therefore choose this vector to be
$\mathbf{S}_{\mathrm{BH}}$.
This information allows us (because the
cross product contains this information in the sign it returns) to
calculate twist angles in the range $[0,2\pi]$ with the following
formula:
\begin{equation}
  \sigma(r) = \tan^{-1} \left[\:|\mathbf{J}_{\rm{BH}} \times \mathbf{J}_{\rm{Disk}}(r)| \:, \: \mathbf{J}_{\rm{BH}} \cdot \mathbf{J}_{\rm{Disk}}(r) \right]
\end{equation}
and then select the directional angle by:
\begin{displaymath}
  \sigma(r) = \left\{
    \begin{array}{lr}
      \sigma(r) & : \,\mathbf{S}_{\rm{BH}} \, \cdot
      \left(\mathbf{S}_{\rm{BH}} \times
        \mathbf{J}_{\rm{Disk}}(r)\right) \geq 0\\
      2 \pi -\sigma(r) & : \,\mathbf{S}_{\rm{BH}} \, \cdot
      \left(\mathbf{S}_{\rm{BH}} \times
        \mathbf{J}_{\rm{Disk}}(r)\right) < 0
    \end{array}
  \right.
\end{displaymath}

We also compute the instantaneous precession and nutation rates for
both the BH spin vector $\mathbf{S}$ and the total disk angular
momentum vector $\mathbf{J}_{\rm{Disk}}$ (which we calculate by
summing the components of all shells). For the BH, we compute its
precession and nutation about the $z$-axis, while the total precession
and nutation of the disk angular momentum vector
$\mathbf{J}_{\rm{Disk}}$ is calculated about the BH spin axis. 

\begin{figure}[t]
  \centering
  \includegraphics[scale=1.0]{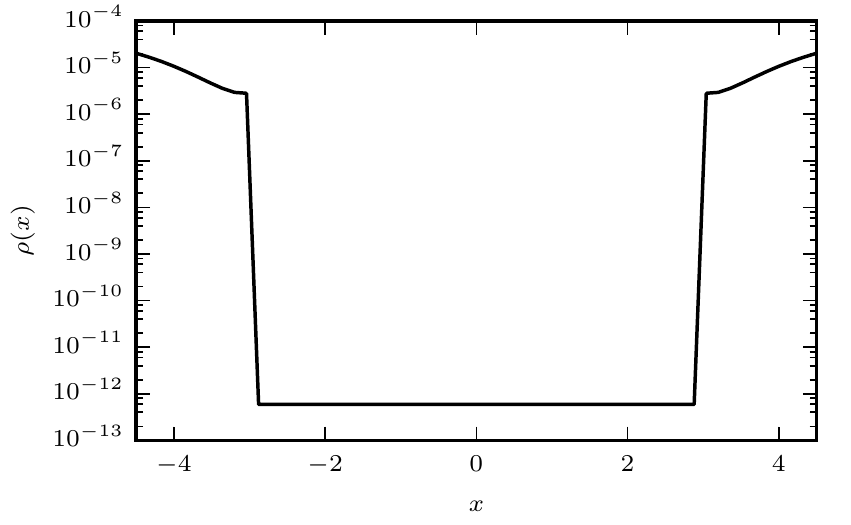}
  \caption{Innermost region of the initial rest-mass density profile
    along the $x$-axis for model {\tt C1Ba0b0}. The disk ends at $3.03
    $ and the AH is initially located at $x=0.46$.}
  \label{fig:BH_fill_rho}
\end{figure}

\section{Initial data and setup}
\label{sec:ID}

Our initial setup is a thick, self-gravitating axisymmetric accretion
disk in equilibrium around a rotating BH. Such a system is built
following the approach laid out in~\cite{Stergioulas2011} that we
briefly describe below. The solver to build the initial data
(ID) first computes models of self-gravitating, massive tori around
non-rotating BHs. Then, for our simulations of tilted disks around rotating
BHs we retain the hydrodynamical content of a model and replace the spacetime by 
a tilted Kerr spacetime in quasi-isotropic coordinates.  The generated initial data is then
interpolated onto Cartesian coordinates and evolved with the
\verb:Einstein Toolkit:. 
We are not aware of a method to generate self-consistent data for
self-gravitating tilted disks around Kerr black holes and therefore
resort to this method. In order to keep the constraint violations as
low as possible, we use disk models with rest masses of only a few
percent of the central BH mass and with small to moderate BH spins.
When replacing the original spacetime of the initial data with the
tilted Kerr spacetime in the computational domain, both the torus
rest-mass $M_0$ and the torus gravitational mass $M_{\mathrm{T}}$ will change
somewhat, due to the fact that the volume element of the spacetime $\sqrt{-g}$, 
where $g$ is the determinant of the four-dimensional metric 
$g_{\mu\nu}$, as well as the lapse, have changed. 
In order to assess the difference in $M_0$ and $M_{\mathrm{T}}$ due to the 
replacement of the spacetime we calculate both quantities 
as in~\cite{Stergioulas2011}, as follows:
\begin{equation}
  M_0 = \int \, \rho\,u^t\,\sqrt{-g}\,d^3x = \int \, \sqrt{\gamma} \, W \, \rho\, d^3x,
\end{equation}
and
\begin{equation}
\begin{aligned}
  M_{\mathrm{T}} & = \int \, (-2\, T^t_t+T^{\mu}_{\mu}) \, \alpha \, \sqrt{\gamma}\, d^3x.
\end{aligned}
\end{equation}
We find the largest difference  in $M_0$ and $M_{\mathrm{T}}$ in model 
{\tt C1Ba03b30}, where we obtain the following fractional differences:
$\Delta\, M_0\,\approx -3\%$ and $\Delta\, M_{\mathrm{T}}\,\approx -3\%$.

\subsection{Self-gravitating accretion disks}

Our ID are self-gravitating disks around a Schwarzschild BH in
quasi-isotropic (QI) coordinates~\cite{Stergioulas2011}. The ID is
fully described by the four metric potentials $\lambda(r,\theta)$,
$B(r,\theta)$, $\alpha(r,\theta)$ and $\omega(r,\theta)$ that
characterize the metric of a stationary, axisymmetric spacetime in
spherical polar QI coordinates,
\begin{equation}
  ds^2 = -e^{2 \nu} dt^2 + e^{2 \alpha} (d\bar{r}^2 + \bar{r}^2 d\phi^2) + \frac{B^2}{e^{2 \nu}} \bar{r}^2 sin^2\theta (d\phi -\omega dt)^2\,,
\end{equation}
where the isotropic radius $\bar{r}$ is given in terms of the areal
radius by
\begin{equation}
  \bar{r} = \frac{1}{2} \left(r - M + \sqrt{r^{2} - 2Mr}\right)\,,
\end{equation}
where $M$ is the mass of the BH.  The ID also provides the pressure
$p$ and orbital angular velocity $\Omega=u^{\phi}/u^t$, 
measured by an observer at infinity at rest.
The evolution time is measured in terms of the initial orbital period 
at the radius of the initial maximum of the rest-mass density, $\rho_c$. 
These two quantities, together with the EOS, are sufficient to obtain the remaining
hydrodynamical quantities. The ID is constructed using the polytropic
EOS described in \Eref{eq:poly_eos}.
The components of the
3-velocity in Cartesian coordinates are given by
\begin{equation}
  v^i = \left(y \left(\frac{\omega - \Omega}{\alpha} \right),-x \left(\frac{\omega - \Omega}{\alpha} \right),0 \right)
\end{equation}
where $\alpha$ is the lapse function. 
The Lorentz factor is calculated directly from the 3-velocity
\begin{equation}
  W = \frac{1}{\sqrt{1-v_iv^i}}= \frac{1}{\sqrt{1-(\omega-\Omega)^2 \frac{B^2}{e^{4 \nu}}(x^2+y^2)}}\,.
\end{equation}

The matter content of the ID only fills the computational domain up to
the event horizon of the central BH.  In our simulations we do not
excise the BH but rather treat it as a puncture. We noticed that when
evolving the BH as a puncture with matter content without excision,
the simulations were not long-term stable and failed at early stages
of the evolution due to errors in the matter evolution at the location
of the puncture.  The reason is a very rapid pile up of matter at this
location, which leads to non-physical values of the hydrodynamical 
variables at the puncture. These lead to failures in the \verb:GRHydro: 
evolution code during the conversion from conserved to primitive variables.
In \cite{Faber2007} a successful method to evolve hydrodynamics in the
presence of punctures was presented. This method checks for unphysical
values of the hydrodynamical variables at the immediate vicinity of
the puncture 
and resets them to physical values. In practice it checks for the 
positivity of the conserved energy ($\tau \geq 0$) and an upper bound for 
$|S|^2$:
\begin{equation}
 \gamma^{i j} S_i S_j \leq \tau (\tau + 2 \rho).
\end{equation}
In our simulations,
we use a similar treatment inside the BH horizon, namely we specify a
fraction of the volume of the AH in which we reset the matter fields
to atmosphere values and the stress-energy tensor $T^{\mu \nu}$ to
zero, which is essentially the same technique as described 
in~\cite{Reisswig2013}. 
In this way, we avoid using a moving excision boundary for the
hydrodynamical variables as in~\cite{Loeffler06a}. 
From the point of view of the evolution it is
safe to do this, as the evolution of the hydrodynamics inside the
horizon cannot influence the evolution outside the horizon. In fact,
we checked that the fraction of the AH in which we apply such
atmosphere resetting has no effect on the evolution, as expected. 
We checked this by applying the 
atmosphere reset for different fractions of the AH. In all cases, 
the evolution outside the AH was unaffected, confirming that the 
region inside the AH is causally disconnected. This has been 
employed for instance in the so-called ``turduckening`` of initial
BH data~\cite{Brown2009}. For
the practical implementation of this approach we use a spherical
surface that contains the shape of
the AH and apply the atmosphere method in a fraction of the minimum
radius of that surface. Our procedure is illustrated in
Fig.~\ref{fig:BH_fill_rho} where we plot the initial density profile
along the $x$-axis for model {\tt C1Ba0b0} of our sample (see
Table~\ref{table:models} below). The AH is initially located at
$x=0.46$ and there exists a smooth density profile across the
horizon. The remaining hydrodynamical quantities are treated in the
same way.  By using this approach we are able to evolve matter fields
in the presence of punctures for very long times without the need for
moving hydrodynamical excision zones.

\begin{table*}
  \begin{center}
    \caption{\label{tab:initial-models} Main characteristics of our
      initial models. From left to right the columns indicate the name
      of the model, the BH mass $M_{\mathrm{BH}}$,  the disk-to-BH mass ratio $q$, defined as the ratio of the irreducible mass of the
    central Schwarzschild BH and the total gravitational mass of the
    torus of the original initial data, the inner and outer
      disk radii $r_{\rm{in}}$ and $r_{\rm{out}}$, the maximum
      rest-mass density $\rho_{\mathrm{c}}$, the polytropic constant of the
      EOS $K$, the orbital period
      $P_{\mathrm{orb}}$ in code units and (milliseconds) and orbital frequency $f_{\mathrm{orb}}$
      at the radius of the initial $\rho_{\mathrm{c}}$, the
      specific angular momentum profile in the equatorial plane of the
      disk $l$ (in terms of
      the Schwarzschild radial coordinate $R$), the BH spin parameter
      $a$, and the initial tilt angle $\beta_0$. All models are
      evolved with a $\Gamma$-law ideal fluid EOS with
      $\Gamma=4/3$. Models indicated with a {\tt *} were simulated
      using half the canonical resolution of $\Delta
      x=0.02$. See main text for details.}
    \label{table:models}
    \begin{ruledtabular}
      \begin{tabular}{lccccccccccc}
        Name & $M_{\mathrm{BH}}$  & $q$ & $r_{\rm{in}}$ & $r_{\rm{out}}$ &  $\rho_{\mathrm{c}}$ & $K$ &
        $P_{\mathrm{orb}}$ & $f_{\mathrm{orb}}$  & $l$-profile  & $a$ & $\beta_0$ \\
        & & &  &  & ($ \times 10^{-5}$) & & & (Hz) & & & (deg) \\
        \hline
        {\tt D2a01b0}         &1.0002 & $0.044$ & 3.42 & 30.1 & $1.05$
        &$0.323$  & 150 (0.74)& 1360 & 3.75 (const)& 0.1 & 0 \\
        {\tt D2a01b5}     &1.0002  & $0.044$ & 3.42 & 30.1 & $1.05$ &$0.323$  & 150 (0.74)& 1360 & 3.75 (const)& 0.1 & 5\\
        {\tt D2a01b15}      &1.0002 & $0.044$ & 3.42 & 30.1 & $1.05$ &$0.323$  & 150 (0.74)& 1360 & 3.75 (const)& 0.1 & 15 \\
        {\tt D2a01b30}      &1.0002 & $0.044$ & 3.42 & 30.1 & $1.05$ &$0.323$ & 150 (0.74)& 1360 & 3.75 (const)& 0.1 & 30 \\
        {\tt D2a05b5}     &1.0002  & $0.044$ & 3.42 & 30.1 & $1.05$ &$0.323$ & 150 (0.74)& 1360 & 3.75 (const)& 0.5 & 5 \\
        {\tt D2a05b15}      &1.0002 & $0.044$ & 3.42& 30.1 & $1.05$ &$0.323$ & 150 (0.74)& 1360 & 3.75 (const)& 0.5 & 15 \\
        {\tt D2a05b30}      &1.0002 & $0.044$ & 3.42& 30.1 & $1.05$ &$0.323$ & 150 (0.74)& 1360 & 3.75 (const)& 0.5 & 30 \\
        {\tt C1Ba0b0}       &0.9569 & $0.160$ & 3.03 & 22.7 & $5.91$ &$0.180$
        & 157 (0.77)& 1300 & 3.67 (const)& 0.0 & 0 \\
        {\tt C1Ba01b5}     &0.9569& $0.160$  & 3.03 & 22.7 & $5.91$ &$0.180$ & 157 (0.77)& 1300 & 3.67 (const)& 0.1 & 5 \\
        {\tt C1Ba01b15}     &0.9569& $0.160$ & 3.03 & 22.7 & $5.91$ &$0.180$ & 157 (0.77)& 1300 & 3.67 (const)& 0.1 & 15 \\
        {\tt C1Ba01b30}     &0.9569& $0.160$ & 3.03 & 22.7 & $5.91$ &$0.180$ & 157 (0.77)& 1300 & 3.67 (const)& 0.1 & 30 \\
        {\tt C1Ba03b5*}     &0.9569& $0.160$  & 3.03 & 22.7 & $5.91$ &$0.180$ & 157 (0.77)& 1300 & 3.67 (const)& 0.3 & 5 \\
        {\tt C1Ba03b15*}     &0.9569& $0.160$ & 3.03 & 22.7 & $5.91$ &$0.180$ & 157 (0.77)& 1300 & 3.67 (const)& 0.3 & 15 \\
        {\tt C1Ba03b30*}     &0.9569& $0.160$ & 3.03 & 22.7 & $5.91$ &$0.180$ & 157 (0.77)& 1300 & 3.67 (const)& 0.3 & 30 \\
        {\tt NC1a0b0*}      & 0.9775&  $0.110$ & 3.60 & 33.5 & $1.69$
        &$0.170$ & 242 (1.19)& 843 & 3.04 $R^{0.11}$& 0.0 & 0 \\
        {\tt NC1a01b5*}   & 0.9775&  $0.110$ & 3.60 & 33.5 & $1.69$  &$0.170$
        & 242 (1.19)&  843&  3.04 $R^{0.11}$ & 0.1 & 5 \\
        {\tt NC1a01b15*}   & 0.9775& $0.110$ & 3.60 & 33.5 &  $1.69$ &$0.170$
        & 242 (1.19)&  843&  3.04 $R^{0.11}$ & 0.1 & 15 \\
        {\tt NC1a01b30*}   & 0.9775& $0.110$  & 3.60 & 33.5 & $1.69$ &$0.170$
        & 242 (1.19)&  843&  3.04 $R^{0.11}$ & 0.1 & 30 \\
        {\tt NC1a03b5*}   & 0.9775&  $0.110$ & 3.60 & 33.5 & $1.69$  &$0.170$
        & 242 (1.19)&  843&  3.04 $R^{0.11}$ & 0.3 & 5 \\
        {\tt NC1a03b15*}   & 0.9775& $0.110$ & 3.60 & 33.5 &  $1.69$ &$0.170$
        & 242 (1.19)&  843&  3.04 $R^{0.11}$ & 0.3 & 15 \\
        {\tt NC1a03b30*}   & 0.9775& $0.110$  & 3.60 & 33.5 & $1.69$ &$0.170$
        & 242 (1.19)&  843&  3.04 $R^{0.11}$ & 0.3 & 30 \\

      \end{tabular}
    \end{ruledtabular}
  \end{center}
\end{table*}

%
\begin{figure}[t]
  \centering
  \includegraphics[scale=0.37]{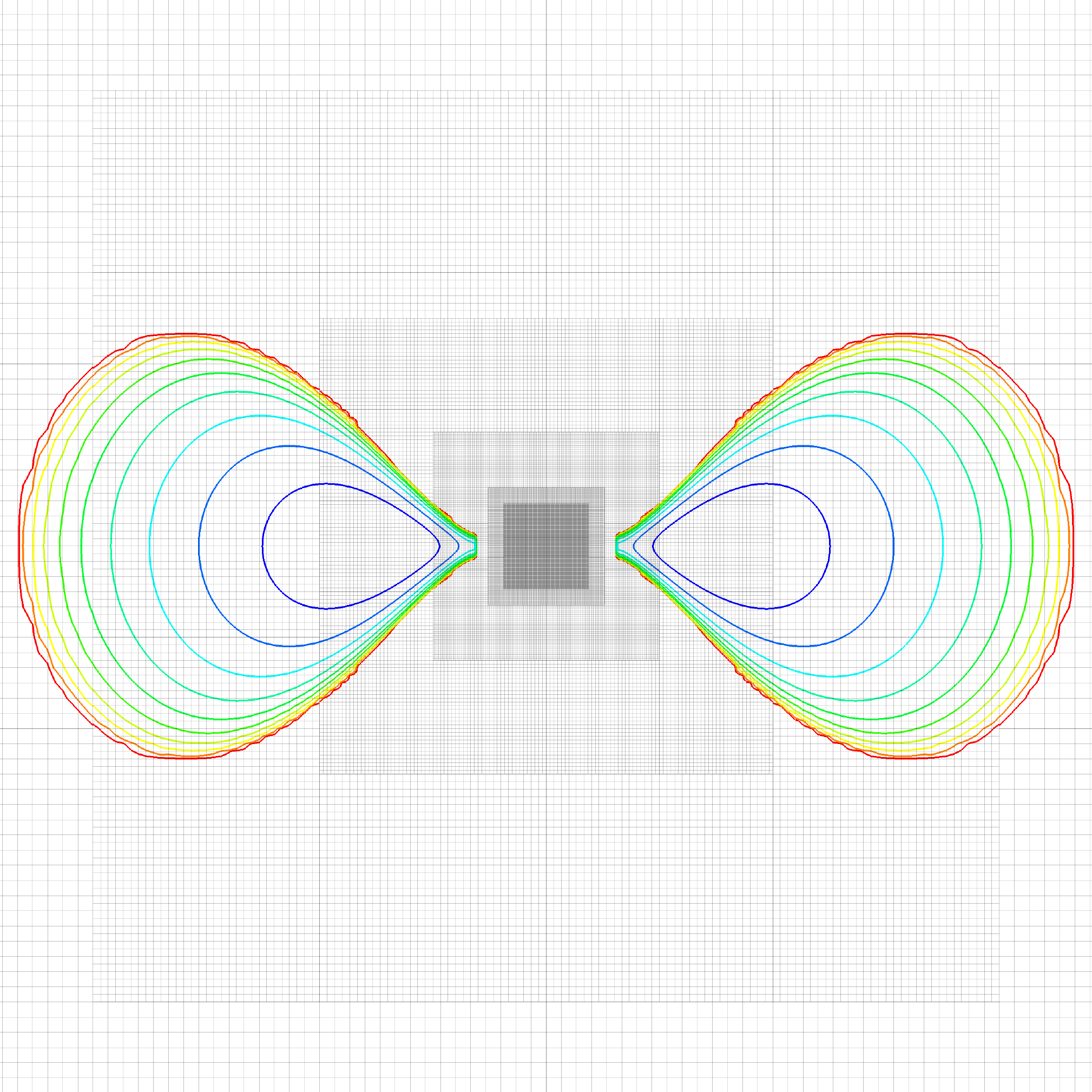}
  \caption{Contour plot in the $xz$ plane of the initial rest-mass
    density profile in the disk for model {\tt C1Ba0b0}. The innermost
    contour (blue line) corresponds to a density of $2.1\times
    10^{-6}$ while the outermost contour (red line) corresponds to
    $5.9\times 10^{-10}$ (code units).  The AMR grid structure,
    showing the 6 innermost mesh refinement levels (out of 13), is
    shown as well in the figure. The box has a size of $46$. 
    See main text for details on the actual resolution of the
    different grids.}
  \label{fig:rho_contour_initial}
\end{figure}
%

\subsection{Tilted Kerr spacetime in improved quasi-isotropic
  coordinates}

We set up a Kerr black hole tilted about the $x$-axis by an angle
$\beta_0$ in the improved quasi-isotropic coordinates proposed by
\cite{Liu2009}. In those coordinates, the radius of the 
horizon does not shrink to zero in the extreme Kerr limit, but
approaches $r = M/4 > 0$. The initial mass $M$ of the Kerr BH is
chosen to be equal to the value used in the ID calculation of the
self-gravitating torus. The spin parameter $a$ varies for different
runs. The list of 21 models we use in our investigation is summarized
in Table~\ref{table:models}.

We perform the rotation of the BH about the $x$-axis in the following
way: we first rotate by an angle $\beta_0$ the coordinates $x,y,z$ to
the tilted coordinates,
\begin{equation}
  \begin{pmatrix}
    \hat{x} \\
    \hat{y} \\
    \hat{z}
  \end{pmatrix}
  = \begin{pmatrix}
    1 & 0 & 0 \\
    0 & \cos\beta_0 & -\sin\beta_0 \\
    0 & \sin\beta_0  & \cos\beta_0 \\
  \end{pmatrix}
  \begin{pmatrix}
    {x} \\
    {y} \\
    {z}
  \end{pmatrix} \,.
\end{equation}
We then calculate the spatial metric $\gamma_{ij}$ and extrinsic
curvature $K_{ij}$ by performing coordinate transformations of the
respective expressions given in \cite{Liu2009} (in spherical polar
coordinates) to our tilted Cartesian coordinates $\hat{x}^{i}$. We set
the initial shift $\beta^i$ to $0$ and choose the following initial
lapse profile:
\begin{eqnarray}
  \alpha = \frac{2}{1+(1+\frac{M}{2 r})^4}\,.
\end{eqnarray}
We find that this initial lapse profile greatly helps with the
conservation of irreducible mass and spin during the initial gauge
transition the system undergoes. The bigger the difference between the
initially chosen lapse profile and the profile the system attains
after the transition to the puncture coordinates, the larger the
errors in irreducible mass and spin.  As all quantities have been set
up in the tilted coordinates $\hat{x}^{i}$, we also need to apply
rotations to vectors and tensors, in order to obtain the correct
components of the shift, spatial metric and extrinsic curvature in the
original $x^{i}$ coordinate system that is going to be used in the
simulations.  We obtain the components of the shift vector with
\begin{equation}
  \boldsymbol{\beta} = \mathbf{R} \,\boldsymbol{\hat{\beta}} \,,
\end{equation}
and the components of the spatial metric and extrinsic curvature with
\begin{eqnarray}
  \boldsymbol{\gamma} &=& \mathbf{R} \, \boldsymbol{\hat{\gamma}} \, \mathbf{R}^\mathsf{T} \,,
  \\
  \boldsymbol{K} &=& \mathbf{R} \, \boldsymbol{\hat{K}} \, \mathbf{R}^\mathsf{T} \,,
\end{eqnarray}
where $\mathbf{R}$ is the rotation matrix rotating from $\hat{x}^i$ to
$x^i$.
Our choice of the initial BH spin magnitude is currently 
limited by our computational method and is therefore somewhat 
smaller than spins expected in certain astrophysical scenarios. 
For example, as the simulations of misaligned BH-NS mergers
of~\cite{Foucart2011,Foucart2013,Kawaguchi2015} 
have shown,
accretion disks form in these systems only for relatively high ($>0.7$) 
initial BH spins; see also~\cite{Foucart2012} for disk mass
predictions from  BH-NS mergers. These initial spins in astrophysical 
merger scenarios are
larger than the spins investigated in our study  
$a \in [0,0.5]$. The reasons for our choice of smaller spins
are the very demanding resolution requirements when evolving 
highly spinning BHs, see, e.g.~\cite{Lousto2012}. Furthermore,
spin conservation seems to be affected non-linearly with 
higher spin magnitude~\cite{Marronetti2008}, which we have
also observed in~\cite{Mewes2015a}. 
The choice of astrophysically more realistic BH spin magnitudes would
have been prohibitively expensive (see, e.g.~\cite{Lousto2012} for details)
and is beyond the scope of
our study of self-gravitating accretion disks around tilted Kerr
BHs. Nevertheless, our results with dimensionless spin values of 0.5 
should provide a first qualitative understanding of what may 
happen at dimensionless spin values of 0.7.
We also note that the existence of the 'mass gap' in BH masses in
compact mergers has been questioned in~\cite{Kreidberg2012}.
While the peak of BH masses in these new considerations is
still around $8\, \mathrm{M}_{\odot}$, very small BH masses in mergers
are at least not ruled out as in previous studies. 
Furthermore, the BH mass predictions population synthesis models
are based on assumptions about the exact
details of the supernova explosion mechanism, see for 
instance~\cite{Belczynski2012},
where no mass gap is found using a delayed supernova explosion 
model assuming a 100-150 ms instability growth time.
As we explained in the introduction, the smaller the BH mass, 
the smaller the BH spin needs to be in order to tidally disrupt the 
NS during the merger in order to leave a thick accretion torus 
around the central BH. It therefore seems that some of the combination of 
BH spins and mass-ratios studied in this work (while having been 
primarily motivated on computational grounds) are a possible 
(although maybe not the most likely) outcome from BH-NS mergers.
Based on these considerations, model {\tt D2}, the lightest in our study,
is the most astrophysically relevant model of the three models studied.  
We choose the initial tilt angle $\beta_0 \in [5^{\circ},30^{\circ}]$. 
In the simulations of tilted BH-NS mergers of~\cite{Foucart2011}, the authors 
found that for initial inclinations $<40^{\circ}$,
the mass of the resulting accretion disk would be around $10-15\%$ of
the initial neutron star mass, whereas higher initial inclinations produced a
sharp drop in the mass of the resulting disk. For those those pre-merger
inclinations $<40^{\circ}$, the resulting tilt of the post-merger disk
was around $10^{\circ}$.  
In the recent simulations of tilted BH-NS mergers of~\cite{Kawaguchi2015},
the authors report a post-merger disk with a tilt $\approx
20-30^{\circ}$,
for an initial tilt of $\approx 60^{\circ}$.  
These findings seem to be in accordance with the probability
distributions of post-merger tilt angles for BH-NS mergers obtained 
in~\cite{Stone2013}, which show a sharp cut-off in the probability 
distribution for post-merger tilt angles $>50^{\circ}$.


\subsection{Numerical setup}
\label{subsection:num_setup}

The use of mesh-refinement techniques is of fundamental importance in our
simulations where different physical scales need to be accurately resolved.
For this reason, we use the \texttt{Carpet} driver that
implements a vertex-centered AMR scheme adopting
nested grids~\cite{Schnetter-etal-03b}. 

Figure \ref{fig:rho_contour_initial} shows isocontours of the initial
rest-mass density profile for a representative disk model ({\tt
  C1Ba0b0}) in a vertical cut ($xz$-plane) together with the initial
innermost AMR grid structure. In our simulations, we use 13 levels of
mesh refinement, with the outer boundary placed at $4096$. We perform
a large series of simulations using two different resolutions, high
resolution runs with a highest resolution of $\Delta x=0.02$ and
simulations with half this resolution, see
Table~\ref{table:models}. Every refinement level except the innermost
one is half the size of the preceding level. The innermost refinement
level, as can be seen in Fig.~\ref{fig:rho_contour_initial}, is more
than half the size of the next refinement level. All simulations have
been performed using a CFL factor of $0.25$. In the outermost $6$
refinement levels, we use the same time step as in level $7$, in order
to prevent instabilities in the spatially varying damping parameter
$\eta$. Using the same (smaller) time step in the outer levels is
necessary, even though the spatially varying $\eta$ parameter 
of~\cite{Alic2010} takes into account the time step limitation of the $\eta$ 
parameter described in~\cite{Schnetter2010a}. 
Details of the accuracy and convergence properties of our 
simulations are provided in Appendix~\ref{sec:convergence}. 
   
\section{Results}
\label{sec:results}

\subsection{Surface plots}

\begin{figure}
  \centering
  \includegraphics[scale=0.205]{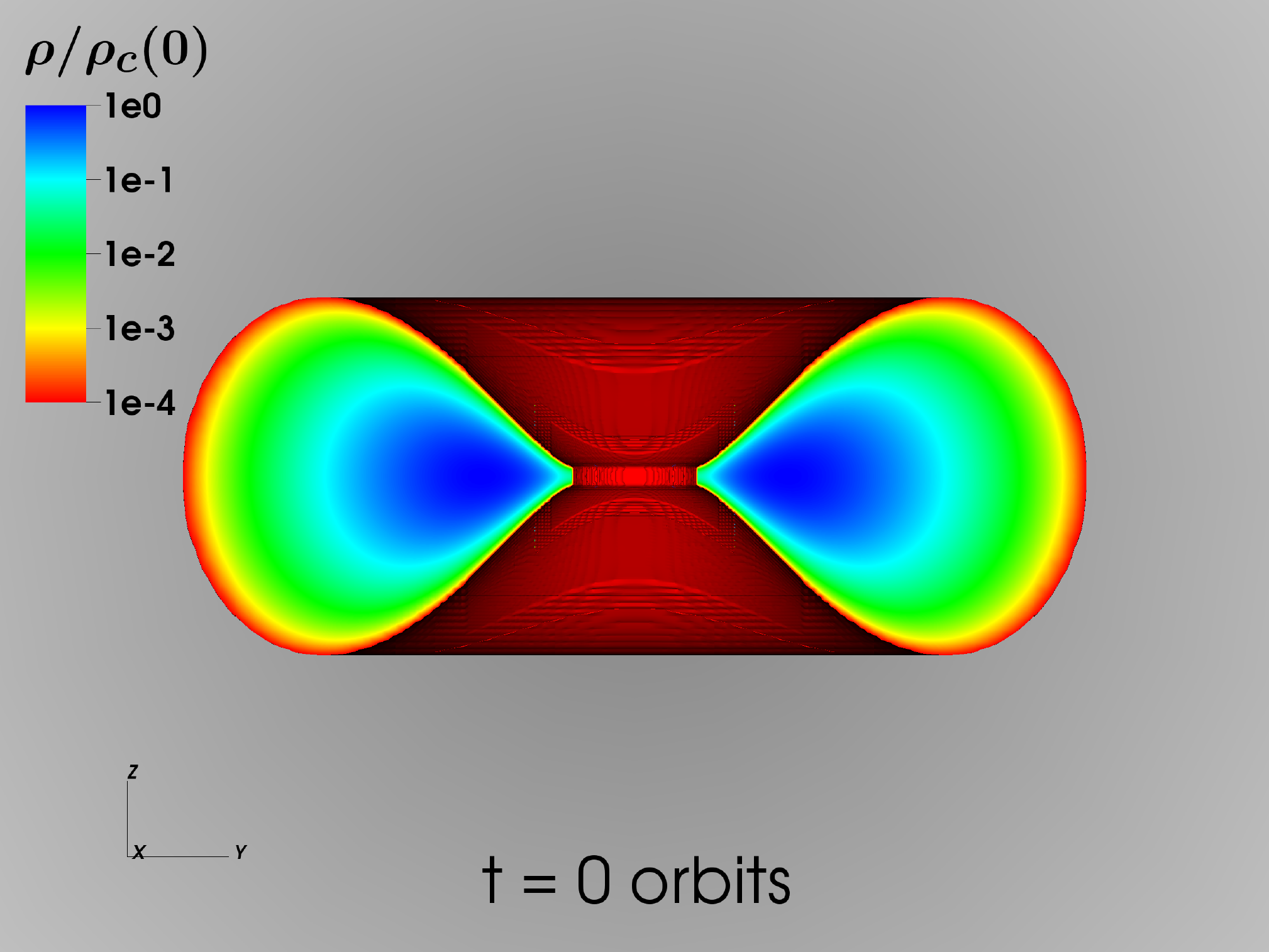}
  \includegraphics[scale=0.205]{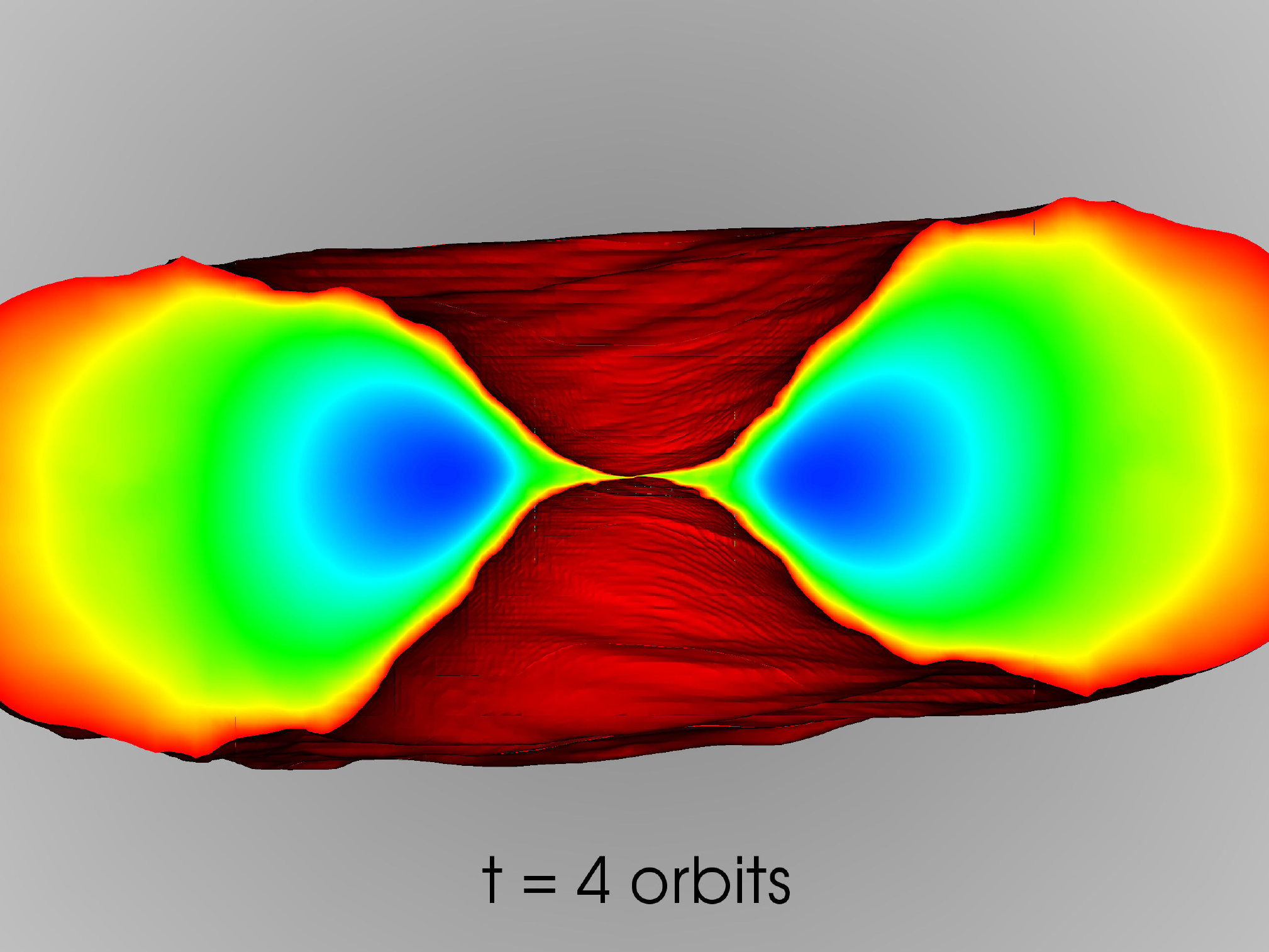}
  \includegraphics[scale=0.205]{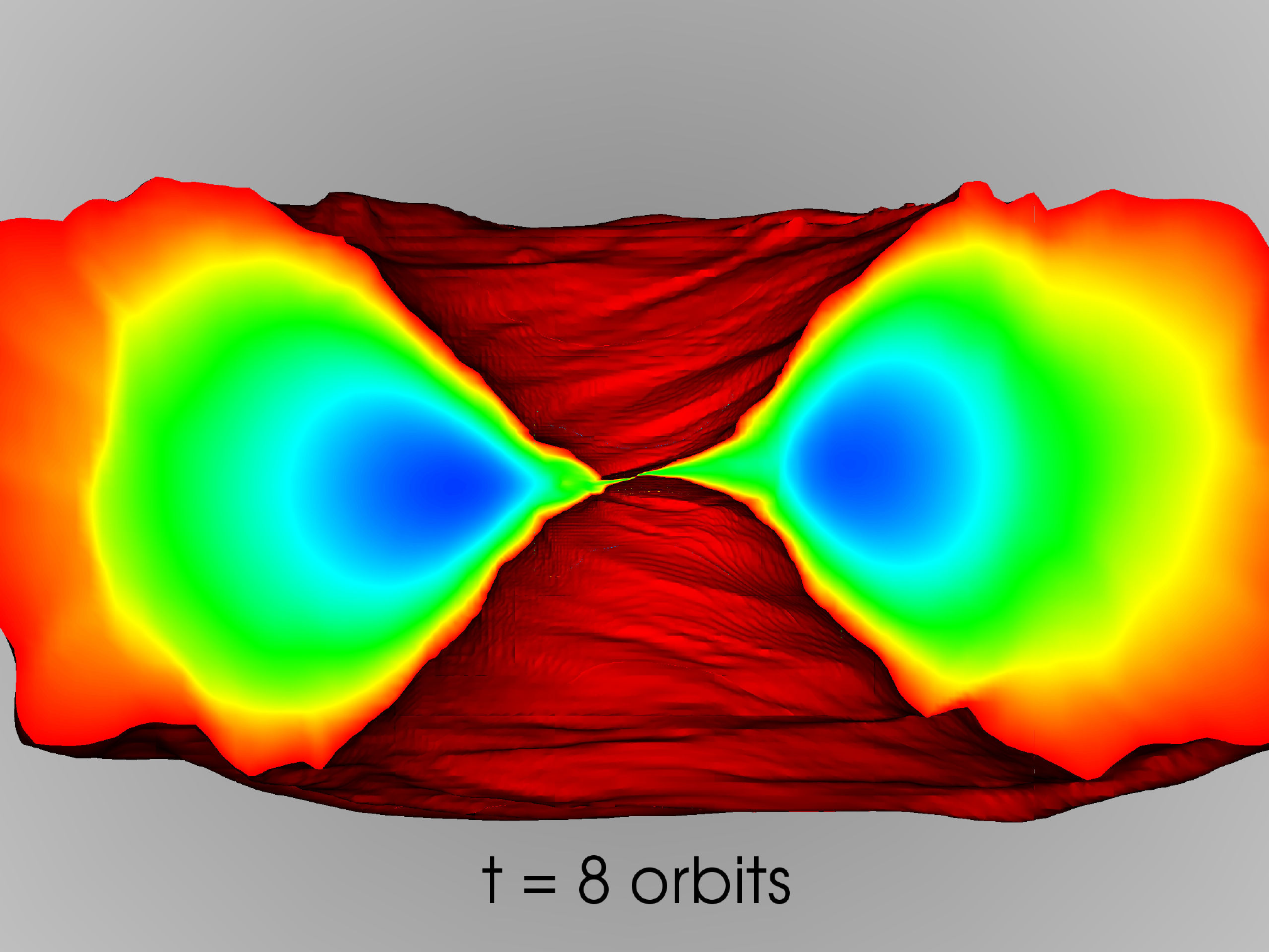}
  \includegraphics[scale=0.205]{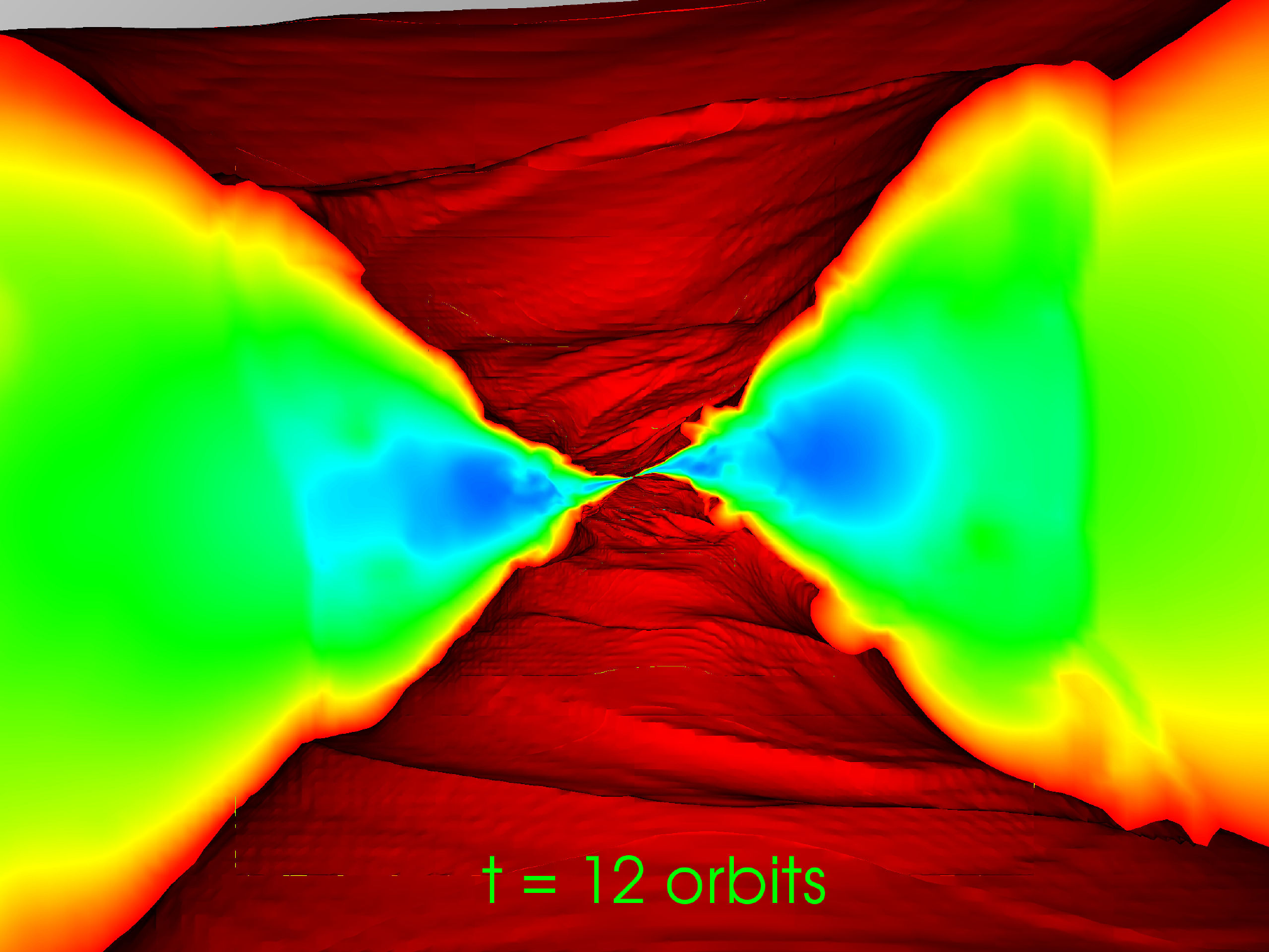}
  \includegraphics[scale=0.205]{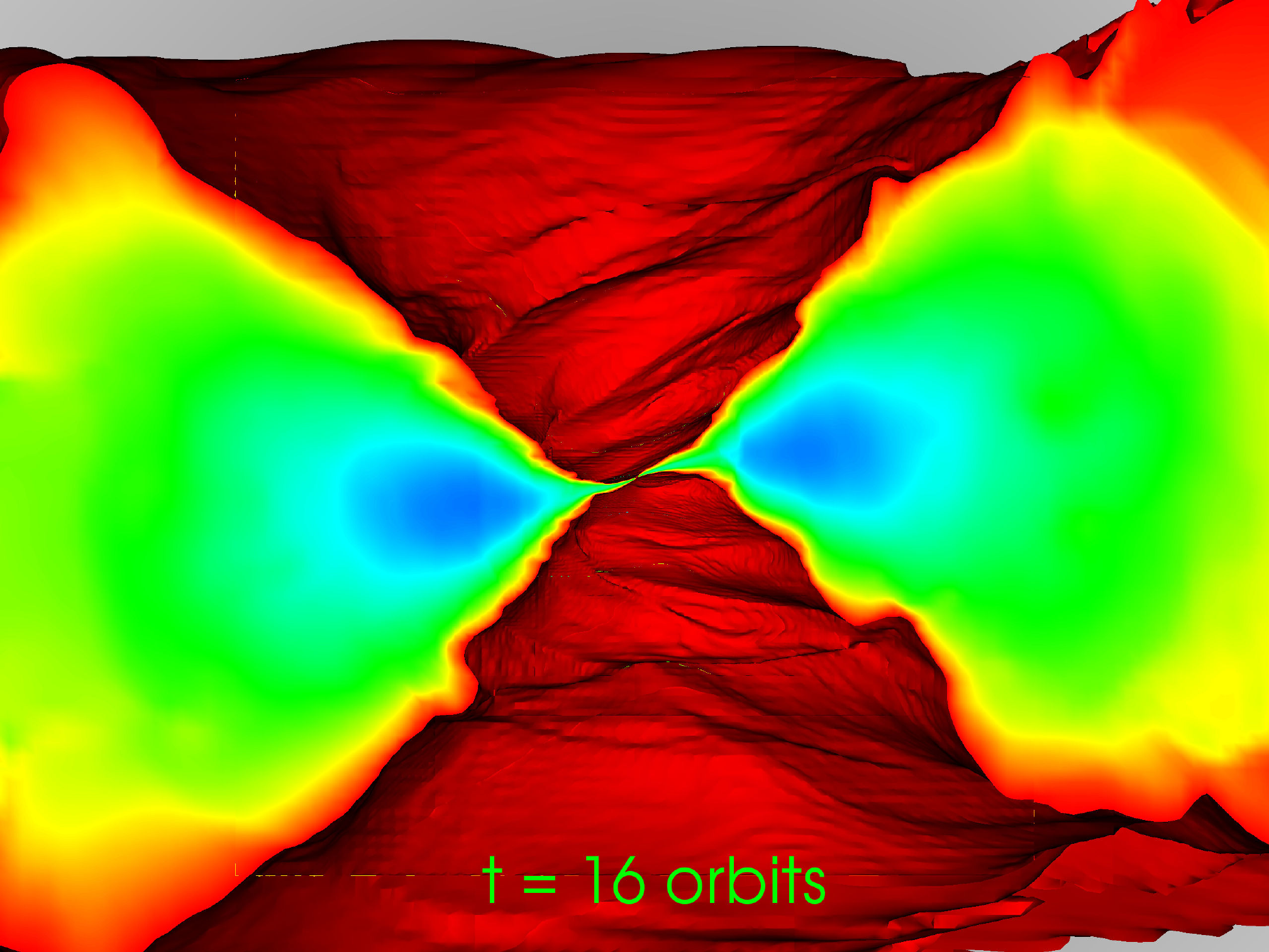}
  \includegraphics[scale=0.205]{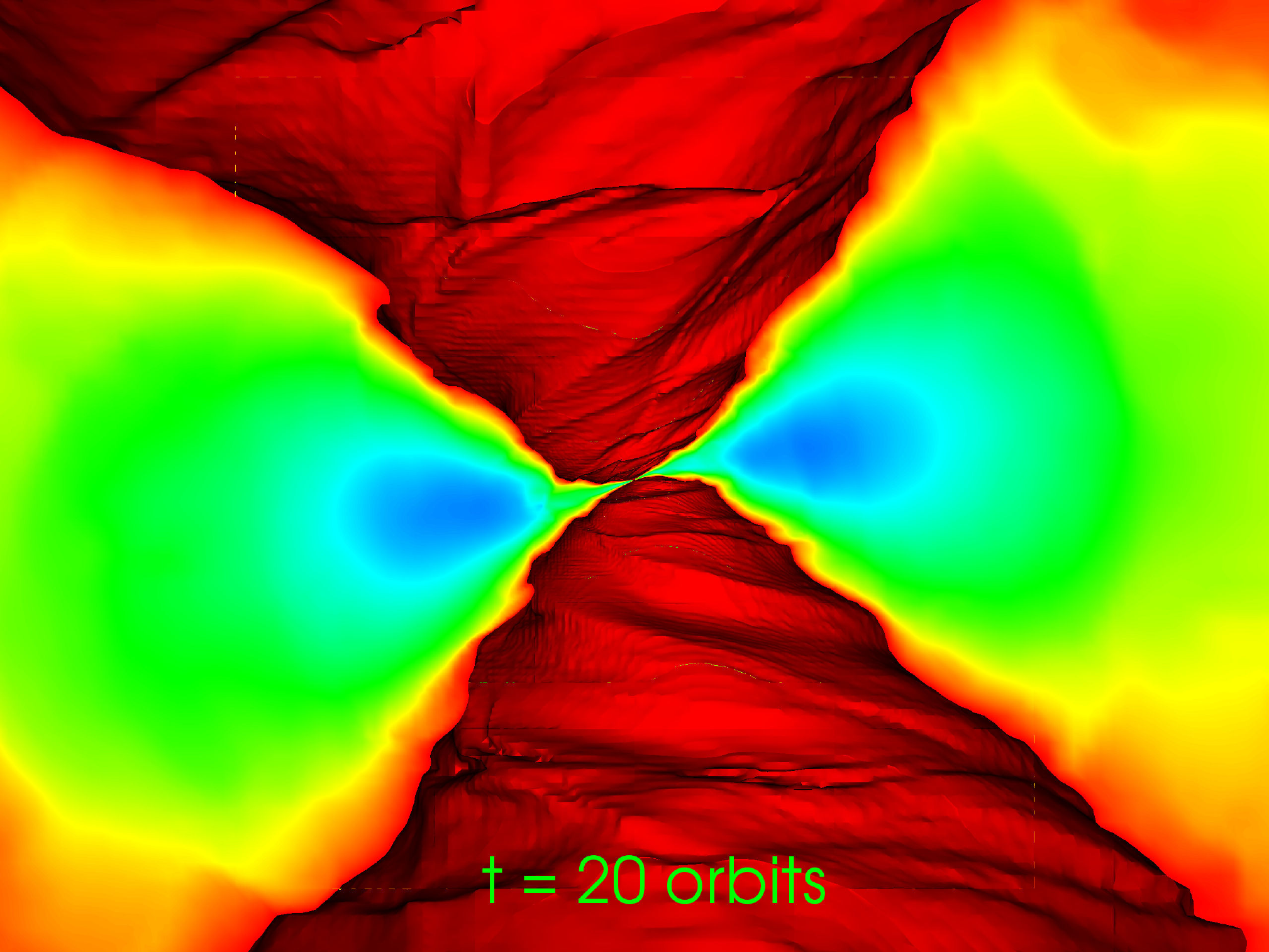}
  \caption{Surface plots of the logarithm of the rest-mass density for
    model {\tt C1Ba01b30} at six different snapshots of the evolution,
    $t/t_{\rm{orb}}=0, 4, 8, 12, 16$ and 20. The domain shown in all
    panels is $60$ across the $y$-axis and $45$ across
    the $z$-axis. The surface of the disk is set at a density of
    $10^{-4}\rho_c$ and half the disk is removed for visualization
    purposes. The BH lies at the center of each panel.  The evolution
    shows the development of two accretion streams on to the BH as the
    disk inclination increases during the evolution with respect to the initial 
    equatorial plane.}
  \label{fig:C1Ba01b30}
\end{figure}
   
We start describing the morphology and dynamics of a representative
model of our sample. The evolution of model {\tt C1Ba01b30} is shown
in Fig.~\ref{fig:C1Ba01b30}. This figure displays surface plots of the
logarithm of the rest-mass density at six different stages of the
evolution. The final panel shows the morphology of the system after
the disk has completed 20 orbits around the central BH.  The
interpolation of the ID from QI spherical polar coordinates to
Cartesian coordinates, along with the inclusion of a moderate BH spin
in the originally non-rotating BH, results in a significant
perturbation of the equilibrium model that triggers a phase of
oscillations of the torus around its equilibrium. These oscillations
are present throughout the simulation and, as the disk is initially
filling its Roche lobe, they induce a small accretion process of
matter through the cusp towards the BH. This does not reduce
significantly the total rest-mass of the torus as we show below.

The oscillations of the disk are damped due to numerical viscosity and
also by the formation of shock waves that propagate inside the disk
and convert kinetic energy into thermal energy.  These
outward-propagating shocks, also found in the simulations
of~\cite{Korobkin2011}, are produced by in-falling 
material along the funnel walls which reaches inner high-density regions 
and bounces back.
One such shock wave can be clearly seen on the right
portion of the disk in the 12 orbits panel of
Fig.~\ref{fig:C1Ba01b30}.
The initial tilt of the
BH spin ($\beta_0=30^{\circ}$) twists and warps the innermost parts of
the disk closer to the BH. This gradual effect becomes evident on the
time series shown in Fig.~\ref{fig:C1Ba01b30} and by the final 20
orbits the disk is significantly misaligned with the (horizontal)
$y$-axis.

\begin{figure*}
  \centering

   \includegraphics[scale=1.0]{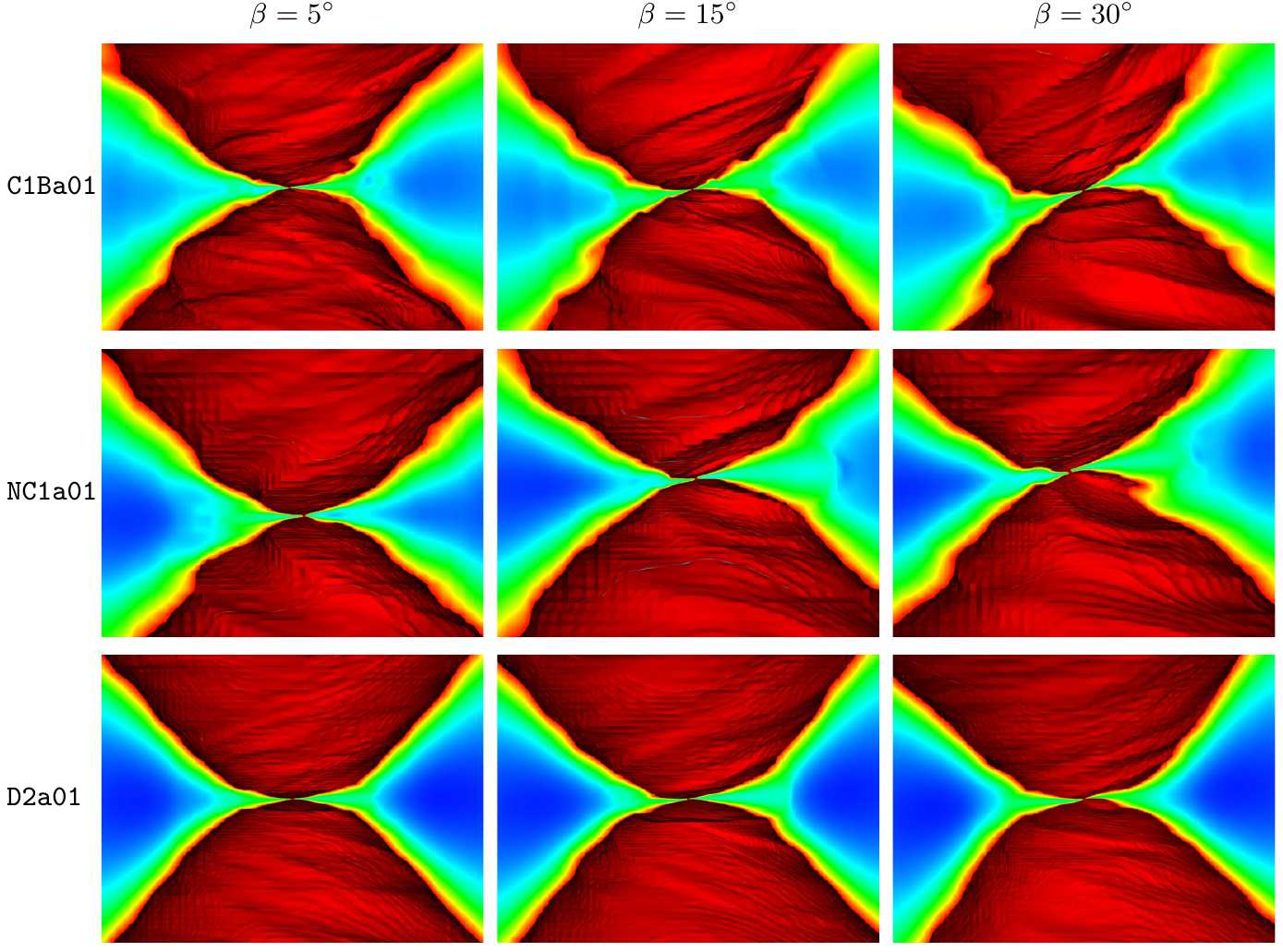}
   \caption{Surface plots of the (normalized) rest-mass density at
        the final time of the evolution $t/t_{\rm{orb}}=20$ for models
        {\tt C1B} with $a=0.1$ (top row), {\tt NC1} with $a=0.1$
        (middle row), and {\tt D2} with $a=0.1$ (bottom row). From left
        to right the columns correspond to initial tilt angles
        $\beta_0=5^{\circ}, 15^{\circ}$, and $30^{\circ}$,
        respectively.  The domain shown in all panels is $20$  
        across the $y$-axis and $15$ across the $z$-axis. The
        color palette and corresponding normalized density used are
        the same as in Fig.~\ref{fig:C1Ba01b30}.}
   \label{fig:surface-plot-comparison}

\end{figure*}

\begin{figure*}
 \centering
 
  \includegraphics[scale=1.0]{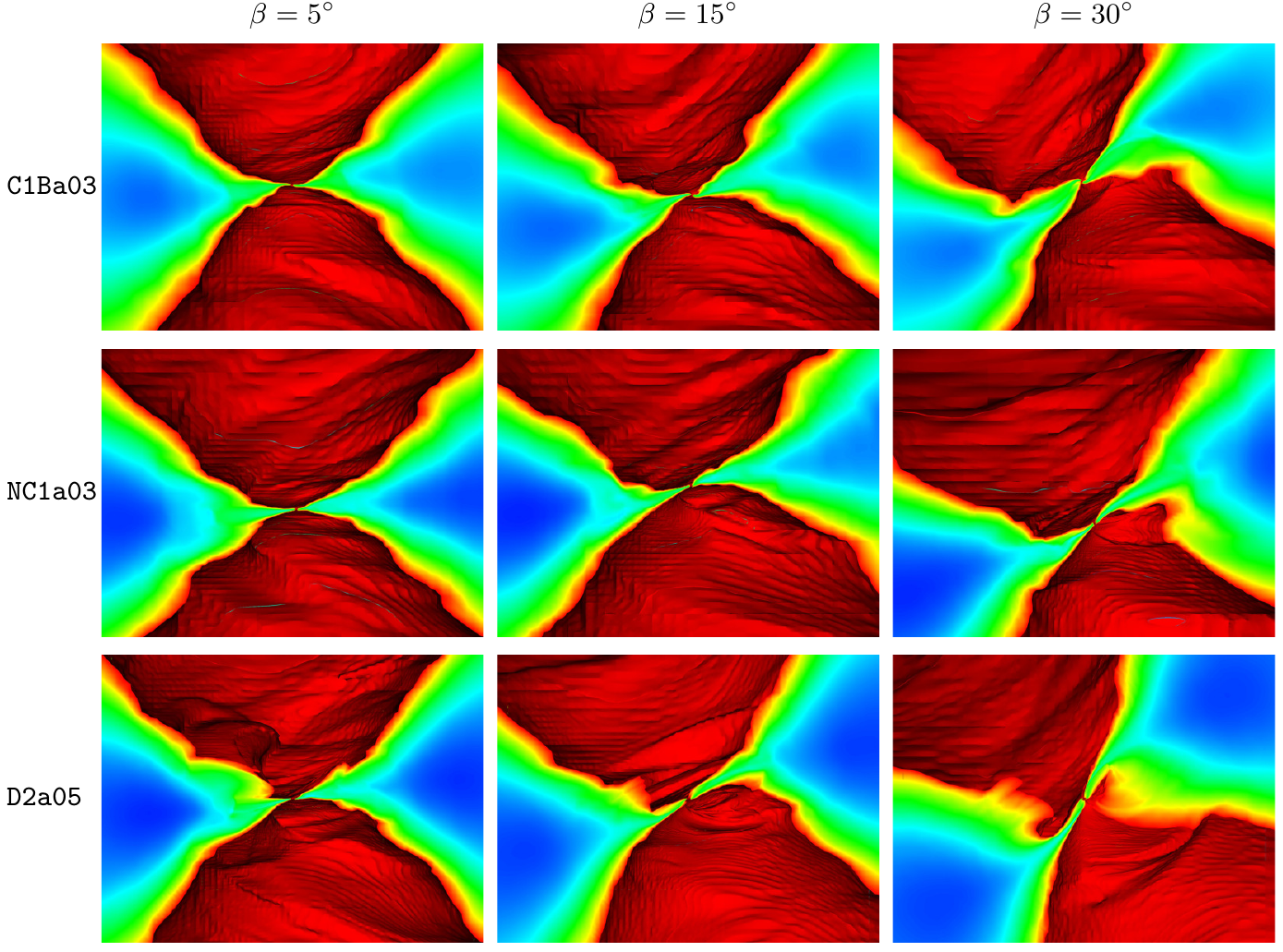}
   \caption{Surface plots of the (normalized) rest-mass density at
        the final time of the evolution $t/t_{\rm{orb}}=20$ for models
        {\tt C1B} with $a=0.3$ (top row), {\tt NC1} with $a=0.3$
        (middle row), and {\tt D2} with $a=0.5$ (bottom row). From left
        to right the columns correspond to initial tilt angles
        $\beta_0=5^{\circ}, 15^{\circ}$, and $30^{\circ}$,
        respectively.  The domain shown in all panels is $20$ 
        across the $y$-axis and $15$ across the $z$-axis. The
        color palette and corresponding normalized density used are
        the same as in Fig.~\ref{fig:C1Ba01b30}.}
   \label{fig:surface-plot-comparison_high_spin}
  
\end{figure*}

   In Fig.~\ref{fig:surface-plot-comparison} we show surface plots
of the rest-mass density for the models {\tt C1B}, {\tt NC1} and
{\tt D2} for the lower spin simulations ($a=0.1$) at the final time
of the evolution $t/t_{\rm{orb}}=20$. The effects of the initial BH
tilt on the morphology of the disk at the final time are more
clearly pronounced for the heavier models {\tt C1B} (top row) and
{\tt NC1} (middle row).  Although the spin of the BH is very low
and the spacetime is therefore nearly spherically symmetric, the
initial tilt angle nevertheless causes a different evolution.
In contrast, the significantly lighter model {\tt D2} is seen to be hardly
affected by the initial tilt by the end of the evolution, at
least for these simulations with low initial spin.
Similarly, in Fig.~\ref{fig:surface-plot-comparison_high_spin}, 
we plot the same rest-mass density surface plots as in 
Fig.~\ref{fig:surface-plot-comparison} for all our models with higher 
initial spin. Comparing the evolution of the disk morphology
with the lower initial spin figure, we see that there are now significant
changes in the morphology when increasing the initial tilt angle
$\beta_0$ for all models. The effects of evolving the disk in the tilted 
Kerr spacetime are now more pronounced, as the higher spin causes 
a significant deviation from spherical symmetry in the spacetime. 
The growth of non-axisymmetric modes associated with the PPI (see
below) in the two more massive models {\tt C1B} and {\tt NC1}
causes an alignment of the overall disk angular momentum 
vector with the BH spin. This interesting effect will be discussed in
Section~\ref{subsection:BH-precession} below, where we comment on
the BH precession and nutation properties.

\begin{figure*}

  \centering
  \includegraphics[scale=1.0]{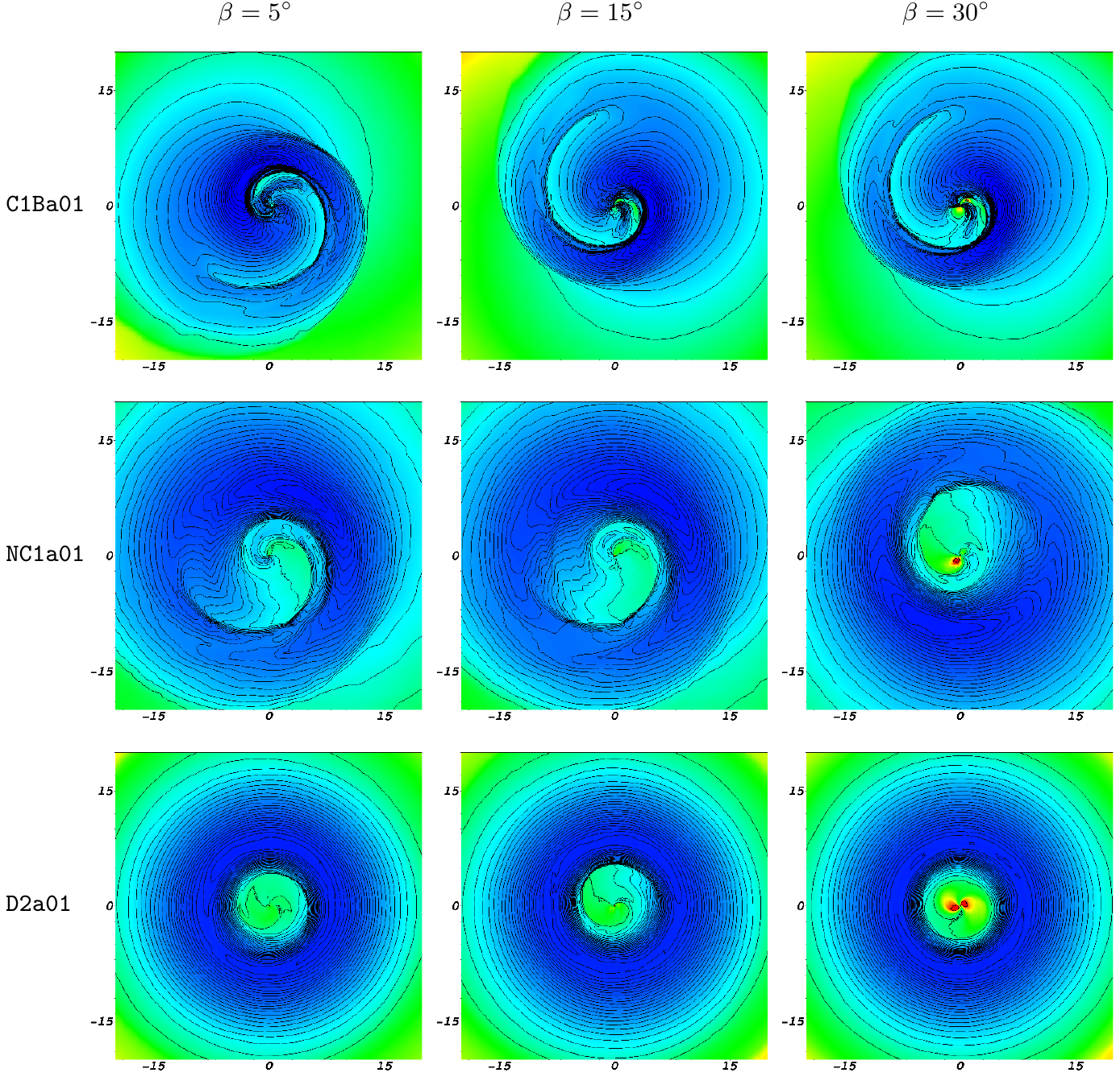}
  \caption{Linearly spaced isocontours in plots showing the
       logarithm of the rest-mass density in the plane perpendicular
        to the total angular momentum vector of the disk. From left to
        right the rows show models {\tt C1Ba01b5}, {\tt C1Ba01b15} and
        {\tt C1Ba01b30} (top), {\tt NC1a01b5}, {\tt NC1a01b15} and {\tt
          NC1a01b30} (middle), and {\tt D2a01b5}, {\tt D2a01b15}, and
        {\tt D2a01b30} (bottom). The color palette and corresponding
        normalized density used are the same as in
        Fig.~\ref{fig:C1Ba01b30}. The morphological structures shown in
        the different panels reflect the dominant PPI mode at the
        time of maximum non-axisymmetric mode amplitude. See main text for
        further details.}
   \label{fig:equatorial-plane-a01}
\end{figure*}

\begin{figure*}
  \centering
  \includegraphics[scale=1.0]{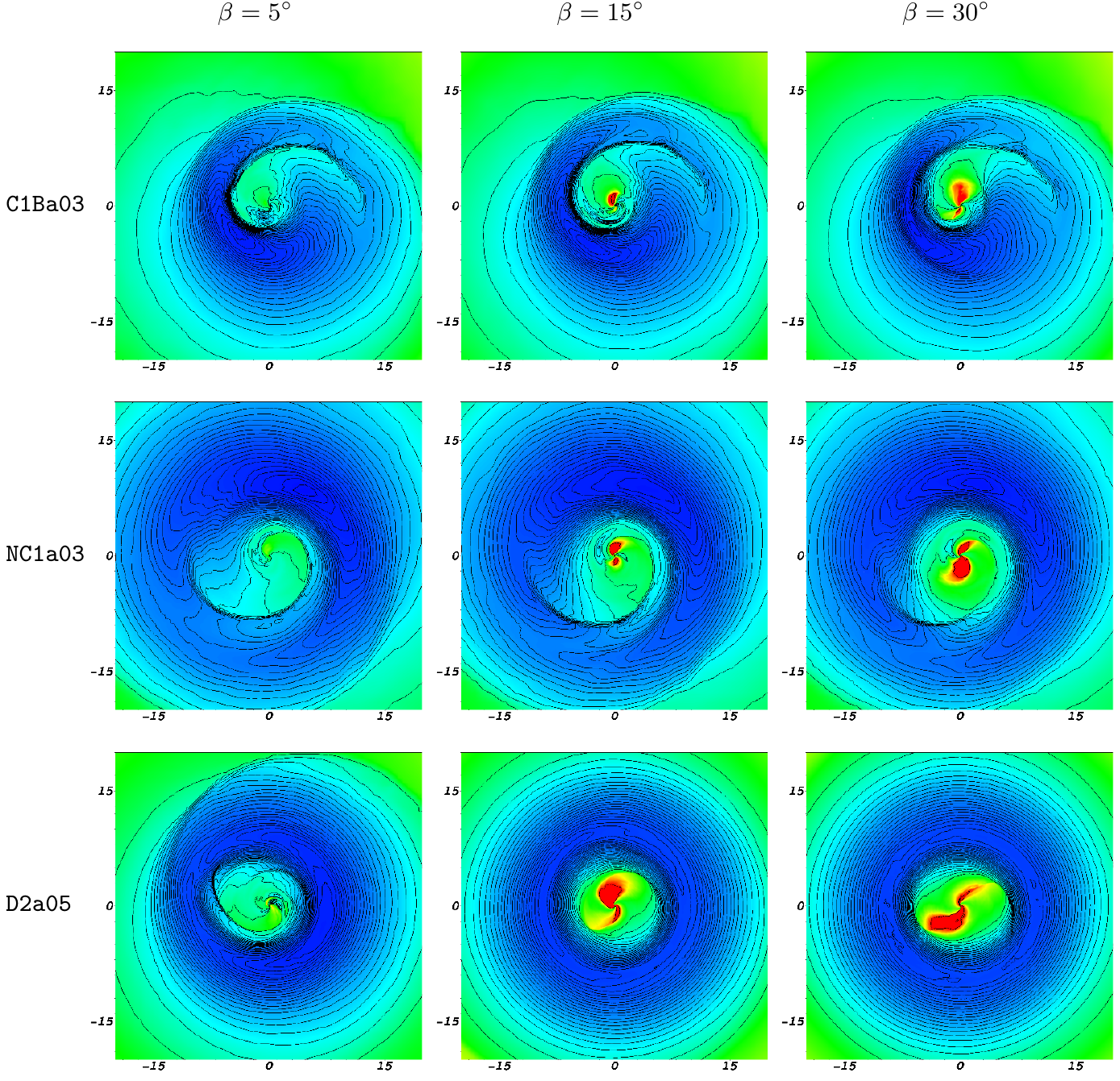}
   \caption{Linearly spaced isocontours in plots showing the
        logarithm of the rest-mass density in the plane perpendicular
        to the total angular momentum vector of the disk. From left to
        right the rows show models {\tt C1Ba03b5}, {\tt C1Ba03b15} and
        {\tt C1Ba03b30} (top), {\tt NC1a03b5}, {\tt NC1a03b15} and {\tt
          NC1a03b30} (middle), and {\tt D2a05b5}, {\tt D2a05b15}, and
        {\tt D2a05b30} (bottom). The color palette and corresponding
        normalized density used are the same as in
        Fig.~\ref{fig:C1Ba01b30}. The morphological structures shown in
        the different panels reflect the dominant PPI mode at the
        time of maximum non-axisymmetric mode amplitude. See main text for
        further details.}
      \label{fig:equatorial-plane-a03}  
\end{figure*}

   In Figures~\ref{fig:equatorial-plane-a01} and
   \ref{fig:equatorial-plane-a03} we plot linearly spaced isocontours
   and the logarithm of the rest-mass density in the plane
   perpendicular to the total angular momentum vector of the
   disk. This kind of ``equatorial" plots allows for a better
   visualization of the possible growth of non-axisymmetric structures
   in the disk. We note that due to the twisting and warping of the
   disk as a result of the initial BH tilt, the choice of the plane on
   to which project the isocontours for an adequate visualization is
   not straightforward and careless choices (as e.g.~the simple choice
   of the equatorial $xy$-plane) may hide important pieces of
   information on the dynamics and morphology. 
   The top row of Fig.~\ref{fig:equatorial-plane-a01} corresponds to models 
   {\tt C1Ba01b5} (left), {\tt C1Ba01b15} (middle), and {\tt C1Ba01b30}
   (right), and all of them are displayed at the time at which the
   growth of the fastest growing nonaxisymmetric structure in the disk
   saturates (see Section~\ref{subsection:PPI-modes} below where we
   discuss the mode growth of the PPI for the different models of our
   sample). The middle row of Fig.~\ref{fig:equatorial-plane-a01}
   corresponds to models {\tt NC1a01b5} (left), {\tt NC1a01b15}
   (middle), and {\tt NC1a01b30} (right), also shown at the time the
   fastest growing mode has maximum amplitude, while the bottom row
   shows models {\tt D2a01b5} (left), {\tt D2a01b15} (middle), and
   {\tt D2a01b30} (right), which are displayed at the final time of
   the evolution ($t=20t_{\rm{orb}}$) when the amplitudes of the
   corresponding PPI modes (if present) are largest. Correspondingly
   Fig.~\ref{fig:equatorial-plane-a03} shows the same type of
   isocontour plots as Fig.~\ref{fig:equatorial-plane-a01} but for the
   models with higher initial spin. The top row of
   Fig.~\ref{fig:equatorial-plane-a03} corresponds to models {\tt
     C1Ba03b5} (left), {\tt C1Ba03b15} (middle), and {\tt C1Ba03b30}
   (right), the middle row of Fig.~\ref{fig:equatorial-plane-a01}
   corresponds to models {\tt NC1a03b5} (left), {\tt NC1a03b15}
   (middle), and {\tt NC1a03b30} (right) and the bottom row shows
   models {\tt D2a05b5} (left), {\tt D2a05b15} (middle), and {\tt
     D2a05b30} (right). As in Fig.~\ref{fig:equatorial-plane-a01}, the
   times shown are at either saturation of the fastest growing
   non-axisymmetric structure or when the maximum mode amplitudes are
   largest.

   The inspection of the different panels displayed in
   Figs.~\ref{fig:equatorial-plane-a01} and
   \ref{fig:equatorial-plane-a03} shows noticeable differences in the
   flow morphology. On the one hand all six models {\tt C1B} (top rows
   in both figures), corresponding to initial BH spins of $a=0.1$ and
   $a=0.3$, respectively, exhibit a very prominent spiral arm once the
   stationary accretion phase is reached. This structure, which is
   visible for all three tilt angles considered, $\beta=5^{\circ},
   15^{\circ}$ and $30^{\circ}$, is associated with the growth of the
   $m=1$ non-axisymmetric PPI mode, which is the fastest-growing mode
   and has the largest amplitude (see also
   Fig.~\ref{fig:ppi-modes-a01b30}).  The middle rows, corresponding
   to the six models {\tt NC1}, show the development of a spiral arm
   as well, however not as clearly pronounced as in the case of {\tt
     C1B}.  On the other hand the isodensity contours of all three
   models {\tt D2} displayed in the bottom row of
   Fig. \ref{fig:equatorial-plane-a01} for a BH spin $a=0.1$ barely
   show any morphological deviation from the initial axisymmetry. 
   Only the innermost regions closest to the BH show some
   slight non-axisymmetric structure, particularly for the model with
   the largest tilt angle ($\beta=30^{\circ}$) shown on the right
   panel. Model {\tt D2} is considerable lighter than model {\tt C1B},
   and model {\tt NC1} has an initial non-constant specific angular
   momentum profile and according to~\cite{Kiuchi2011} more massive
   tori with $j=$ const profiles favor the appearance of the PPI with
   respect to less massive tori (see in particular Figs.~2 and 3(a)
   of~\cite{Kiuchi2011} for their model {\tt C1} which has very
   similar properties regarding size and BH-to-disk mass ratio to our
   model {\tt C1B}). Our results seem to only partially confirm those
   previous findings as there seems to be some weak dependence of a
   moderate $m=2$ PPI growth with increasing tilt angles.

   The dependence on the BH spin seems however to be more significant
   for the light model {\tt D2}, as shown in the panels at the bottom
   row of Figures~\ref{fig:equatorial-plane-a01} and
   \ref{fig:equatorial-plane-a03}. For the higher spin runs shown in
   Fig.~\ref{fig:equatorial-plane-a03}, all three panels show
   non-axisymmetric morphological features. All three snapshots
   correspond to the same final time of the evolution
   ($t=20t_{\rm{orb}}$) and the dynamical differences between the
   three models are only due to the initial BH tilt angle. For
   $\beta=5^{\circ}$ (left panel) the $m=1$ spiral structure is
   visible and dominant but, at the same time, the $m=2$ mode seems to
   have reached an almost similar amplitude (see the bottom panel of
   Fig.~\ref{fig:ppi-modes-a01b30}). Such $m=2$ feature becomes
   clearly dominant the larger the BH inclination becomes, as depicted
   in the central regions of the middle and right panels. 

   \subsection{Maximum rest-mass density evolution}

\begin{figure}
  \centering
  \includegraphics[scale=1.0]{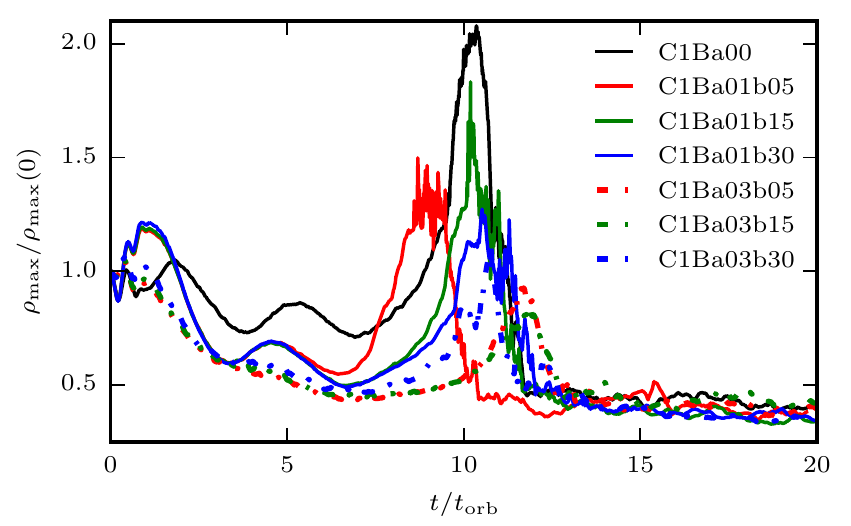}
  \\
  \vspace{-0.57cm}
  \includegraphics[scale=1.0]{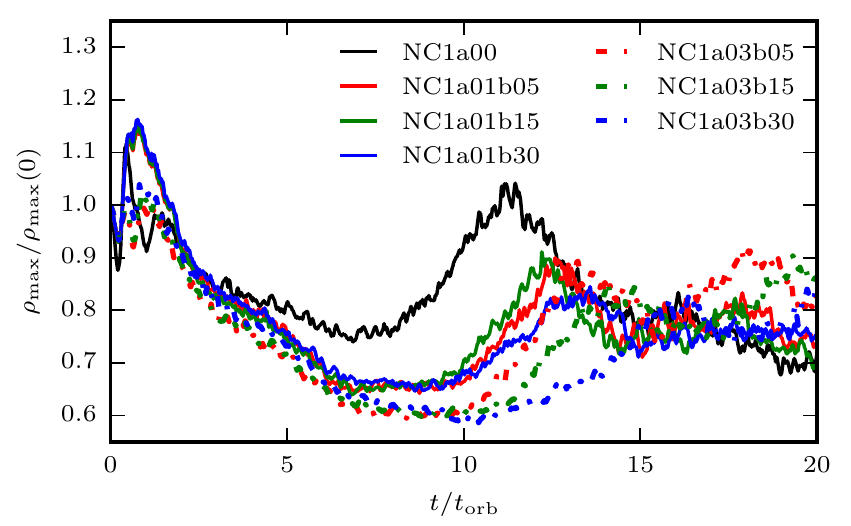}
  \\
  \vspace{-0.57cm}
  \includegraphics[scale=1.0]{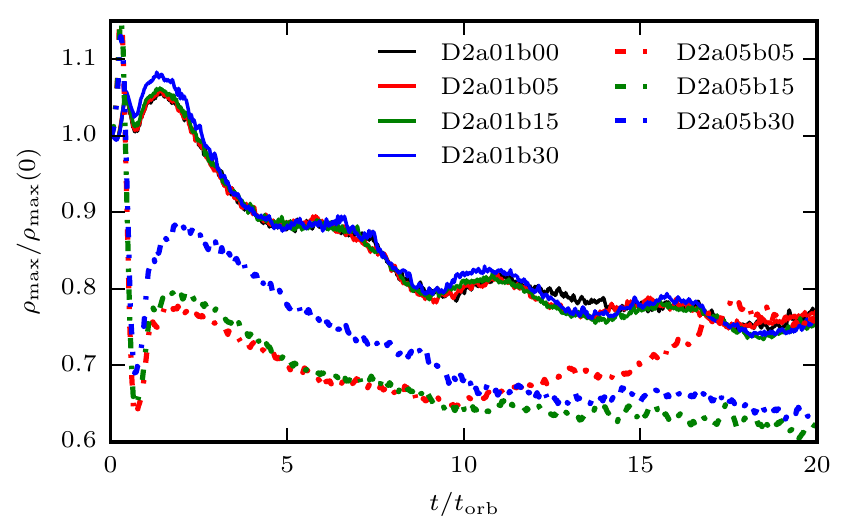}
  \caption{Evolution of the maximum rest-mass density, normalized by
    its initial value, for models {\tt C1B} (top), {\tt NC1} (middle)
    and {\tt D2} (bottom). The dashed curves correspond to the high
    spin models of our sample.}
  \label{fig:rho-max}
\end{figure}

   We next show the evolution of the maximum rest-mass density in the
   disks in Fig.~\ref{fig:rho-max}, normalized to the corresponding
   initial central density of each model (see Table~\ref{table:models}).   
   For model {\tt D2} (bottom panel), the effect of tilting the BH has
   very little effect on the evolution of $\rho_{max}$ for a BH spin
   of $a=0.1$. However, for $a=0.5$, the evolution of $\rho_{max}$
   differs for the different initial tilt angles. Model {\tt D2a05b5}
   exhibits an upward drift from $10$ orbits onwards to bring
   $\rho_{max}$ to the level of the lower spin runs towards the end of
   the simulation. This might be connected to the growth of the m=1
   non-axisymmetric mode in this model, as discussed in the previous
   section.
   In the initial stages of the
   evolution, $\rho_{max}$ is higher for larger tilt angles, a
   dependence that is also observed in the accretion rate (see
   Section~\ref{subsection:mdot}). The maximum rest-mass density stays
   below its initial value for almost the entire evolution for both
   models {\tt D2} and {\tt NC1} (middle panel). This is very
   different in model {\tt C1B} (top panel) where the occurrence and
   saturation of the PPI has a drastic effect on the evolution of
   $\rho_{max}$.  When the PPI starts growing, a situation which is
   accompanied by an exponential growth of the accretion rate,
   $\rho_{max}$ grows to up to twice its initial value, reaching the
   highest value for the untilted model {\tt C1Ba00}. It
   subsequently settles to a value that is very similar for all
   initial spins and tilt angles, but the time of growth and the shape
   of the ``lump" in the density evolution are different for the
   different spins and initial tilt angles. Model {\tt NC1} shows
   features from both models {\tt D2} and {\tt C1B} described above.
   Similarly to model {\tt D2}, $\rho_{max}$ stays below its initial
   value for almost the entire evolution, with model {\tt NC1a00}
   being an exception where it grows above the initial value
   briefly. There is, as in models {\tt C1B}, a rapid growth in the
   central density, albeit somewhat smaller for the non-constant
   angular momentum disks, associated with the growth of the PPI in
   all the models. The magnitudes at which the mode growth saturates
   are similar, while the time at which the growth sets in depends on
   the initial spin and tilt angle.

\begin{figure}
  \centering
  \includegraphics[scale=1.0]{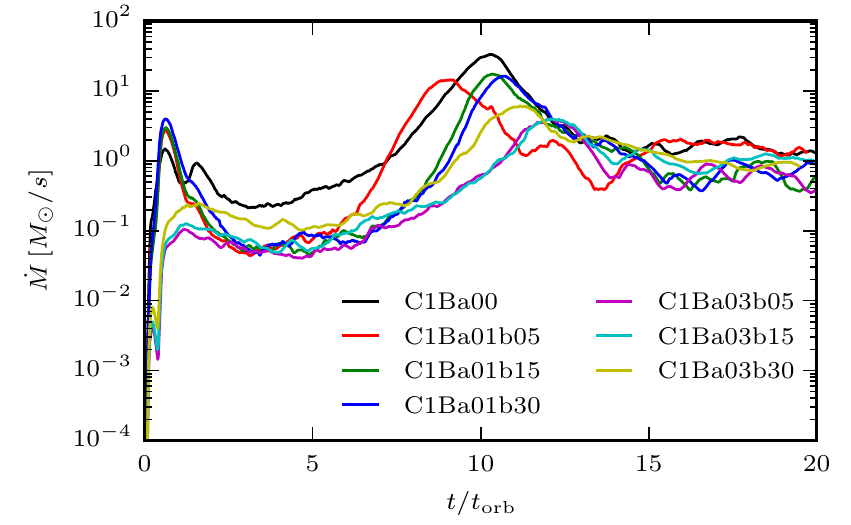}
  \\
  \vspace{-0.57cm}
  \includegraphics[scale=1.0]{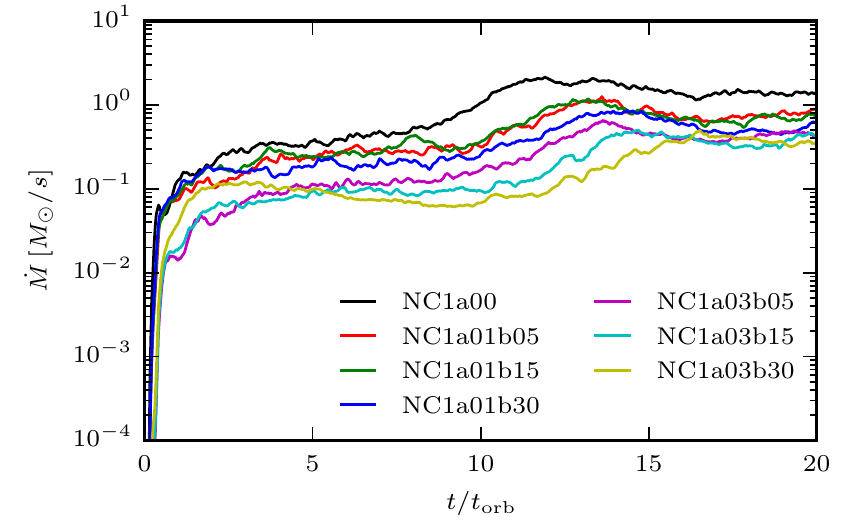}
  \\
  \vspace{-0.57cm}
  \includegraphics[scale=1.0]{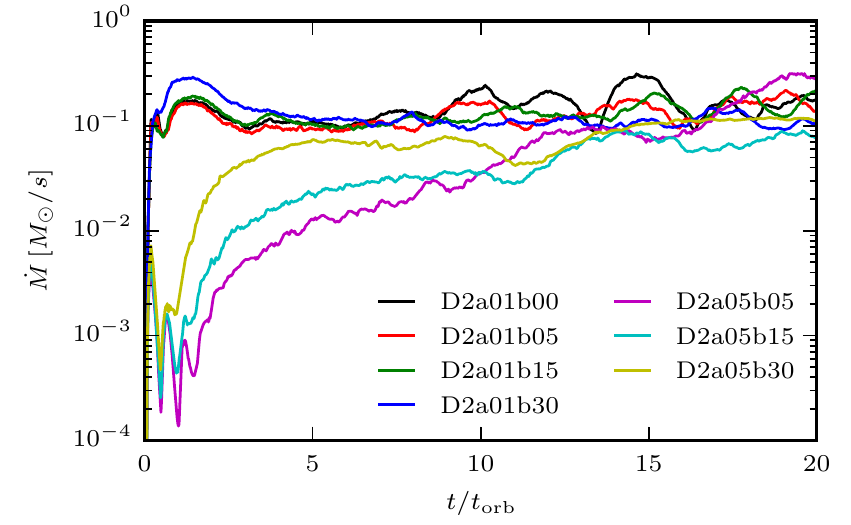}
  \caption{Evolution of the accretion rate in $M_{\odot}/s$ 
  for models {\tt C1B} (top), {\tt NC1} (middle) and {\tt D2}
    (bottom).}
  \label{fig:mdot}
\end{figure}

\subsection{Evolution of the accretion rate}
\label{subsection:mdot}
Fig.~\ref{fig:mdot} shows the 
time evolution of the mass accretion
rate for all models of our sample, in units of $\mathrm{M}_\odot/s$.
The colors used for the different models as well as the line style
follow the same convention as in the rest-mass density evolution
figure (Fig.~\ref{fig:rho-max}).  The top panel displays the mass flux
for models {\tt C1B}, the middle panel for models {\tt NC1}, and the
bottom panel for models {\tt D2}. The mass flux is computed as the
instantaneous flux of matter across the AH surface which is parameterized 
in spherical coordinates by the polar angle $\theta$ and  azimuthal angle  
$\phi$ and is given by:
\begin{eqnarray}
  \dot{M}= 2\pi r^2 \int_0^{2\pi} \int_0^{\pi} D \, v^r \, \sin \theta \, d\phi \, d\theta \,,
\end{eqnarray}
where $r$ is the radius of the AH and $v^r$ is the radial velocity 
of the fluid crossing the sphere. All three panels of Fig.~\ref{fig:mdot} show 
that after a transient initial phase the mass accretion rate is seen to tend 
asymptotically to a fairly constant value. 

In the previous fixed spacetime simulations of~\cite{Fragile2005} it
was found that the tilt and the twist of the disks are not strongly
responsive to $\dot{M}$. Our results for a dynamical spacetime
validate those earlier findings, as the final dispersion found in
$\dot{M}$ once this quantity reaches steady-state values is relatively
small irrespective of the BH spin and inclination angles
($\dot{M}\,[\mathrm{M}_\odot/s]\,\in[0.2,1.5]$
for models {\tt C1B} and {\tt NC1}, and $[0.06,0.2]$ for model
{\tt D2}). However, the transition to the
steady-state does seem to show a clear dependence on the BH spin. We
find that the initial $\dot{M}$ values are consistently smaller the
larger the BH spin, about 2 orders of magnitude for models {\tt D2}
after $\sim$1 orbit and about half that value for the more massive
disks {\tt C1B} and {\tt NC1}. Another trend, which is most clearly
seen in model {\tt D2}, is the initial dependence of the accretion
rate on the tilt angle as well. We find that the higher the initial
tilt angle, the higher the accretion rate in the early stages of the
evolution. This is because the innermost stable circular orbit (ISCO)
of a rotating Kerr BH has an ellipsoidal shape, attaining the largest
size at the equatorial plane of the BH. For prograde orbits such as
those in our work, the higher the BHs spin, the smaller the size of
the ISCO and the closer can the matter orbit around the BH before
falling in. When tilted, this centrifugal barrier becomes however
less and less effective and the accretion rate becomes therefore
higher for larger inclinations.

\begin{figure}
  \centering
  \includegraphics[scale=1.0]{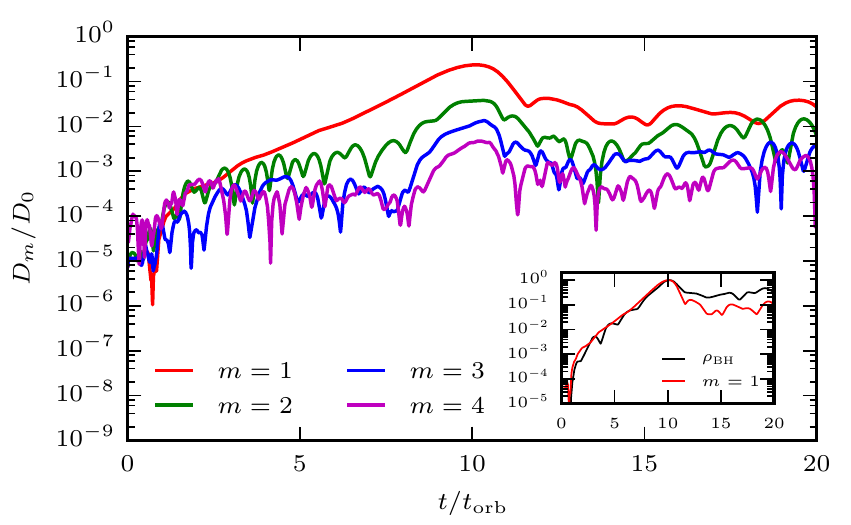}
  \\
  \vspace{-0.57cm}
  \includegraphics[scale=1.0]{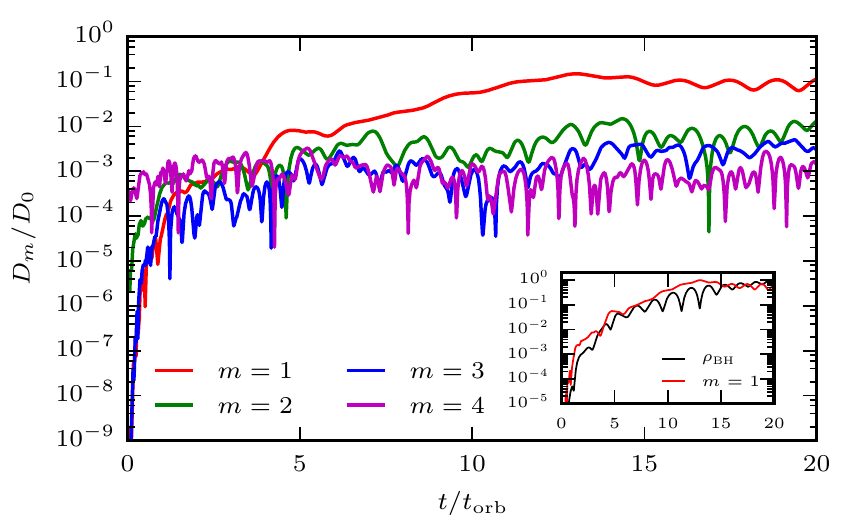}
  \\
  \vspace{-0.57cm}
  \includegraphics[scale=1.0]{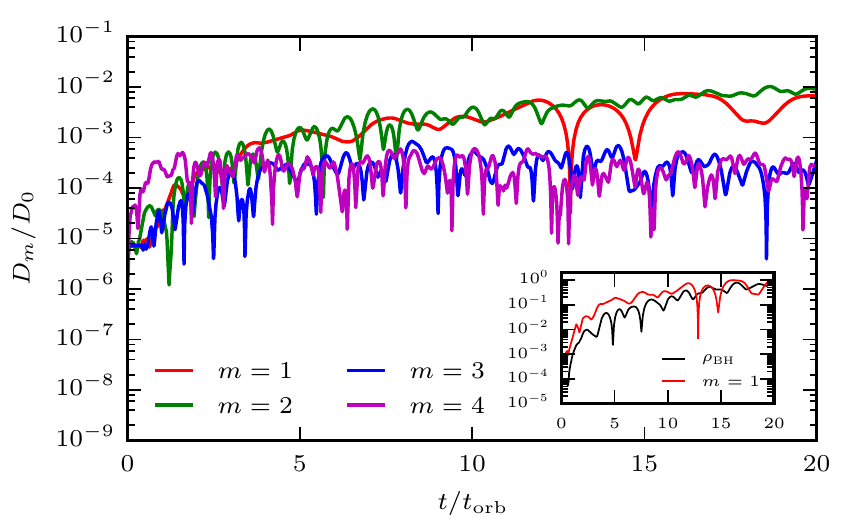}
  \caption{Evolution of the first four non-axisymmetric (PPI) modes
   scaled by the total rest-mass density ($D_0$) for models {\tt C1Ba01b30} (top), {\tt
      NC1a01b30} (middle) and {\tt D2a01b30} (bottom). The inset shows
    the evolution of the $m=1$ mode and the polar distance of the
    centre of the BH from the origin, both having been rescaled with
    their respective maximum values attained during the evolution.}
  \label{fig:ppi-modes-a01b30}
\end{figure}

A common feature present in the mass-flux plots of the simulations
of~\cite{Fragile2005} was a late-time ``bump" which the authors
attributed to the reflection of an outgoing wave within the disk at
the outer edge. That feature does not seem to be so apparent in our
models apart from maybe model {\tt C1B} which shows a bump around
$t\sim10t_{\rm{orb}}$ somewhat similar to that reported
by~\cite{Fragile2005}. However, in our interpretation, the phase of
rapid growth in $\dot{M}$ in less than about 5 orbits seems to have a
different origin. This conclusion is reached by comparing the time at
which the growth of the PPI saturates (discussed in
Figs.~\ref{fig:ppi-modes-a01b30} and~\ref{fig:ppi-modes} in the
following section) with the saturation time of the accretion rate for
model {\tt C1B}, which yields a significant close agreement. We note
that this peak in the accretion rate is the highest of all our models,   
$\sim 30 \rm{M}_{\odot}/\rm{s}$, lasting for less than about 1 ms.
Therefore, we identify the exponential growth and subsequent
saturation of the mass accretion rate with the growth and saturation
of the PPI in our models.

The total rest mass accreted during the evolution ranges within 
$M_{\mathrm{acc}}\,\in[0.022,0.033]$, $[0.005,0.013]$ and
$[0.0006,0.002]$ for models {\tt C1B}, {\tt NC1} and {\tt D2},
respectively. The accretion is facilitated by 
the outward transport of angular momentum, which we describe 
in section~\ref{subsection:ang_mom_transport} below. The 
growth of the PPI, to which we attribute the high peak accretion rates, is very
effective in transporting angular momentum outwards. 
While we have performed our study without considering magnetic 
fields, it is known that magnetic fields are very effective at transporting 
angular momentum in accretion disks~\cite{Balbus1991},
via the development of the so-called magnetorotational 
instability~\cite{Velikhov1959,Chandrasekhar1960}. Not taking into
account the effects of magnetic fields in accretion disks seems to 
underestimate the accretion rate~\cite{Gold2014}, see also recent 
GRMHD simulations of BH-NS mergers~\cite{Paschalidis2015,Kiuchi2015}
and the accretion rates reported therein.

\subsection{PPI growth in the disk and BH movement}
\label{subsection:PPI-modes}
In their seminal paper on the 
PPI~\cite{Papaloizou1984}, the authors showed that vertically thick, 
radially slender tori 
with constant specific angular momentum profiles are unstable against
the development of global non-axisymmetric modes. 
Two of our models, {\tt C1B} and 
{\tt D2}, have constant specific angular momentum profiles, while 
model {\tt NC1} has a radial dependence of $l \sim R^q$, where
$q=0.11$. 
In~\cite{Zurek1986}, the authors showed that tori with non-constant
angular momentum profiles might become PP-unstable as well.
In order to check for the onset and the growth of non-axisymmetric 
instabilities, we monitor the mode amplitudes computed using Eq. (36),
which is basically a Fourier decomposition of the azimuthal
distribution of the density in the disk~\cite{Zurek1986}.
In simulations of accretion tori susceptible to the development of the PPI 
performed on spherical grids, it is customary to induce the instability with small 
initial non-axisymmetric density or velocity perturbations.
As outlined 
{\it any} non-axisymmetric perturbation will trigger the growth of the 
instability provided the torus is not stable against it. We do not 
actively seed the PPI in any of our models. 
Instead, there are two main sources
of perturbations in our simulations, both connected to the fact that 
we evolve the BH-disk system in a Cartesian grid. First, the 
exact axisymmetry of the initial data is lost when we interpolate
the initial data onto the Cartesian grid of the evolution, introducing
small, non-axisymmetric perturbations in all interpolated quantities. This
is due to the fact that in Cartesian grids neighboring cells lie on
concentric shells only in the limit of infinite resolution. 
The second
source of perturbation is the so-called junk-radiation leaving the BH 
while the gauge evolves from the values specified by the initial data to
the puncture gauge attained during the evolution. Here, we need to
distinguish between models evolved around Schwarzschild BHs 
and the models evolved around tilted Kerr BHs. For the former, 
the junk-radiation leaving the BH initially should
be axisymmetric. Indeed, when analyzing this initial burst of
radiation with the Weyl scalar $\Psi_4$, we find that the 
amplitudes of the $\Psi^{l,0}_4$ multipoles (which are axisymmetric) 
are the largest. Nevertheless, the initial junk-radiation has also 
non-vanishing amplitudes in the other $\Psi^{l,m}_4$ multipoles.
Again, the reason for the presence of these non-axisymmetric 
$\Psi^{l,m}_4$ multipoles seems to be connected to the evolution 
in the Cartesian grid, and specifically to the existence of mesh
refinement boundaries, which are \emph{not} axisymmetric even
in the continuum limit (see~\cite{Zlochower2012} for interference 
patterns produced by the radiation crossing mesh refinement 
boundaries). Finally, in the case of the disk being around a tilted 
Kerr BH, the junk-radiation reaching the disk is now truly non-axisymmetric even 
without taking into account the effects of the Cartesian grid and the mesh refinement
boundaries, as the axis of rotation is tilted with 
respect to the $z=0$ plane.  


We show in Fig.~\ref{fig:ppi-modes-a01b30} the evolution of the
amplitude of the first four non-axisymmetric modes $D_m \, (m=1-4)$ in
the disk for models {\tt C1Ba01b30} (top), {\tt NC1a01b30} (middle)
and {\tt D2a01b30} (bottom). Those amplitudes are computed using
Eq.~(\ref{eq:mode-calculation}).  In addition, in the inset of all three plots we
also show as a black solid curve the evolution of the polar distance
of the center of the BH from the origin of the computational grid,
which we label $\rho_{\mathrm{BH}}$.

Model {\tt C1Ba01b30} is the one that most clearly is found to develop
the PPI, as signaled by the exponential growth of the $m=1$ mode.
Modes $m=2-4$ also show a phase of exponential growth lasting for
about 2 orbital periods (between $t\sim 8-10 \, t_{\rm{orb}}$) but the
saturation amplitudes reached are 1 to 2 orders of magnitude smaller
than for $m=1$.  As mentioned in the previous section, the growth of
the PPI is accompanied by an exponential growth of the mass accretion
rate (see Fig.~\ref{fig:mdot}). For this model the PPI clearly
saturates at around $t\sim 10\,t_{\mathrm{orb}}$, when the amplitude
of the $m=1$ mode quickly drops by an order of magnitude. After that,
the amplitude of all the modes we have analyzed remains rather
constant. The strength of the $m=2-4$ modes becomes similar at late
times while the $m=1$ mode still retains a significantly larger
amplitude.

The polar distance of the BH from the origin for model {\tt C1Ba01b30}
closely follows the behavior of the $m=1$ fastest growing PPI mode.
This distance is found to grow at the same actual rate as the $m=1$ 
mode, because the circular motion of the $m=1$ overdensity hump 
(well visible in the top-rightmost panel of 
Fig.~\ref{fig:equatorial-plane-a01}) causes the
BH to start moving in a spiral trajectory. After mode saturation the
polar distance of the BH from the origin remains almost constant,
reflecting a similar trend in the $m=1$ evolution. This spiral motion 
of the BH as a result of the development of the PPI has also been 
observed in the simulations of~\cite{Korobkin2011}.

For the non-constant angular momentum model {\tt NC1a01b30} (middle
panel) the growth rate of the $m=1$ mode is smaller, although it is
still the dominant mode, as in model {\tt C1Ba01b30}. The growth of
the mode saturates somewhat later at around $t\sim
12.5\,t_{\mathrm{orb}}$ but, contrary to the evolution of {\tt
  C1Ba01b30}, it does not drop significantly after saturation,
remaining fairly constant thereafter.  Furthermore, it remains the
dominant mode at late times while the three remaining modes attain
similar amplitudes. The evolution of the polar distance of the BH from
the origin shows a small secular drift after the saturation of the
PPI.

\begin{figure}
  \centering
  \includegraphics[scale=1.0]{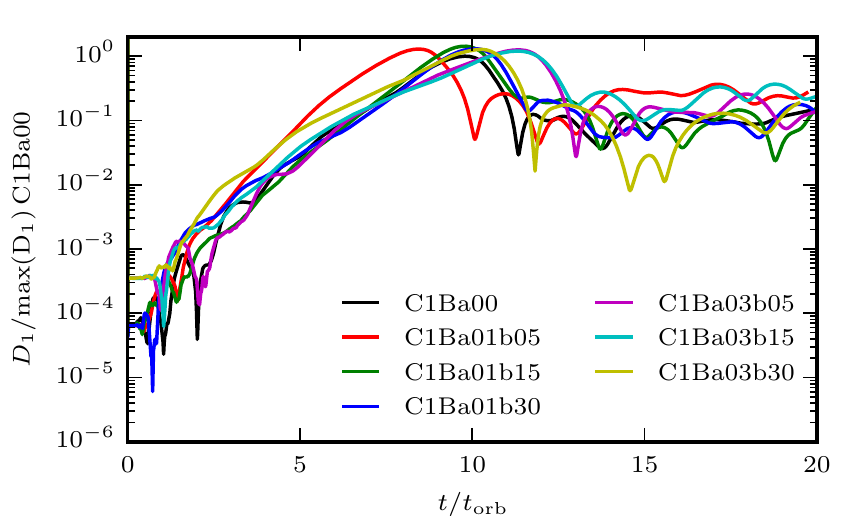}
  \\
  \vspace{-0.57cm}
  \includegraphics[scale=1.0]{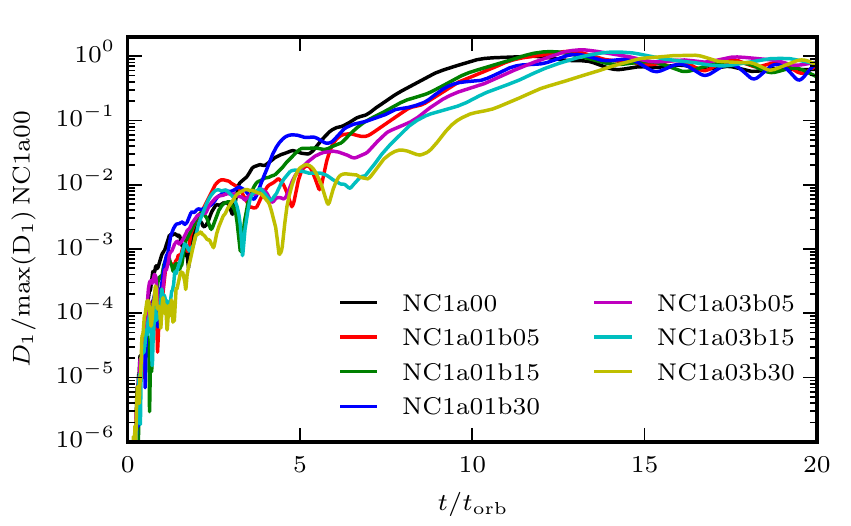}
  \\
  \vspace{-0.57cm}
  \includegraphics[scale=1.0]{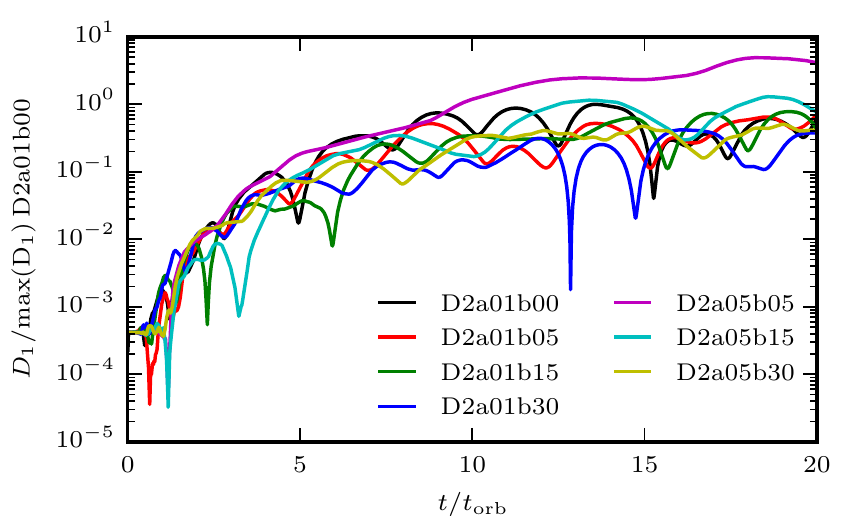}
  \caption{Evolution of the non-axisymmetric $m=1$ mode for all
    models, namely {\tt C1B} (top), {\tt NC1} (middle) and {\tt D2}
    (bottom), rescaled to the maximum amplitude of the $D_1$ mode of
    the respective untilted models.}
  \label{fig:ppi-modes}
\end{figure}
   
On the other hand, model {\tt D2a01b30} (bottom panel) shows a very
distinct behavior. The amplitude of all four modes and the polar
distance of the BH remain below the values attained in models {\tt
  C1Ba01b30} and {\tt NC1a01b30}, typically in the range between 1 to
2 orders of magnitude. Furthermore, the $m=1$ and $m=2$ modes grow at
approximately the same rate and to similar final values during the
evolution. The growth rate of $\rho_{\mathrm{BH}}$ is very similar to
the growth rate of the modes $m=1$ and $m=2$.  We also note that at
early times the $m=4$ mode shows the largest amplitude (purple line;
not so clearly noticeable in the previous two models), perhaps an
artifact of the Cartesian grid we use in our
simulations. Nevertheless, the amplitude of this mode at late times is
sufficiently smaller than the dominant amplitudes (mostly that of the
$m=1$ mode) to safely consider its effect on the dynamics
negligible. From the mode analysis, we can therefore say that model
{\tt D2} is essentially PP-stable. We will return to this point in
section~\ref{subsection:ang_mom_transport}, where we analyze 
the angular momentum transport in the models.

\begin{figure*}
  \centering
  \includegraphics[scale=0.67]{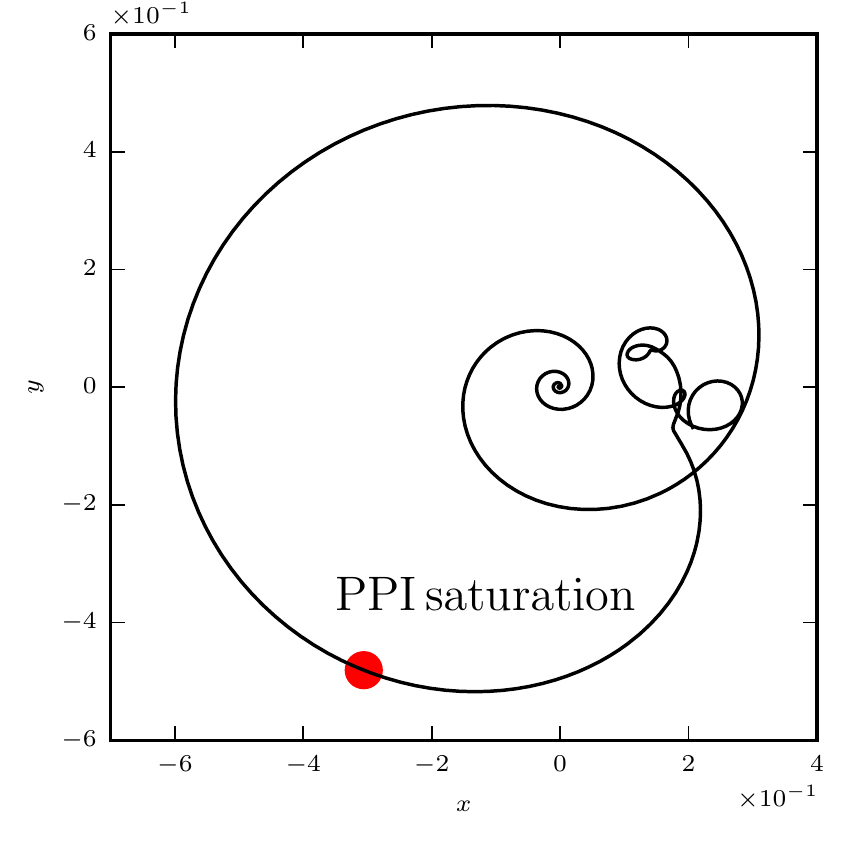}
  \includegraphics[scale=0.67]{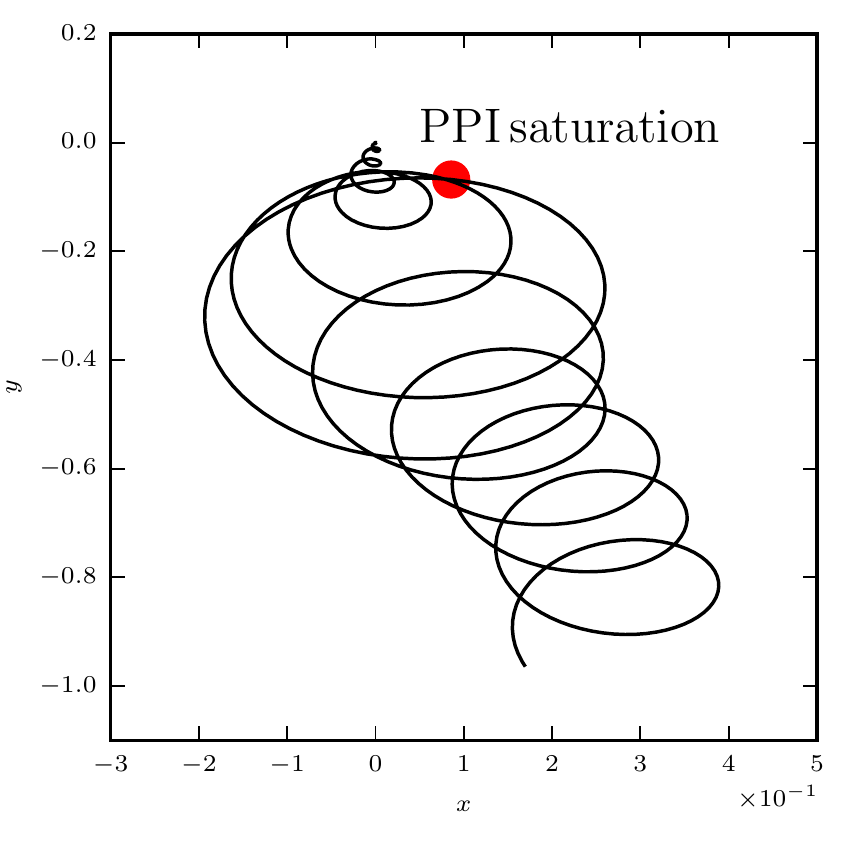}
  \includegraphics[scale=0.67]{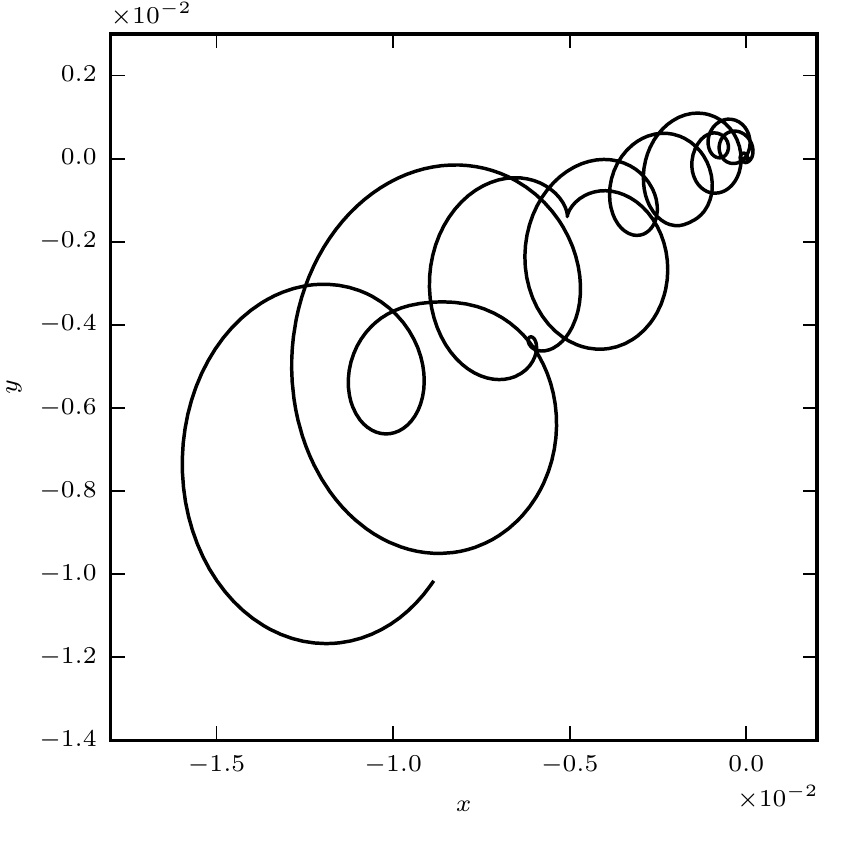}
  \caption{Position of the BH center on the $xy$ plane for models {\tt
      C1Ba01b30} (left), {\tt NC1a01b30} (middle) and {\tt D2a01b30}
    (right). For models {\tt C1Ba01b30} and {\tt NC1a01b30}, the red dot indicates 
    the point in the trajectory when the PPI saturates. Note that the scale ratio is 
    not preserved in the figures for clarity in the visualization and that the length 
    scale is significantly different in every panel.}
  \label{fig:BH-xy-movement}
\end{figure*}
   
The results we have just described connect the exponential growth and
saturation of the mass accretion rate with the growth and saturation
of the PPI in the models where it has unambiguously taken place.  To
further justify this, we plot in Fig.~\ref{fig:ppi-modes} the
evolution of the $m=1$ mode for all models considered in our
sample. Focusing on the evolution of the models {\tt C1B} displayed
in this figure (top panel) we can clearly see that the temporal order
of the saturation of the mode growth closely coincides with the
various peaks present in the mass accretion rate evolutions shown in
Fig.~\ref{fig:mdot} (compare curves of the same color). The
comparison among the three types of models also shows that the
saturation mode growth for models {\tt C1B} is only slightly larger
than for models {\tt NC1}, their late-time values being fairly
similar, but about 2 orders of magnitude larger than in models {\tt
  D2}. The slope of the mode growth is also steeper in models {\tt
  C1B}, while the behavior of the mode evolution in the case of the
light tori {\tt D2} almost seems to indicate that the development of
non-axisymmetric instabilities is only marginal for such models. A
striking feature worth highlighting from the middle panel of
Fig.~\ref{fig:ppi-modes} is the non-existing dependence of the late time
$m=1$ mode amplitude on the BH spin and tilt angle for the
non-constant angular momentum tori {\tt NC1}.

To demonstrate the connection between the growth of non-axisymmetric
modes in the disk with the movement of the central BH, we plot in
Fig.~\ref{fig:BH-xy-movement} the position of the BH center on the
$xy$ plane for models {\tt C1Ba01b30} (top), {\tt NC1a01b30} (middle)
and {\tt D2a01b30} (bottom). This figure shows several interesting
features connected to the growth of the modes in the tori. 
The motion of the BH is caused by the formation 
of a ``quasi-binary system" between the BH and the $m=1$ overdensity 
lump. It is this binary motion that causes the long-term emission of 
GWs reported in~\cite{Kiuchi2011}. We will return to this in the discussion of 
the GW emission connected to the PPI in section~\ref{subsection:GW}.
For model
{\tt C1Ba01b30}, the development of the dominant $m=1$ structure in
the disk exerts a small kick in the BH. As a result the BH starts
moving in a spiral trajectory until the PPI saturates and the mode
amplitude drops, at which point the BH is roughly located at
$(x,y)=(-0.3,-0.5)$. After saturation, the $xy$ position of the BH is
a combination of linear and circular motion. In model {\tt NC1a01b30},
the linear motion is stronger from the beginning, causing a linear
shift in the growing spiral motion. Upon saturation of the PPI, the
circular motion continues with a smaller radius, in accordance to the
evolution of the $m=1$ mode which remains at a rather constant value
after saturation and up until the end of the evolution. Finally, for
model {\tt D2a01b30} we have a superposition of circular and linear
motion again, as the modes $m=1$ and $m=2$ are of almost equal
magnitude during the evolution. Here, the overall distance covered by
the BH is significantly smaller than for the heavier models {\tt
  C1Ba01b30} and {\tt NC1a01b30}, as the mode amplitudes are much
smaller for {\tt D2} and the disk is also lighter.

We argue that the non-axisymmetric instability we observe in our models 
is indeed the PPI based on the azimuthal mode evolution in the disk and 
the development of the $m=1$ overdensity lump. However, we do not claim 
to have presented a direct, undeniable proof that the instability observed 
in our simulations is the analytical PPI studied in~\cite{Papaloizou1984}. 
We also note that there exist other instabilities in self-gravitating 
disks which display spiral arm formation in the disk or even disk 
fragmentation, such as the axisymmetric local Toomre 
instability~\cite{Toomre1964}, but as~\cite{Christodoulou1992} have shown, 
the Toomre instability is unlikely to affect radially slender accretion 
tori such as the ones considered in our study.

\subsection{PPI saturation and black hole kick}
\label{subsection:BH-kick}

\begin{figure}[t]
  \centering
  \includegraphics[scale=0.37]{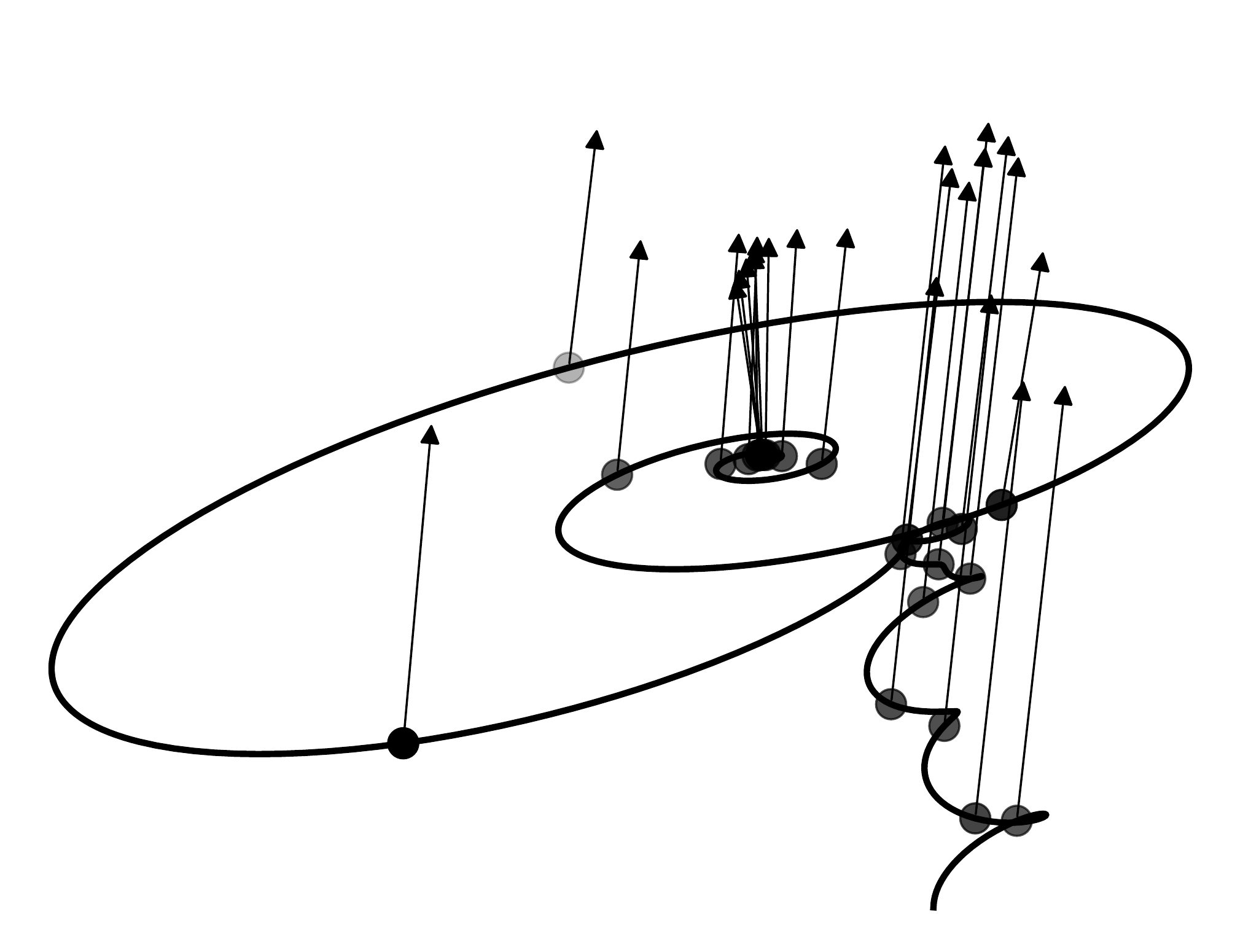}
  \caption{Full 3D trajectory of the BH along with its spin vectors
    for model {\tt C1Ba01b30}. The magnitude of the spin vectors has
    been rescaled with a constant factor for visualization purposes.}
  \label{fig:BH-3D-trajectory}
\end{figure}

\begin{figure}[t]
  \centering
  \includegraphics[scale=1.0]{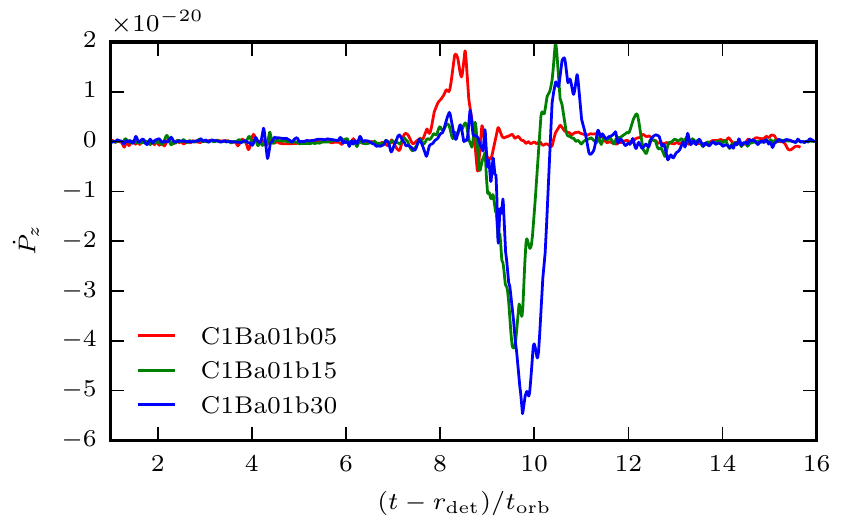}
  \caption{Radiated linear momentum in the $z$-direction 
    for models {\tt C1Ba01}.}
  \label{fig:rad_lin_mom}
\end{figure}

The complete 3D trajectory of the BH and its spin vector is plotted
for model {\tt C1Ba01b30} in Fig.~\ref{fig:BH-3D-trajectory}. In this
plot each small sphere with an attached arrow corresponds to the
position of the BH and the magnitude (scaled for better visualization)
and direction of its spin vector. The positions are plotted
at equal time intervals, so we see that the BH is moving much faster
during the final stages of the PPI than before its occurrence or after
its saturation. We can also see that there is a movement in the
vertical direction, which is caused by the fact that the ``binary" 
motion of the BH and $m=1$ overdensity lump does not lie in the 
equatorial plane anymore due to the initial tilt. While the PPI 
is growing, the BH-torus system therefore moves up and down in an 
oscillatory fashion. As we have seen in the $m=1$ mode plots for models 
{\tt C1B} (see the top panel of  fig.~\ref{fig:ppi-modes}), the saturation 
of the PPI is very fast. The rapid saturation of the PPI (and the 
corresponding destruction of the $m=1$ overdensity lump) result in 
a mild vertical kick of the BH+torus system. To see that this is indeed 
connected to the saturation of the PPI, we plot in Fig.~\ref{fig:rad_lin_mom} 
the linear momentum radiated away by gravitational waves in the 
$z$-direction for models {\tt C1Ba01}. From the plot, we can 
clearly see that the time of maximum emission of linear momentum 
corresponds exactly to the time the PPI saturates for each model. 
The radiated linear momentum has been calculated with the 
{\tt pyGWAnalysis} package~\cite{Reisswig2011a}.

\subsection{BH precession and nutation}
\label{subsection:BH-precession}

Contrary to the test-fluid simulations
of~\cite{Fragile2005,Fragile2007a}, as we are evolving the full
spacetime, we can monitor the response of the BH to the evolution of
the disk in the initially tilted configuration. In particular we can
measure the total precession of the BH spin about the $z$-axis and its
rate, as well as the inclination of the spin with respect to the
$z$-axis and the corresponding nutation.

\begin{figure}
  \centering
  \includegraphics[scale=1.0]{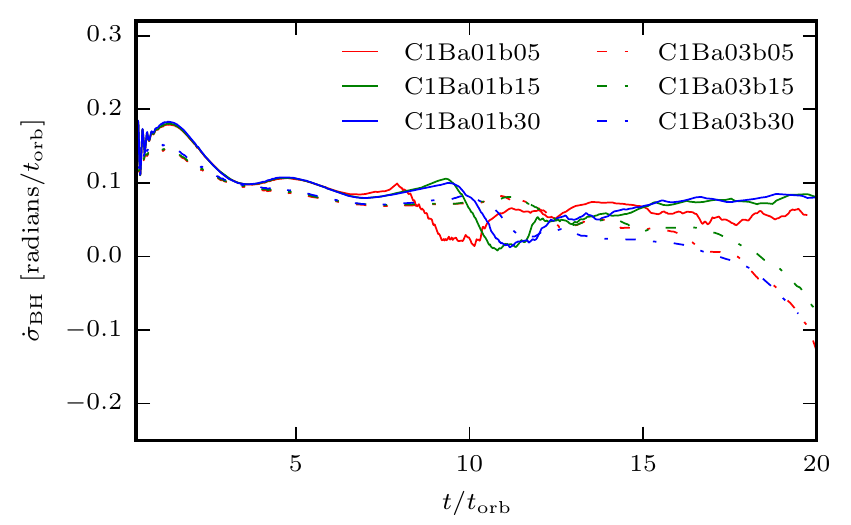}
  \\
  \vspace{-0.57cm}
  \includegraphics[scale=1.0]{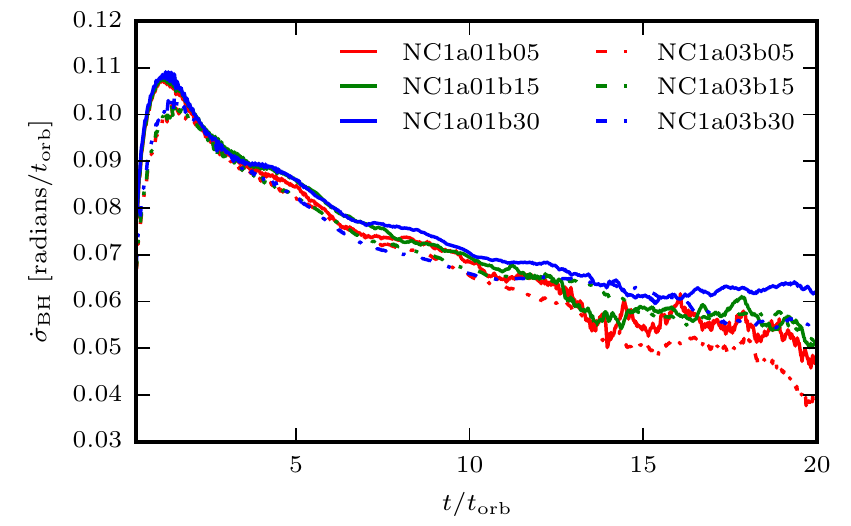}
  \\
  \vspace{-0.57cm}
  \includegraphics[scale=1.0]{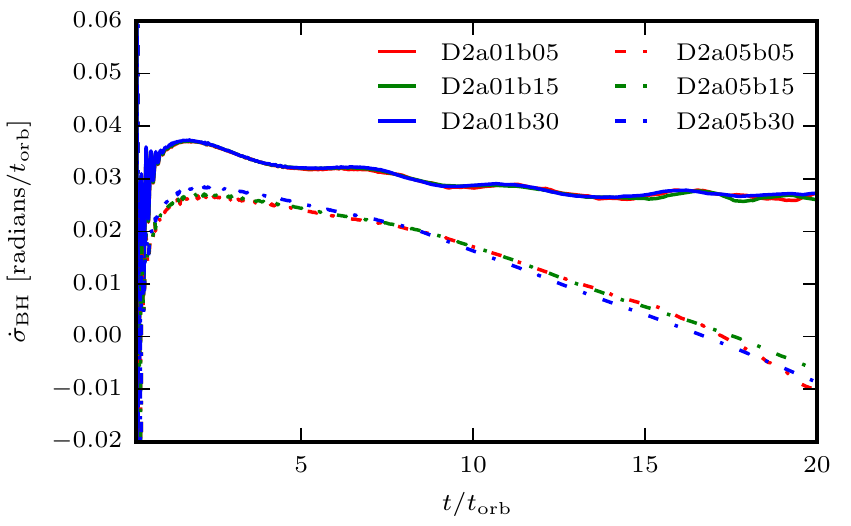}
  \caption{Evolution of the precession rate of the BH spin about the
    $z$-axis for models {\tt C1B} (top), {\tt NC1} (middle) and {\tt
      D2} (bottom) in units of $\mathrm{radians}/t_{\mathrm{orb}}$.}
  \label{fig:BH_precession_rate}
\end{figure}

\begin{figure}
  \centering
  \includegraphics[scale=1.0]{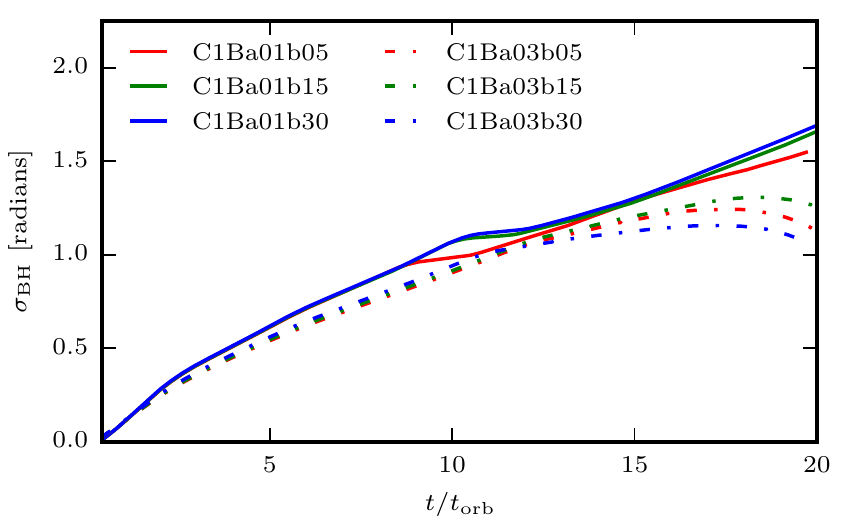}
  \\
  \vspace{-0.57cm}
  \includegraphics[scale=1.0]{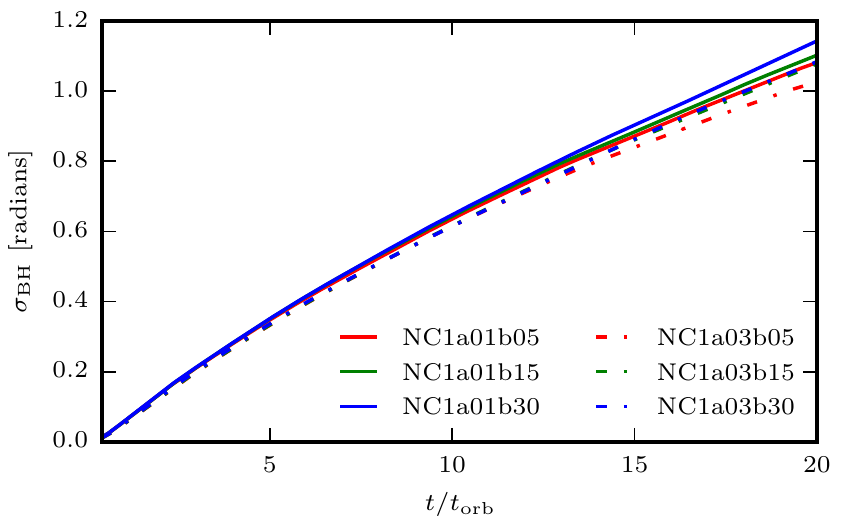}
  \\
  \vspace{-0.57cm}
  \includegraphics[scale=1.0]{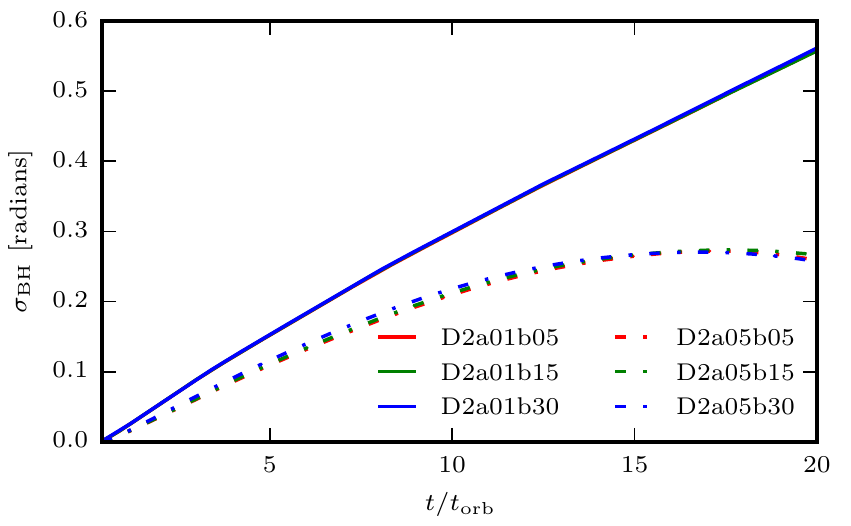}
  \caption{Total precession of the BH spin about the $z$-axis for
    models {\tt C1B} (top), {\tt NC1} (middle) and {\tt D2} (bottom)
    in units of $\mathrm{radians}$.}
  \label{fig:BH_precession_total}
\end{figure}

In Fig.~\ref{fig:BH_precession_rate} we plot the evolution of the
precession rate $\dot{\sigma}_{\mathrm{BH}}$ of the central BH about
the $z$-axis for models {\tt C1B} (top), {\tt NC1} (middle) and {\tt
  D2} (bottom panel) in radians per orbital period. The first thing to
mention is that the constant angular momentum disks {\tt D2} and {\tt
  C1B} show a clear dependence of the precession rate with the initial
BH spin magnitude. Namely, for models {\tt D2}, the larger the spin the
smaller the precession rate, and for models {\tt C1B} smaller
spins lead to higher precession rates. However, for the non-constant
angular momentum torus {\tt NC1} the precession rate is very similar
irrespective of the spin magnitude, particularly at the early stages
($t<10\,t_{\rm{orb}}$).
In models {\tt D2} the evolution of $\dot{\sigma}_{\mathrm{BH}}$ is
very smooth, and changes sign for the higher spin runs after about 17
orbits, whereas it remains fairly constant for the lower spin
simulations.

A fact worth emphasizing is that there is essentially no dependence of
the precession rate on the tilt angle for constant angular momentum
tori.  This is particularly true for models {\tt D2} but also models
{\tt C1B} show this lack of dependence up until the growth of the PPI.
In all models {\tt C1B} we observe a prominent modulation of the
precession rate when the PPI enters the final stages of its growth and
then saturates, as well as the change in sign for the higher spin
runs. The magnitude of the precession rate is about an order of
magnitude higher compared to models {\tt D2}, which we attribute to the
higher disk-to-BH mass ratio models {\tt C1B} possess. The evolution
of the BH precession rate for models {\tt NC1} seems to fit in between
the two other models, the magnitude of the precession rate is
somewhere in between but not very different evolutions are observed for
the two BH spins considered. We also note
from Fig.~\ref{fig:BH_precession_rate} that models {\tt C1Ba01} 
and {\tt D2a01} attain a fairly constant precession rate 
by the end of our simulations, namely $\sim20\, \mathrm{Hz}$ and 
$\sim6\, \mathrm{Hz}$ for models {\tt C1Ba01} and {\tt D2a01}, respectively. 

\begin{figure}
  \centering
  \includegraphics[scale=1.0]{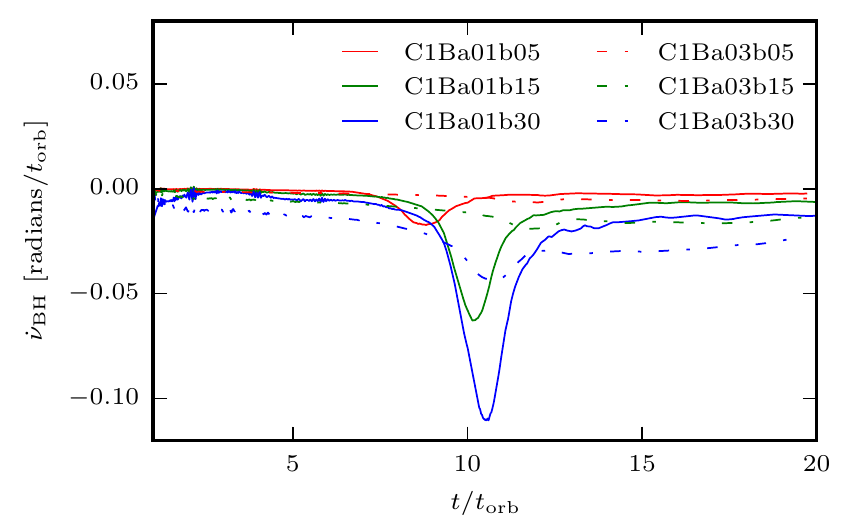}
  \\
  \vspace{-0.57cm}
  \includegraphics[scale=1.0]{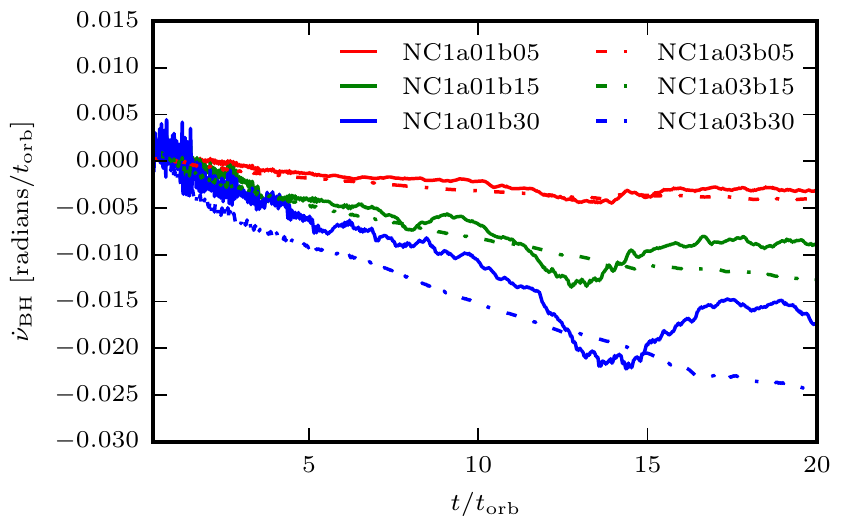}
  \\
  \vspace{-0.57cm}
  \includegraphics[scale=1.0]{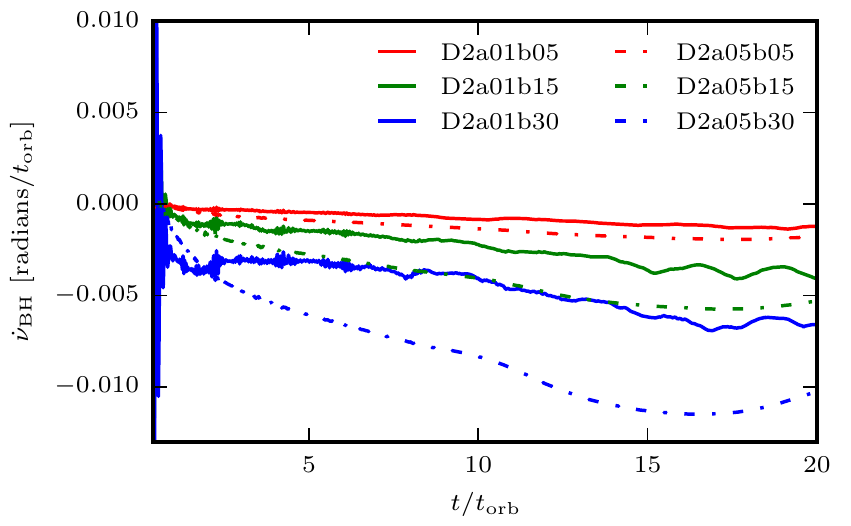}
  \caption{Evolution of the nutation rate of the BH spin about the
    $z$-axis for models {\tt C1B} (top), {\tt NC1} (middle) and {\tt
      D2} (bottom) in units of $\mathrm{radians}/t_{\mathrm{orb}}$.}
  \label{fig:nutation-rate}
\end{figure}

\begin{figure}
  \centering
  \includegraphics[scale=1.0]{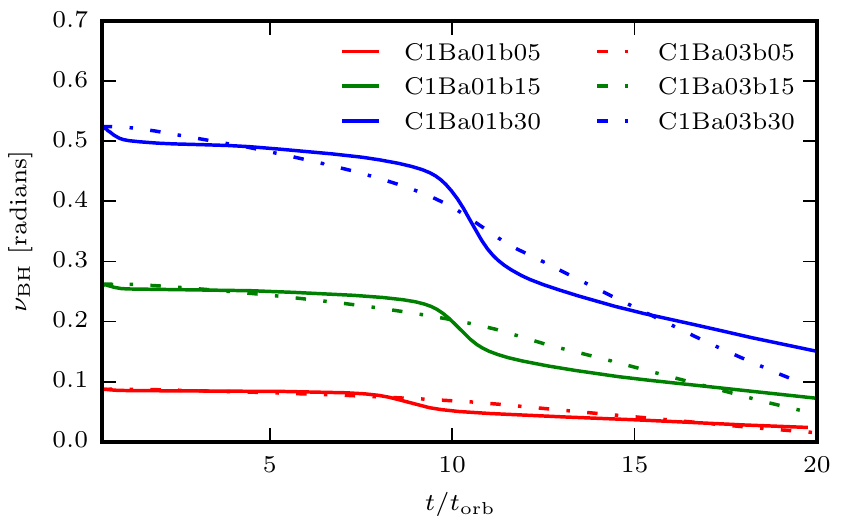}
  \\
  \vspace{-0.57cm}
  \includegraphics[scale=1.0]{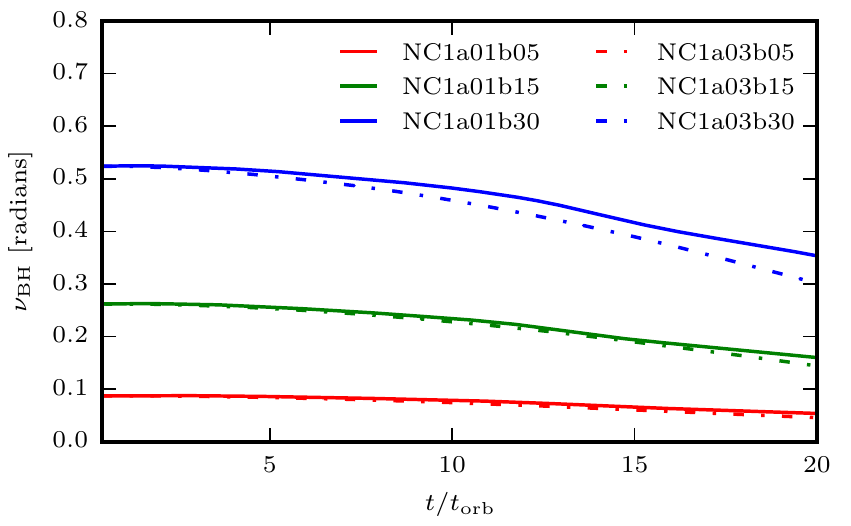}
  \\
  \vspace{-0.57cm}
  \includegraphics[scale=1.0]{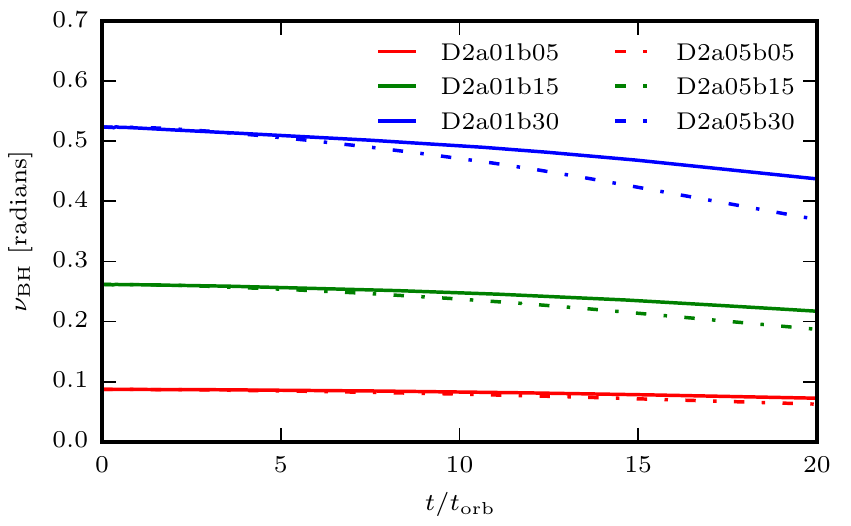}
  \caption{Evolution of the tilt angle of the BH spin with respect to
    the $z$-axis for models {\tt C1B} (top), {\tt NC1} (middle) and
    {\tt D2} (bottom) in units of
    radians.}
  \label{fig:BH-nutation-total}
\end{figure}

The total precession of the BH spin about the $z$-axis is shown in
Fig.~\ref{fig:BH_precession_total}. The lower spin models {\tt
  C1Ba01}, that show the highest BH precession rate, have completed a
quarter of a full precession cycle by the end of our simulations at
$t=20t_{\mathrm{orb}}$. Given that the precession rate for those
models appears to settle to a rather constant value (see
Fig.~\ref{fig:BH_precession_rate}) we would need a prohibitively 
expensive evolution of about $80$ orbital periods for a full precession cycle 
of the central BH for those models. If we assume that the precessing Kerr 
BH radiates GW as a freely precessing rigid body, it will radiate at
$\dot{\sigma}_{\mathrm{BH}}$ and $2\, \dot{\sigma}_{\mathrm{BH}}$
in the $l=2$, $m=1,2$ multipole modes~\cite{Zimmermann1979}. 
This means we would have to evolve the BH-torus system for at least 
one full BH precession cycle in order to see the slow modulation of the GW
signal caused by the BH precession. It remains an interesting open issue 
to see if the upcoming advanced GW detectors will be able to measure 
these GWs in the low tens of Hz.

The imprint of the growth of non-axisymmetric instabilities in the
disk on the BH response can be seen even clearer in
Fig.~\ref{fig:nutation-rate}. This figure shows the evolution of the
nutation rate for all models, that is, the evolution of the temporal
variation of the tilt angle of the BH with the $z$-axis,
$\dot{\nu}_{\rm{BH}}$. 
For models {\tt C1B} (top panel) there is a clear period of alignment of
the BH spin with the $z$-axis around the time when the PPI starts to
grow. This is signaled by the rapid fall and rise of the nutation
rate around the $10\,t_{\rm{orb}}$ mark for the slow spin ($a=0.1$)
BHs, but it is also (less) visible in the more rapidly rotating BHs
(dashed lines).  Upon saturation of the instability the nutation rate
becomes almost constant and fairly close to being again zero for the
$a=0.1$ BH models (see solid lines). This behavior is consistent with
the evolution of the BH tilt angle itself displayed in
Fig.~\ref{fig:BH-nutation-total}, which, for the case of models {\tt
  C1B} with small spins, shows two distinct zones of constant spin
inclination, particularly for models with $\beta_0=5^{\circ}$ and
$15^{\circ}$ (model with $\beta_0=30^{\circ}$ still shows a negative
slope at late times). The transition to zero nutation rate in the
$a=0.3$ {\tt C1B} models after PPI saturation seems to take a longer
time than that displayed in Fig.~\ref{fig:nutation-rate}.

In addition Fig.~\ref{fig:nutation-rate} also shows that the maximum
nutation rate attained increases with increasing initial tilt
angle. This is very clearly seen in models {\tt C1B} but it is also
visible in most of the models {\tt D2} and {\tt NC1}. In particular
the evolution of $\dot{\nu}_{\rm{BH}}$ for the non-constant angular
momentum models {\tt NC1} with $a=0.3$ (dashed curves) does not seem
to reach a local maximum, at least in the timescales we can afford in
our simulations, despite these models are also affected by the
PPI. This might be explained by the fact that the non-axisymmetric
$m=1$ mode does not drop as sharply after the saturation of the PPI in
the models {\tt NC1} as it does in the {\tt C1B} models (see
Fig.~\ref{fig:ppi-modes}). The growth of the mode in the disk seems to
cause an alignment of the BH spin with the $z$-axis. Furthermore, the
nutation rate is much larger for the {\tt C1B} models. Finally, models
{\tt D2}, in the absence of any strong growth of PPI modes, do not
show strong modulation of the nutation rate. The modulation of
this rate is therefore closely connected to the existence
of $m=1$ non-axisymmetric modes in the disk.

The corresponding evolution of the inclination of the BH spin from the $z$-axis is
seen in Fig.~\ref{fig:BH-nutation-total}. The panels show that 
a significant realignment of the BH spin and the $z$-axis
takes place by the end of the simulations in models {\tt C1B}, i.e.~in those 
that develop the PPI most significantly.
For all models we see that the final inclination is smaller the
bigger the initial tilt angle is, for equal spin magnitudes. 
This trend also seems to be more pronounced the 
higher the initial spin magnitude is.  
   
\subsection{Disk twist and tilt}
\label{subsection:disk_precession_nutation}

\begin{figure}
  \centering
  \includegraphics[scale=1.0]{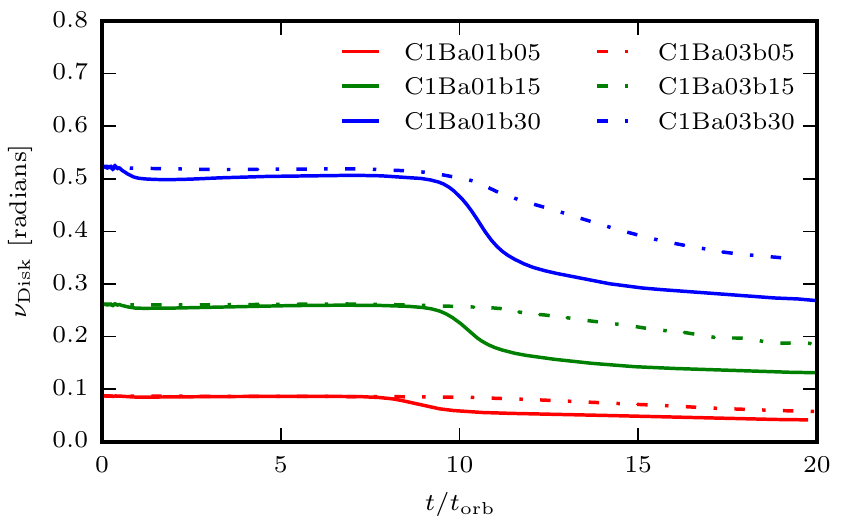}
  \\
  \vspace{-0.57cm}
  \includegraphics[scale=1.0]{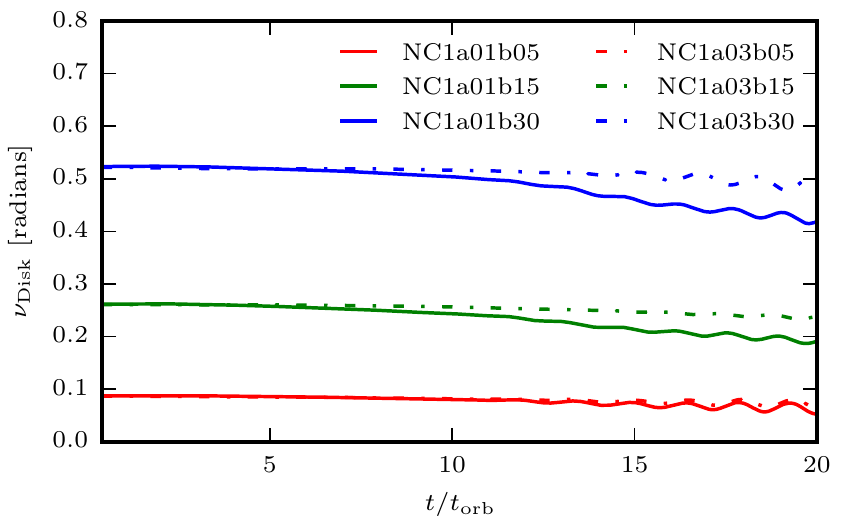}
  \\
  \vspace{-0.57cm}
  \includegraphics[scale=1.0]{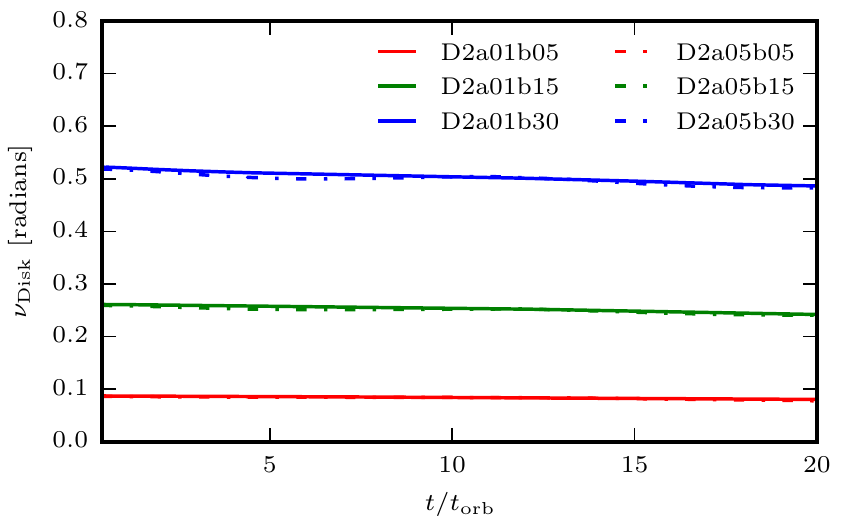}
  \caption{Evolution of the tilt angle $\nu_{\rm{Disk}}$ between then
    BH spin and the total disk angular momentum $J_{\mathrm{Disk}}$
    for models {\tt C1B} (top), {\tt NC1} (middle) and {\tt D2}
    (bottom) in units of $\mathrm{radians}$.}
  \label{fig:disk_global_nutation}
\end{figure}

\begin{figure}
  \centering
  \includegraphics[scale=1.0]{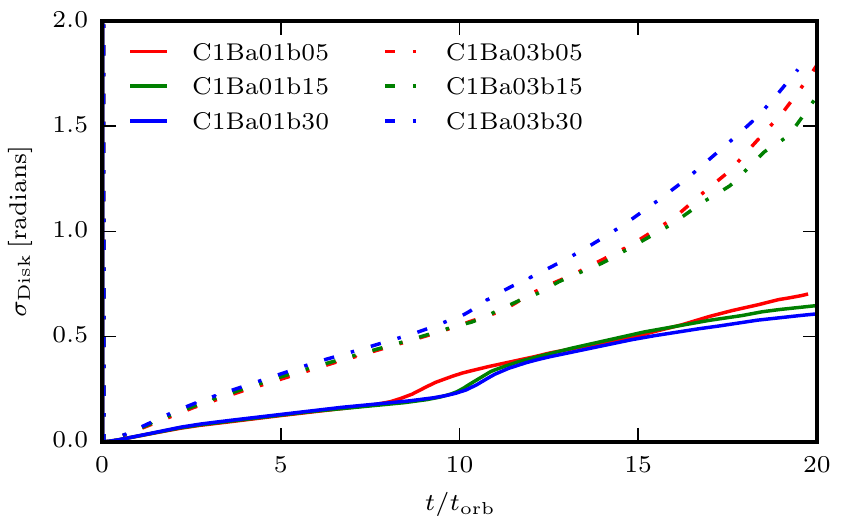}
  \\
  \vspace{-0.57cm}
  \includegraphics[scale=1.0]{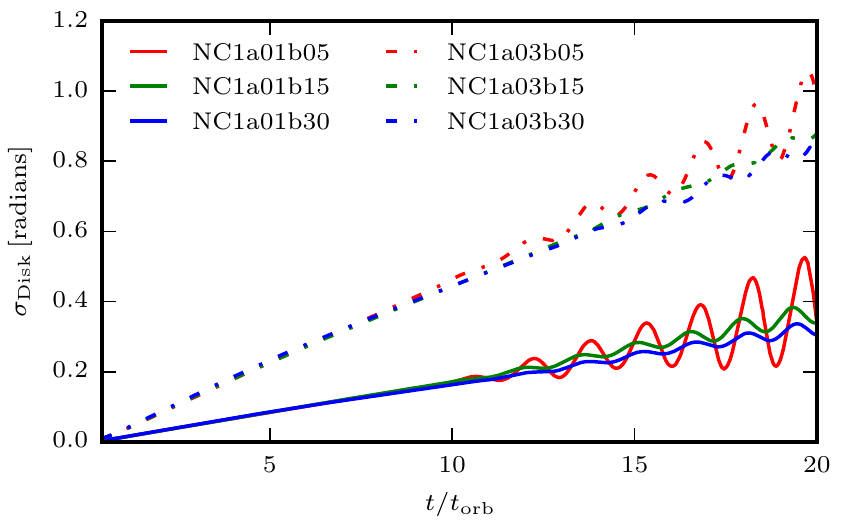}
  \\
  \vspace{-0.57cm}
  \includegraphics[scale=1.0]{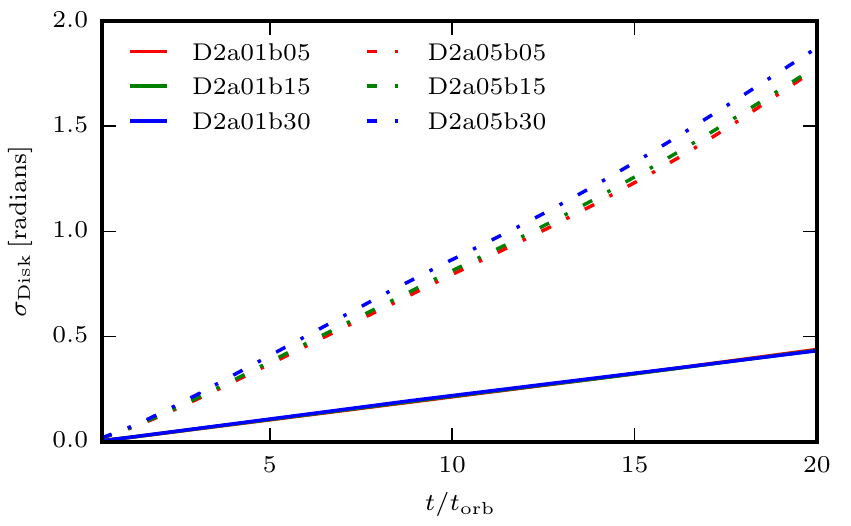}
  \caption{Evolution of the total precession of the total disk angular
    momentum $J_{\mathrm{Disk}}$ around the BH spin for models {\tt
      C1B} (top), {\tt NC1} (middle) and {\tt D2} (bottom) in units of
    $\mathrm{radians}$.}
  \label{fig:disk_global_precession}
\end{figure}

We turn now to describe in this section the response of the disks as they evolve
in the tilted spacetime, using the diagnostics we have introduced in
Section~\ref{subsection:twist_tilt_diagnostics}. In
Figures~\ref{fig:disk_global_nutation}
and~\ref{fig:disk_global_precession}, 
we plot the evolution of the tilt
and precession of the total angular momentum vector of the disk 
$J_{\mathrm{Disk}}$. This vector is
the sum of the angular momentum vectors of each shell, as described in
Section~\ref{subsection:twist_tilt_diagnostics}. We will refer to
these as the global disk tilt and global disk precession around the BH
spin, in order to distinguish them from the analysis of the twist
$\sigma(r)$ and tilt $\nu(r)$ angles in the individual disk shells.
   
\begin{figure*}
 \centering
 \includegraphics[scale=1.0]{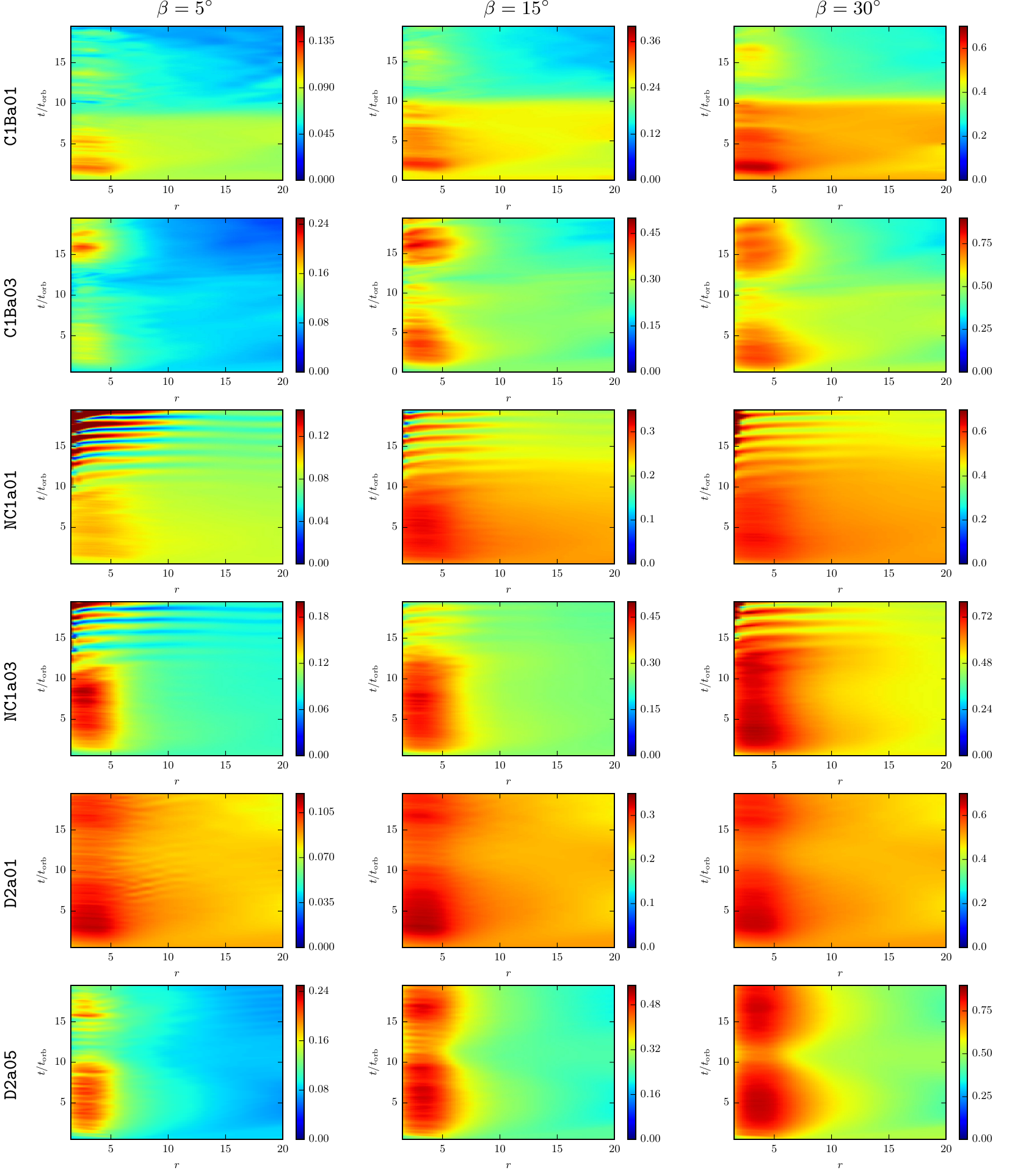}
  \caption{Spacetime $t-r$ diagrams of the radial disk tilt profile $\nu(r)$
        for models {\tt C1Ba01}, {\tt C1Ba03} (top two panels), {\tt
          NC1a01}, {\tt NC1a03} (middle two panels), and {\tt D2a01}
        and {\tt D2a05} (bottom two panels). From left to right the
        panels show the initial tilt angles $\beta_0$ of $5^\circ$,
        $15^\circ$ and $30^\circ$. Each panel shows the radial profile
        of the disk tilt for the entire time evolution.}
   \label{fig:disk_tilt_profiles}
\end{figure*}
   
   The evolution of the global disk tilt, shown in
   Fig.~\ref{fig:disk_global_nutation}, shows resemblances with the
   evolution of the tilt of the BH spin (see
   Fig.~\ref{fig:BH-nutation-total}). For models {\tt C1B} (top panel)
   we see that the development of the PPI not only causes a partial
   realignment of the BH spin with the $z$-axis, but also a
   realignment of the BH spin and disk angular momentum. This
   alignment seems to become more constant towards the end of the
   simulation than the alignment of BH spin and the $z$-axis. We also
   note that the amount of realignment is reversed for the global disk
   tilt for models {\tt C1B}. Here the higher initial spin magnitude
   leads to a lower amount of alignment, in contrast with the
   evolution of the BH tilt. On the other hand, the alignment between
   BH spin and total disk angular momentum for models {\tt NC1}
   (middle panel) is not as pronounced as the realignment of the BH
   spin with the $z$-axis. As in the case of models {\tt C1B}, the
   trend of more alignment with higher spin for the BH tilt is
   inverted for the evolution of the global disk tilt. There are also
   oscillations in the evolution of the global disk tilt for models
   {\tt NC1}, which are not visible in the other two set of
   models. These oscillations seem to be caused by the 
   persistent $m=1$ non-axisymmetric mode in models {\tt NC1}.
   Finally,  for models {\tt D2} (bottom panel), neither the initial spin
   magnitude nor the initial tilt angle seem to affect the
   evolution of the global disk tilt, which is furthermore hardly
   changed during the entire evolution for all initial spins.
The development of the PPI therefore seems to cause an alignment 
of the BH spin with the total disk angular momentum, which is
larger for smaller initial spins. As both $\nu_{\mathrm{BH}}$ and 
$\nu_{\mathrm{Disk}}$ decrease during the growth of the PPI,
it is clear that the BH spin is tilted into the orbital plane of the
disk. 
   
   In Fig.~\ref{fig:disk_global_precession} we plot the global
   precession of the total disk angular momentum vector about the BH
   spin axis for models {\tt C1B} (top), {\tt NC1} (middle) and {\tt
     D2} (bottom panel). For all models, the higher the spin (dashed
   curves), the larger the global precession, as expected. The initial
   tilt angle does not seem to influence strongly the evolution for
   models {\tt C1B} and {\tt D2}, while the evolution for {\tt NC1}
   shows oscillations towards the end of the evolution. Note that
     these oscillations are superimposed on the slow growth of the
     precession, which always has a positive slope, as expected. The 
     oscillations are caused by the wobbling and smaller precession 
     cycles about the direction of the global vector. These oscillations are 
     then visible in the projection of the disk angular momentum vector 
     onto the equatorial plane of the BH. The amplitude
   of the oscillations is larger the smaller the initial tilt angle.
   The non-axisymmetric modes that survive much longer in the case of
   models {\tt NC1} could be causing these oscillations in the global
   disk tilt and precession.
   
   In Fig.~\ref{fig:disk_tilt_profiles} we show spacetime ($t-r$) diagrams of the
   radial disk tilt profile for the entire time evolution for
   all tilted models in order to see if and which disk models become warped
   during their evolution. The initial radial tilt profile is flat and 
   the panels cover the initial radial extent of each model.  For models 
   {\tt C1B} (top two rows) we see that by the end of the simulations
   the radial tilt profile for larger radii is significantly lowered 
   compared to the initial tilt angle. Furthermore, there is a clear 
   distinction in the radial tilt profile before and after the saturation 
   of the PPI for the lower spin models ($a=0.1$), in
   accordance with what we have observed for the global disk angular
   momentum tilt (see Fig.~\ref{fig:disk_global_nutation}). For the 
   higher spin models ($a=0.3$) the tilt profile evolution is broken 
   up into two regions as well. In these cases, the peak close to the 
   origin reemerges after the saturation of the PPI.
   Likewise, models {\tt NC1} (two middle rows) also show a peak close
   to the origin, but contrary to the previous set of models, the late
   time tilt profiles are oscillating in the regions closest to the central
   BH. The oscillations are stronger for both, radii closer to the origin
   and smaller initial BH spin. As described before, we
   believe that these oscillations are a consequence of the long-lasting
   non-axisymmetric $m=1$ mode in the disk. 
   
   The oscillations of the tilt angle close to the origin in models {\tt NC1} might 
   be connected to the so called Kozai-Lidov (KL)  mechanism~\cite{Kozai1962,Lidov1962}, 
   where the eccentricity and inclination of particle orbits around an object
   that is itself in a binary are interchanged in
   an oscillatory fashion. Recently~\cite{Martin2014} performed simulations
   of the KL mechanism in hydrodynamical disks around a component of a
   binary system, finding
   oscillations of the inclination angle. The persistent $m=1$ structure in the
   disks of models {\tt NC1} can be thought of as an eccentric distribution
   of matter. While our disks are not evolved around a BH in a binary
   system, we have nevertheless shown in 
   Section~\ref{subsection:PPI-modes} that the growth of the $m=1$
   non-axisymmetric modes causes the BH to start moving in a
   ``quasi-binary" with the overdensity lump in the disks. As seen
   in Fig.~\ref{fig:BH-xy-movement}, for models {\tt NC1}, the long
   lasting ``planet" causes the BH to move in a continuous elliptic
   fashion in the projection onto the $xy$-plane. We believe this 
   movement of the BH induced by the PPI serves as the correct 
   boundary conditions for the KL mechanism to occur in models
   {\tt NC1}. In~\cite{Martin2014} the authors estimate the timescale of
   the KL oscillations in their disks as being $\tau_\mathrm{KL} \sim 16 \,P_b$, where 
   $P_b$ is the orbital period of the central object in the binary.
   In our case, the periods of the tilt angle oscillation and the 
   orbital motion of the Kerr BH are roughly equal for all models.
   We plan to further investigate the correctness of this
   interpretation of the tilt angle oscillations in a future work.
   
   To complete the analysis of Fig.~\ref{fig:disk_tilt_profiles}, we note that models 
   {\tt D2} (two bottom rows) become clearly warped, with all
   models showing a peak near the origin. The profiles of the low spin
   models {\tt D2a01} are very similar in shape for the three initial
   tilt angles. The peak in models {\tt D2a05} is seen to oscillate
   stronger the smaller the initial tilt angle. Note that for none
   of the models we see an alignment of the inner region of the disk
   with the equatorial plane of the BH (this would mean $\nu(r)=0$
   there).  We therefore find no occurrence of the Bardeen-Petterson
   effect in our simulations, at least in the timescales
   considered. This is in agreement with the results found in the fixed-spacetime 
   inviscid simulations with no angular momentum
   transport of~\cite{Fragile2005} and also with the GRMHD simulations 
   of~\cite{Fragile2007a} that included angular momentum transport 
   driven by the magneto-rotational instability.
   The observed profile and evolution of the disk warp seems to be in 
qualitative agreement with the analytic work on warp propagation as 
bending-waves in tilted thick accretion disks; see, for instance, 
the analytic tilt profiles in~\cite{Lubow2002}, the analytic model 
of linear warp evolution of~\cite{Foucart2014} and the application 
of the linear warp evolution model in~\cite{Franchini2015}.
   We note that while the Bardeen-Petterson effect
   is not manifest in our models, we nevertheless do observe a
   \emph{global} partial realignment of the BH spin with the disk
   angular momentum, caused by the growth of the PPI.
  This alignment of the disk angular momentum with the BH spin has also 
been observed in the post-merger evolution of a tilted accretion disk 
in~\cite{Kawaguchi2015}. The authors conclude that the transport of 
angular momentum by non-axisymmetric shock waves in the disk might be 
responsible for such a Bardeen-Petterson-like effect. Elucidating this
issue deserves further studies. 

\begin{figure*}
  \centering
  \includegraphics[scale=1.0]{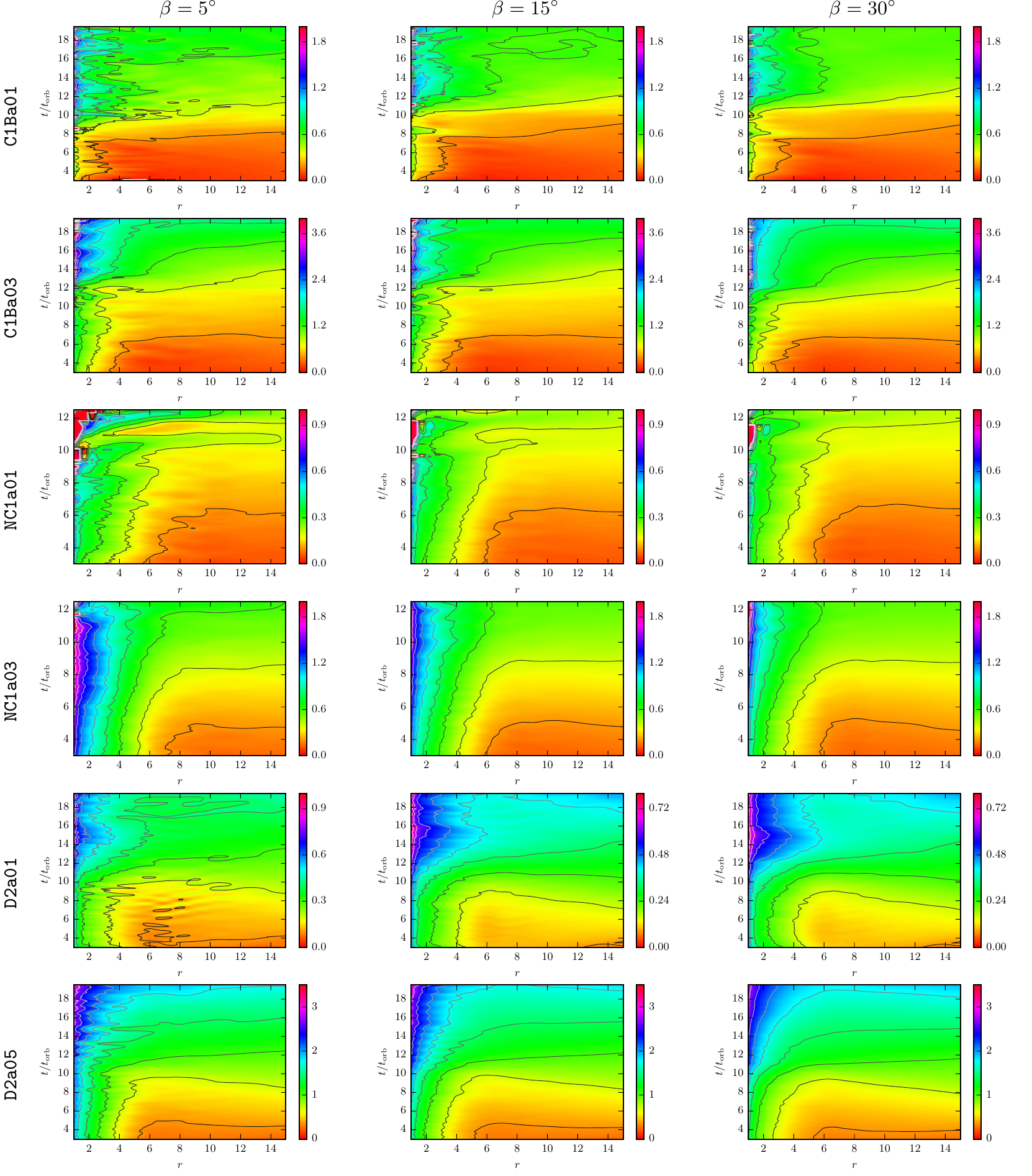}
   \caption{Spacetime $t-r$ diagrams of the radial disk twist profile $\sigma(r)$
        for models {\tt C1Ba01}, {\tt C1Ba03} (top two panels), {\tt
          NC1a01}, {\tt NC1a03} (middle two panels), and {\tt D2a01}
        and {\tt D2a05} (bottom two panels). From left to right the
        panels show the initial tilt angles $\beta_0$ of $5^\circ$,
        $15^\circ$ and $30^\circ$. Each panel shows the radial profile
        of the inner region disk twist from 3 orbits onwards.}
   \label{fig:disk_twist_profiles} 
\end{figure*}

Finally, to check for Lense-Thirring precession, we plot the evolution
of the inner region disk twist $\sigma(r)$ from $t=3$ orbits onwards
for all our tilted models in Fig.~\ref{fig:disk_twist_profiles}. We
see that all models become twisted (as $\sigma(r)$ varies 
with $r$) in the regions for radii up to $5$. The plots further indicate solid
precession of the disks, as the radial profile of $\sigma(r)$ is
increasing smoothly in time almost independently of the radius
in the outer regions of the disk and does not remain $0$ for large
radii. 
This is consistent with the solid-body precession found in the 
fixed spacetime simulations of~\cite{Fragile2005}. 
Models {\tt D2} actually show a slightly growing
twist for larger radii. In this trend, the higher initial spin models
show flatter profiles, while there is some radial modulation of the
profiles for the lower spin simulations. A striking feature of the
plots is the much smaller difference in accumulated twist for large radii
between different initial spin magnitude for models {\tt NC1}. In
models {\tt C1B} and {\tt D2} the higher spin models show a much
higher twist at large radii than their low spin counterparts. 
As the long survival of the $m=1$ mode in the {\tt NC1} models seems to
have an effect on the evolution of the twist, we plot this evolution only up
to the saturation of the PPI for these models in order to properly visualize
the solid body precession of the disk.
   
\subsection{Angular momentum transport}
\label{subsection:ang_mom_transport}

\begin{figure}
  \centering
  \includegraphics[scale=1.0]{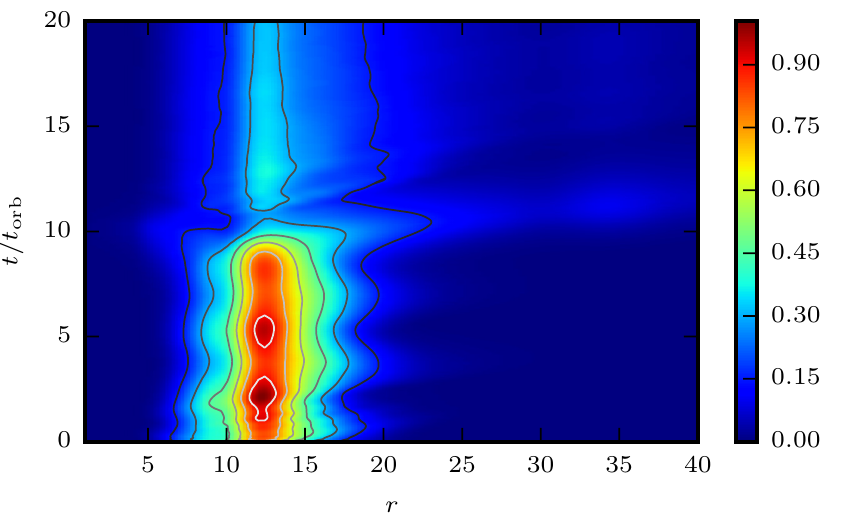}
  \\
  \vspace{-0.57cm}
  \includegraphics[scale=1.0]{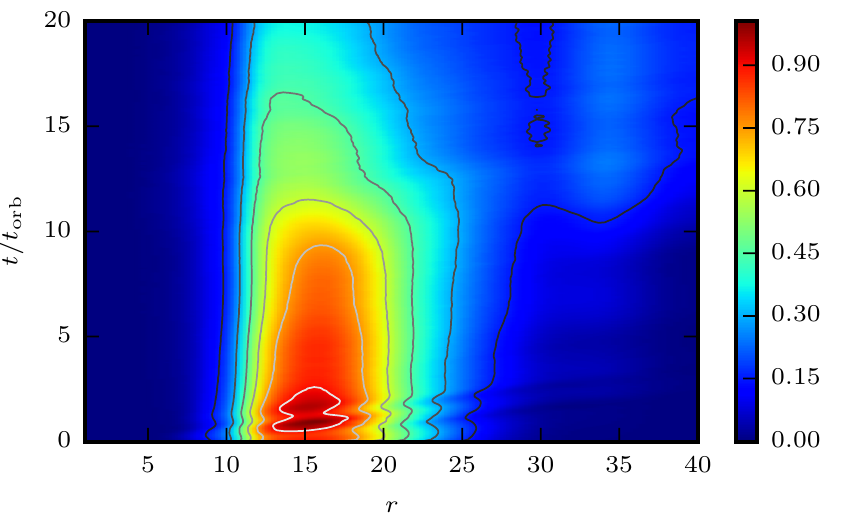}
  \\
  \vspace{-0.57cm}
  \includegraphics[scale=1.0]{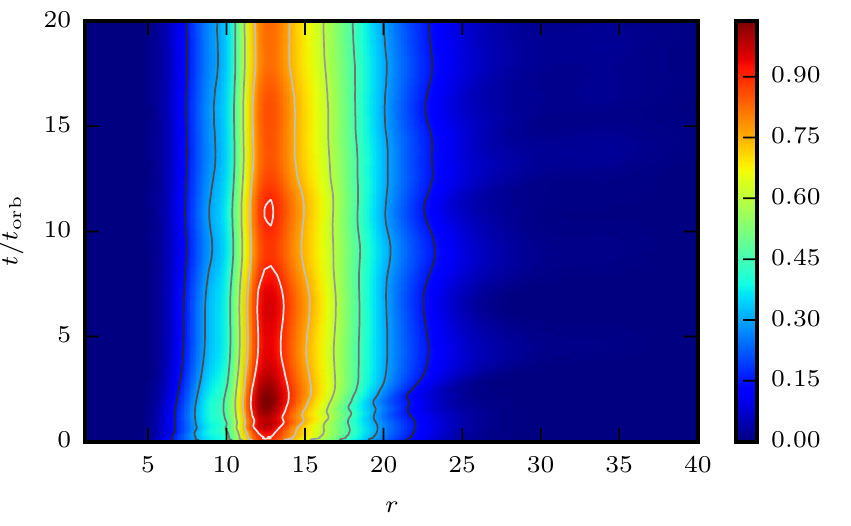}
  \caption{Spacetime $t-r$ diagram showing the time evolution of the
    radial profile of the angular momentum magnitude $\|J(r)\|$ for
    untilted models {\tt C1Ba0b0} (top panel), {\tt NC1a0b0} (middle
    panel) and {\tt D2a01b0} (bottom panel).}
  \label{fig:J_transport}
\end{figure}

In this section we investigate the transport of angular momentum in
the disks. To illustrate the transport
of angular momentum during the evolution of our models, we plot
spacetime $t-r$ diagrams showing the magnitude of the angular momentum
in radial shells, computed using Eq.~(\ref{eq:J_shell}). Such diagrams
are displayed in Fig.~\ref{fig:J_transport}. The magnitude of the angular momentum in
each shell, $\|J(r)\|$, has been rescaled to the global maximum value
attained in any shell during the
evolution. Fig.~\ref{fig:J_transport} shows the evolution of
$\|J(r)\|$ for the three untilted models {\tt C1Ba0b0} (top panel),
{\tt NC1a0b0} (middle panel) and {\tt D2a01b0} (bottom panel). This
figure reveals that compared to the latter two models, model {\tt
  C1Ba0b0} shows a radical redistribution of angular momentum in the
radial direction. This drastic difference is due to the development of
the PPI in model {\tt C1Ba0b0}.  Following the initial perturbation we
see that the contours of $\|J(r)\|$ start growing in a wave-like
manner, transporting angular momentum outwards. This happens up until
the saturation of the PPI at about 10 orbital periods. As noted by~\cite{Balbus2003}
during the growth phase the PPI is thought to be an effective mechanism for
angular momentum transport in thick accretion disks, where a right
combination of rotation and pressure at the boundaries of the disk
allows the growing non-axisymmetric modes to transport angular
momentum outwards.  
In~\cite{Zurek1986}, the authors 
showed as well that the development of nonaxisymmetric instabilities
is closely connected to the transport of angular momentum. The authors
observed that disks with a larger $q$ than the critical value given 
in~\cite{Papaloizou1984}, $q>q_{\mathrm{PP}}=2-\sqrt{3}$, were still unstable,
although the angular momentum transport was slower in those cases.
After saturation, the new
distribution of angular momentum seems PP-stable and does not exhibit
drastic changes for the rest of the evolution.

We can compare the evolution of $\|J(r)\|$ to the development of the
$m=1$ non-axisymmetric mode in the disk, which drops to a value of
around $1\%$ of its maximum value after saturation (see top panel of
Fig.~\ref{fig:ppi-modes}). As described in \cite{Balbus2003}, the
development of the PPI crucially relies on the properties of both the
inner and outer boundaries of the disk.  In \cite{Blaes1987} it was
shown that a small amount of accretion (which modifies the inner
boundary of the disk) was sufficient to saturate the growing PPI.
Furthermore, wider tori are known to be not as violently unstable as
slender tori \cite{Hawley1991}.  In subsections \ref{subsection:mdot}
and \ref{subsection:PPI-modes} we already showed that the accretion
rate and the growth of the $m=1$ non-axisymmetric mode saturate at
the same time, but the $t-r$ diagram of Fig.~\ref{fig:J_transport}
also shows that the outer angular momentum boundary of the disk is
pushed outwards with growing amplitude as well. It therefore seems
that both mechanisms are saturating the PPI simultaneously in model
{\tt C1Ba00}.

Model {\tt NC1a0b0}, displayed in the middle panel of Fig.~\ref{fig:J_transport},
has a non-constant specific angular momentum distribution, and shows a very 
different evolution of $\|J(r)\|$. The inner
region of the disk shows a gentle reduction of angular momentum during
the entire evolution and no sudden and drastic redistribution is found
as occurred in model {\tt C1Ba0b0}. The outer contour grows to
encompass the entire outer domain shown in the plot. Another
difference is the almost complete absence of variability of the inner
region of the disk, which is in contrast to model {\tt C1Ba0b0}, where
oscillatory behavior is seen in the inner boundary of the disk
too. The specific angular momentum profile of model {\tt NC1a0b0} is
between constant and Keplerian, with Keplerian disks being essentially
PP-stable. Model {\tt NC1a0b0} still develops the PPI, albeit in a
more mild manner than model {\tt C1Ba0b0}, with the torus of model
{\tt NC1a0b0} being initially wider.

Finally, model {\tt D2a01b0} (bottom panel), the lightest of our
models, is PP-stable as already shown in subsection
\ref{subsection:PPI-modes}. The corresponding spacetime diagram of $\|J(r)\|$  
shows no pronounced outward transport of angular
momentum as the evolution proceeds, the profile remaining very similar
to its initial configuration. The inner region with a high angular
momentum magnitude is thinned out slightly, and the profile spreads a
bit overall, but the changes are nowhere near as pronounced as for
models {\tt C1Ba0b0} and {\tt NC1a0b0}.

As we have seen in the analysis of the modes, the amplitude of the $m=1$ 
mode drops sharply upon the saturation of the PPI in the {\tt C1Ba00} model.
The subsequent evolution of the angular momentum profile remains more
constant, similar to what we observe for the entire evolution of the 
PP-stable model {\tt D2a01b0}. There is a clear split between the 
stages of the evolution: the first during which the PPI develops and 
saturates, and the subsequent evolution of an essentially PP-stable 
torus. 

Model {\tt NC1a00}, on the other hand, shows a very different behavior: 
as already seen in the mode analysis, the $m=1$ mode amplitude remains
roughly at its saturation value for the rest of the evolution, which is
lower than the values attained by models {\tt C1B}. The 
persistent $m=1$ mode is seen to continuously transport angular momentum 
outwards till the end of the simulation. The restructuring of the disk 
by the transport of angular momentum is more long-lived, compared to the 
drastic change the saturation of the PPI brings about in model {\tt C1Ba00}.
These findings seem to confirm those
found in~\cite{Zurek1986}.
Using the same kind of spacetime diagrams we have also checked the
dependence of the angular momentum transport on the BH spin $a$ and on
the initial tilt angle $\beta_0$ for all models of our sample. We find
that the angular momentum transport shows weak dependence with the
tilt angle (the higher the tilt angle the slowest the
transport) and shows essentially no dependence with the spin, at least
for the moderate values of the BH spin we could afford in our study.
This is a remarkable result, as it demonstrates that models {\tt C1B} and 
{\tt NC1} are PP-unstable for a wide range of initial BH spin magnitudes
and tilt angles. 

\subsection{Gravitational waves}
\label{subsection:GW}
The recent numerical simulations of~\cite{Kiuchi2011} have shown that the 
growth and saturation of the PPI makes BH-torus systems strong 
emitters of large amplitude, quasi-periodic gravitational waves, 
potentially detectable by forthcoming ground-based and spacecraft 
detectors. It was found in particular that the $m = 1$ non-axisymmetric 
structure survives with an appreciable amplitude after the saturation 
of the PPI nonlinear growth, which leads to the emission of 
quasi-periodic GWs with a large amplitude. 
(First estimates through full GRHD simulations of the GW 
detectability by the PPI instability in self-gravitating BH-torus 
systems were obtained in~\cite{Korobkin2011} where the nonlinear saturation phase 
was almost reached - see also~\cite{VanPutten2001} 
for earlier proposals and alternative estimates of the gravitational 
waves emitted by non-axisymmetric instabilities in such systems.)
In this section we turn our attention to analyze the
implications of the PPI in our tilted BH-torus systems regarding the
emission of GWs, namely the dependence of the resulting gravitational
waveforms and spectra on the BH spin magnitude and initial tilt angle.

\begin{figure}
  \centering
  \includegraphics[scale=1.0]{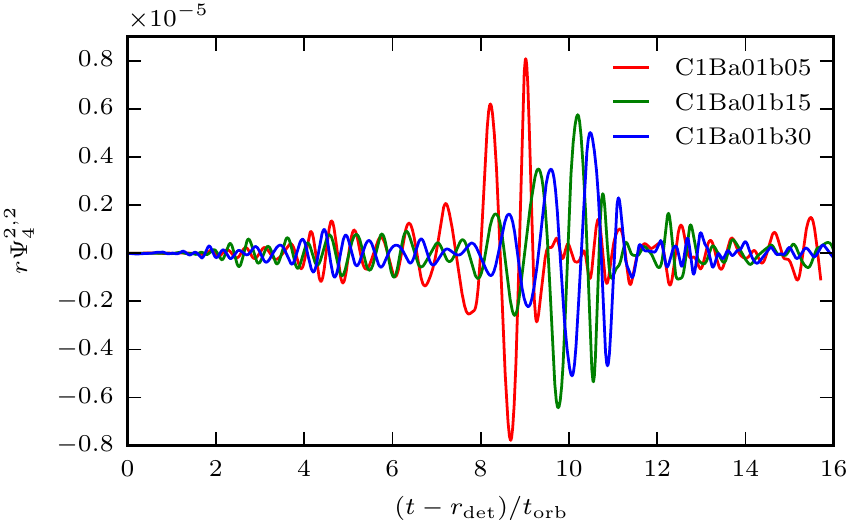}
  \\
  \vspace{-0.45cm}
  \includegraphics[scale=1.0]{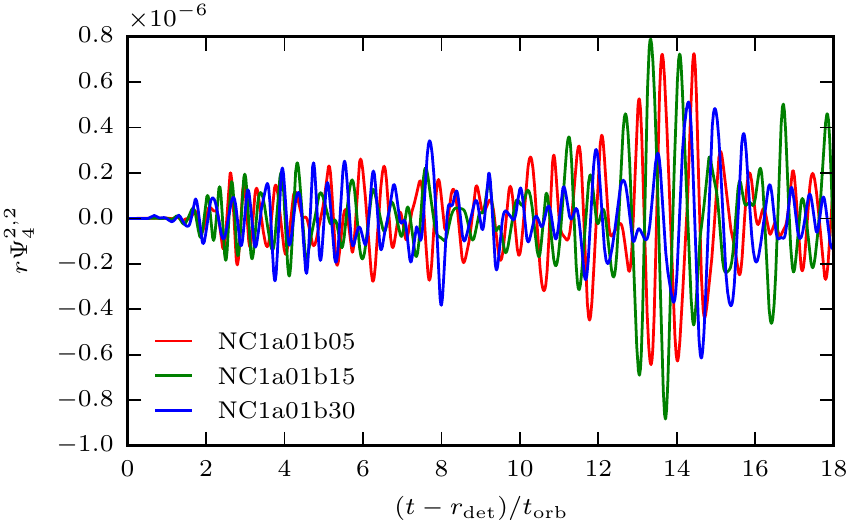}
  \caption{Time evolution of the $l=m=2$ mode of the real part of the
    Weyl scalar $\Psi_4$ multiplied by the extraction radius
    ($r=640$). 
    The top panel shows models {\tt C1Ba01} and the bottom
    panel models {\tt NC1a01}.}
  \label{fig:Re_psi4}
\end{figure}

\begin{figure}
  \centering
  \includegraphics[scale=1.0]{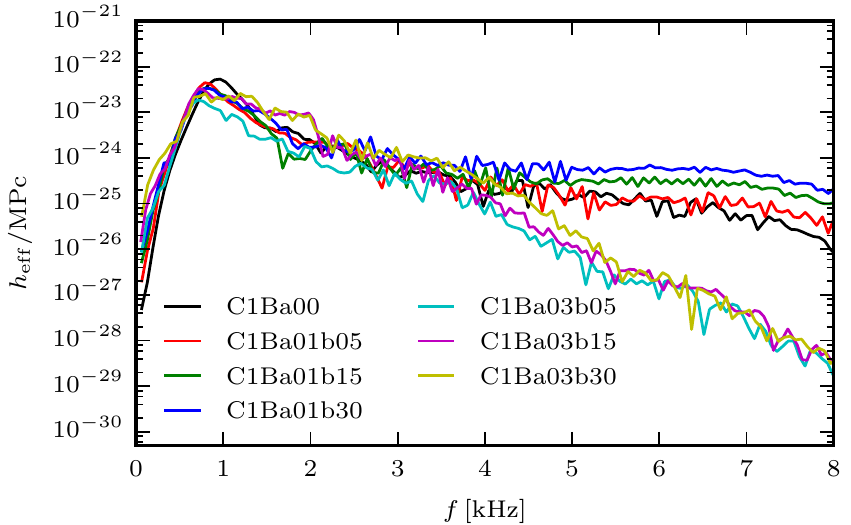}
  \\
  \vspace{-0.45cm}
  \includegraphics[scale=1.0]{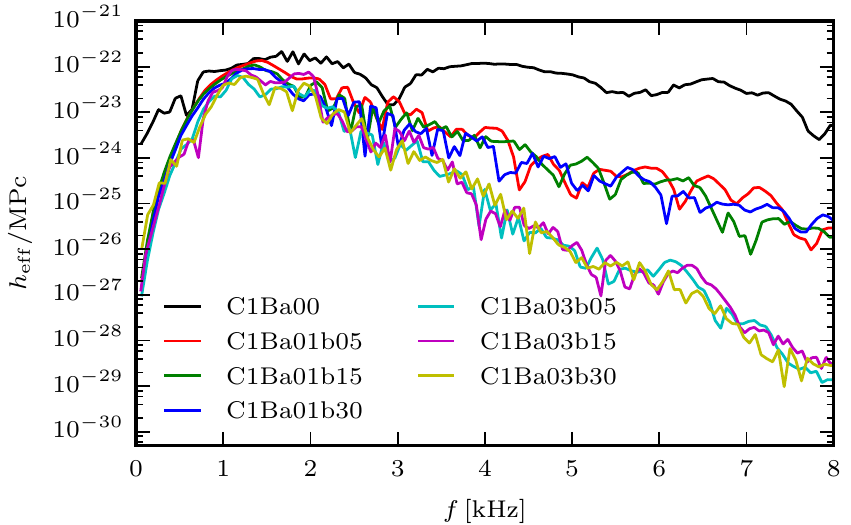}
  \caption{Spectrum of the effective strain for models {\tt C1B}, showing the
    $l=2,m=1$ modes (top) and $l=m=2$ modes (bottom) for the source
    located at 1 Mpc.}
  \label{fig:heff_C1B}
\end{figure}
   
\begin{figure}
  \centering
  \includegraphics[scale=1.0]{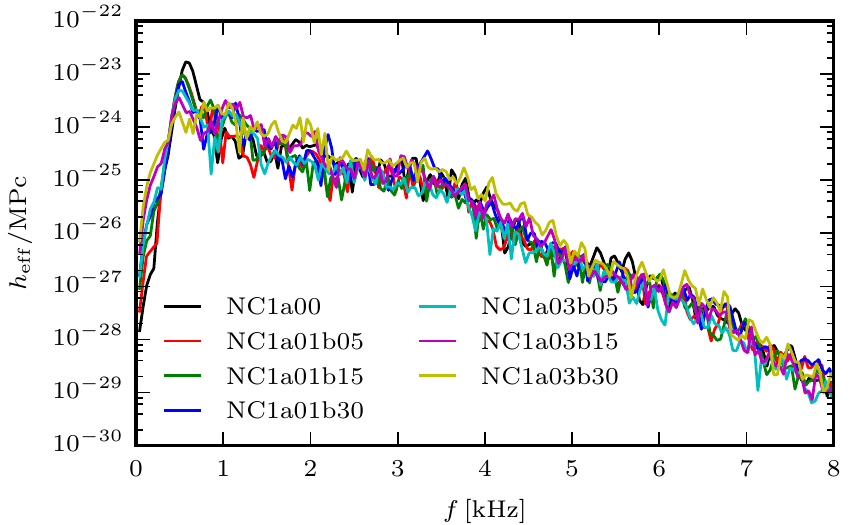}
  \\
  \vspace{-0.45cm}
  \includegraphics[scale=1.0]{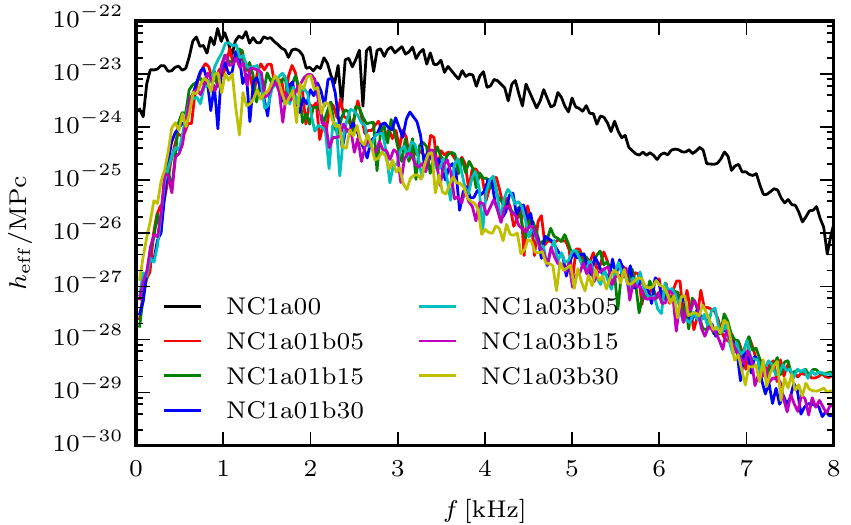}
  \caption{Spectrum of the effective strain for models {\tt NC1}, showing the
    $l=2,m=1$ modes (top) and $l=m=2$ modes (bottom) for the source
    located at 1 Mpc.}
  \label{fig:heff_NC1}
\end{figure}
   
In Fig.~\ref{fig:Re_psi4} we plot the real part of the $(l,m)=(2,2)$
mode of the outgoing part of the complex Weyl scalar $\Psi_4$ for
models {\tt C1Ba01} and {\tt NC1a01} and for all initial tilt angles
$\beta_0$.  This quantity has been computed at an extraction radius
$r=640$. 
Fig.~\ref{fig:Re_psi4} shows that, in agreement
with~\cite{Kiuchi2011}, all models display strong emission of
gravitational waves and that the emission persists for many dynamical
timescales well after the saturation of the PPI.  The peak amplitude
for models {\tt C1Ba01} is about an order of magnitude higher than for
models {\tt NC1a01}. We recall that model {\tt C1B} has a slightly
higher initial disk-to-BH mass ratio than model {\tt NC1} ($0.16$ and
$0.11$, respectively) and that the development of the PPI in models
{\tt C1B} is more pronounced than in the non-constant specific angular
momentum models {\tt NC1} (see
Fig.~\ref{fig:equatorial-plane-a01}). These two features are
responsible for the difference in the peak amplitudes between the two
initial models.  We also note that the peak amplitude in the {\tt
  C1Ba01} models, reached at PPI saturation, is much higher than the
amplitude of the remaining signal, whereas in models {\tt NC1a01} the
variations are not so pronounced.  As we showed in
Fig.~\ref{fig:ppi-modes}, upon PPI saturation the dominant $m=1$ PPI
mode drops to about $1\%$ of its peak value in models {\tt C1B}, while
it remains at a similar strength for models {\tt NC1}.

Fig.~\ref{fig:Re_psi4} also shows that the GW signal
has a weak dependence with the initial tilt angle, particularly for
the non-constant specific angular momentum models {\tt NC1}. The
smallest peak amplitudes are attained for the most tilted BH spacetime
($\beta_0=30^{\circ}$). In any event, the differences found in the
values spanned by the peak amplitudes with regard to $\beta_0$ are not
too significant.

\begin{figure}
  \centering
  \includegraphics[scale=1.0]{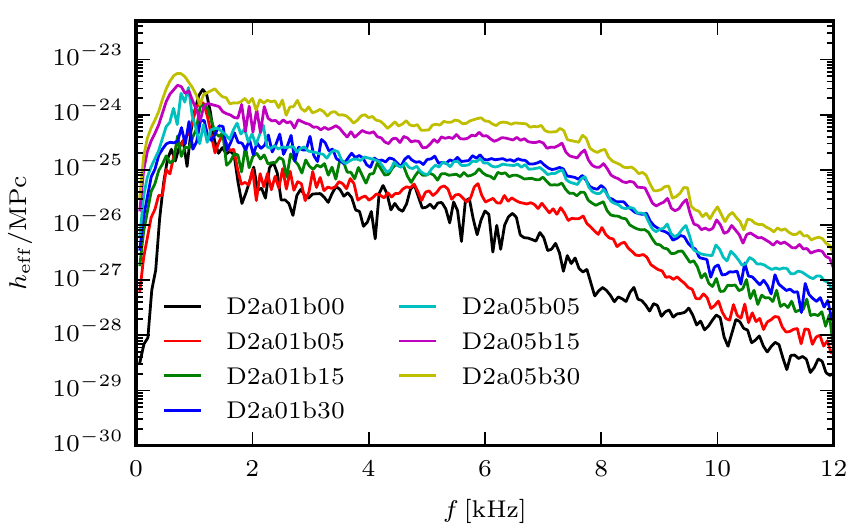}
  \\
  \vspace{-0.45cm}
  \includegraphics[scale=1.0]{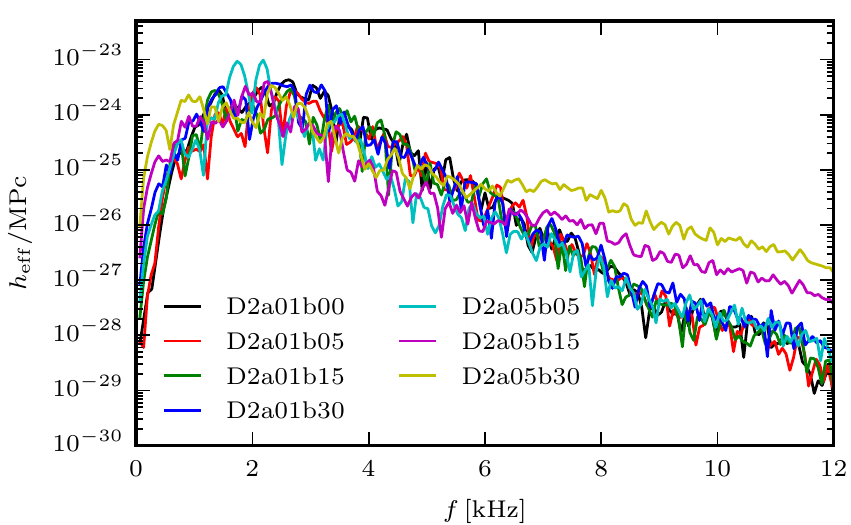}
  \\
  \vspace{-0.45cm}
  \includegraphics[scale=1.0]{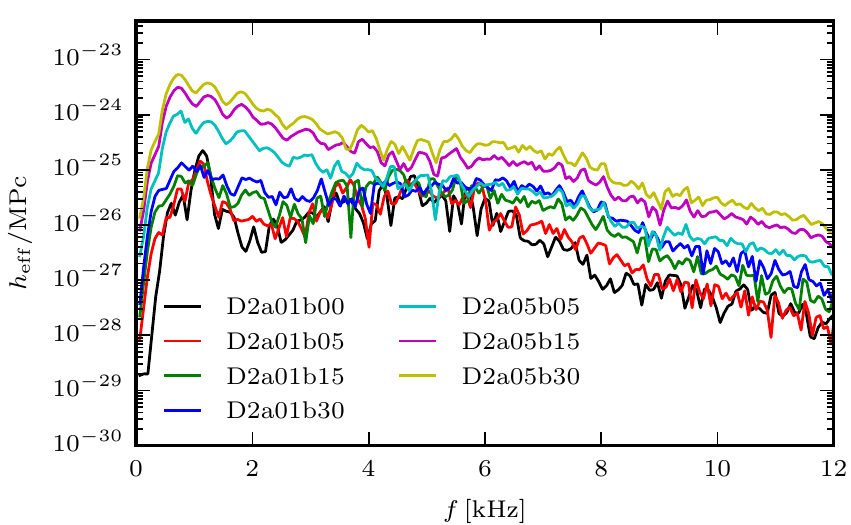}
  \caption{Spectrum of the effective strain for models {\tt D2}, showing the $l=2,m=1$
    modes (top panel) and $l=m=2$ modes (middle panel) and $l=3,m=1$
    modes (bottom panel) for the source
    located at 1 Mpc.}
  \label{fig:heff_D2}
\end{figure}

We use the fixed frequency integration (FFI) described 
in~\cite{Reisswig2011a} in the integration of the $\Psi_4$ data 
to obtain the GW strain. 
In Figs.~\ref{fig:heff_C1B}-\ref{fig:heff_D2} we plot the effective
strain $h_{\mathrm{eff}} / \mathrm{MPc}$ as a function of frequency for
all models, placing the sources at 1 MPc,  
in order to see any possible imprint of the initial spin
magnitude and tilt angle on the GW spectrum.  
As described above, we use the same time-step in the
7 outermost refinement levels ($\Delta t=0.32$ and
0.64 for higher and lower resolution runs, respectively), 
which means that the time-step at the extraction radius of 
$r=640$ is sufficiently small to sample the GW multipole signal 
with a $\delta t$ of 1.28 (2.56) for the higher (lower) resolution
runs, respectively. Due to the Nyquist criterion, we can therefore 
resolve the GW spectra up to $\approx 80$ (40) kHz.
The top panels of Figs.~\ref{fig:heff_C1B} and~\ref{fig:heff_NC1}
correspond to the $(l,m)=(2,1)$ mode and the bottom panels to the
$(l,m)=(2,2)$ mode. The highest power is usually found at a frequency
of about $1$kHz.  For the effective strain of the $(l,m)=(2,1)$ mode
for models {\tt C1B} shown in the top panel of
Fig.~\ref{fig:heff_C1B}, we can clearly identify two trends. On the
one hand there is a difference at high frequencies between the models
with either no or low ($a=0.1$) initial BH spin (models {\tt C1Ba00}
and {\tt C1Ba01}) and the high spin models ($a=0.3$). The spectra for
these two groups show a different slope from $f\sim 2 \mathrm{kHz}$
onwards.  On the other hand, for models {\tt C1Ba00} and {\tt C1Ba01},
an increase in the initial tilt angle leads to an increase in the
power for frequencies $f> 2 \mathrm{kHz}$. This also seems to be
present at the beginning of the spectrum although it is somewhat less
clear.

Correspondingly, the power spectra of the $(l,m)=(2,2)$ mode shown in
the bottom panel of Fig.~\ref{fig:heff_C1B} neatly splits the models
in three groups, according to the initial spin magnitude of the BH.
We find that the slope of the spectra becomes steeper from $f\sim 2
\mathrm{kHz}$ onwards the higher the BH spin.  In all cases, the slope
of the effective strain with the frequency shows essentially no
dependence with the initial tilt angle for the various groups of
models characterized by the same BH spin.  The power in model {\tt
  C1Ba00} is at least an order of magnitude larger for all frequencies
than that of the tilted models. This is because the spiraling movement
in the equatorial plane of both the BH and the $m=1$ overdensity PPI
``planet" emit GWs predominantly in the $l=m=2$ mode which are emitted
in a direction perpendicular to the equatorial plane. This is an
optimal situation for the model with aligned BH and disk spins.
However for the tilted models the ``orbit" of the BH and of the
overdensity lump in the disk is not confined to the equatorial plane
anymore because of the reaction of the disk to the tilted BH spin.
This results in a decrease in the GW power. 
   
The spectra of the non-constant specific angular momentum models {\tt
  NC1} displayed in Fig.~\ref{fig:heff_NC1} is markedly different to
those of constant specific angular momentum tori. Namely, their
dependence on the frequency is similar for all modes (same slope)
irrespective of the initial tilt angle and of the BH spin. No split in
two or three families is found according to the BH spin.  The
effective strain of the $(l,m)= (2,1)$ mode for models {\tt NC1} shows
a plateau in the low frequency part (up to $\sim 0.5 \mathrm{kHz}$)
for models {\tt NC1a03b15} and {\tt NC1a03b30}. The spectra shows a
prominent peak around $0.6 \mathrm{kHz}$ for all models but these
two. For the $l=m=2$ mode we find that the curve for the untilted
model {\tt NC1a00} lies more than an order of magnitude above the
curves corresponding to the initially tilted disks, as seen also for
the {\tt C1Ba00} model in the previous figure.

Finally, in Fig.~\ref{fig:heff_D2} we show
the effective strains of the $(l,m)=(2,1)$ mode (top panel), (2,2)
mode (middle panel) and (3,1) mode (bottom panel) for models {\tt
  D2}. In the spectra of the $(l,m)=(2,1)$ mode there is a clear
variation with the initial tilt and BH spin magnitude, with both
increasing the power of the mode for all frequencies.  Contrary to
model {\tt C1B} and similar to model {\tt NC1}, we find now that model
{\tt D2} does not show the grouping of models depending on the BH
spin, and the associated frequency dependence (i.e.~steeper slopes for
higher spins). As this model is PP-stable and is built with an
initially constant distribution of specific angular momentum, its
dynamics is somewhat between that of the other two models.  The
spectra of the $l=m=2$ mode (middle panel) show in particular a very
different behavior to those shown for models {\tt C1B} and {\tt NC1},
namely that the model with zero BH spin does not show significantly
larger power than the models with $a=0.1$ and $a=0.5$.  Since models
{\tt D2} do not develop the PPI, there is no strong emission of GW in
the $l=m=2$ mode due to the absence of the spiral motion of both BH
and overdensity `planet' that were observed for the other two initial
models. Therefore, the spectrum of the untilted model
{\tt D2a01b0} does not show a higher power density than the rest of the tilted
{\tt D2} models.  The two models {\tt D2a05b15} and {\tt D2a05b30} both show two
peaks in the low frequency part of the spectrum, as well as a flatter
slope between $\sim 4.5$ and $\sim 6.5 \mathrm{kHz}$.

The absence of the PPI in the {\tt D2} models results in a GW emission
where the $l=m=2$ mode is not the dominant mode. To show this, we plot
the effective strain of the $(l,m)=(3,1)$ mode in the bottom panel of
Fig.~\ref{fig:heff_D2}. The power is comparable with that of the
$(l,m)=(2,1)$ and $l=m=2$ modes, which is not the case for models {\tt
  C1B} and {\tt NC1} that both develop the PPI. The plot of the
effective strain of the $(l,m)=(3,1)$ mode shows that same trend as
that of the $(l,m)=(2,1)$ mode, i.e.~the power density increases with
spin magnitude and initial tilt angle. Furthermore, we can see clear
quasi-periodic oscillations in the low frequency part of the spectrum
for models {\tt D2a05}, as well as a developing plateau which becomes
flatter with higher initial tilt angle from $\sim 4.5-6.5
\mathrm{kHz}$.

\begin{figure}
  \centering
  \includegraphics[scale=1.0]{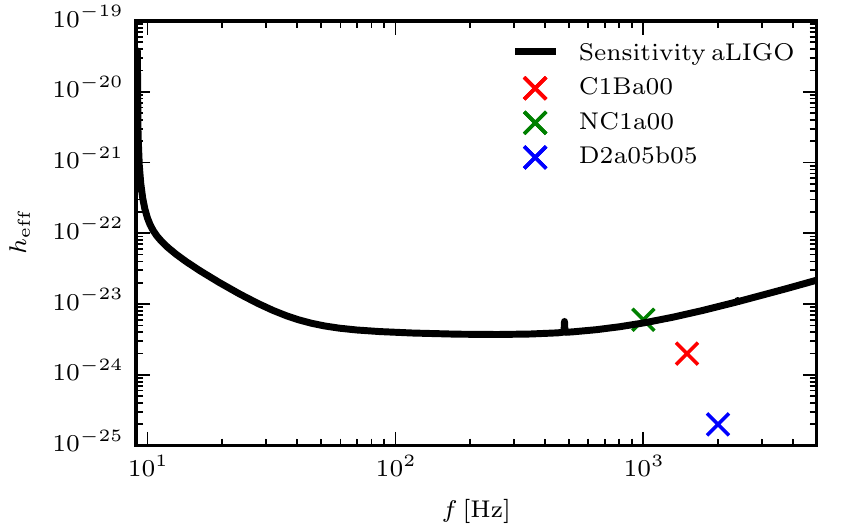}
  \caption{Plot showing the hypothetical peak amplitudes of the effective
    strain, inferred from the
    accretion timescale, for the three different initial models. 
    The sensitivity curve for advanced LIGO with zero-detuning and high-power configuration 
    is shown in black.}
  \label{fig:heff_adligo}
\end{figure}

In order to estimate the detectability of the GWs emitted by these
systems, we follow the analysis in~\cite{Kiuchi2011}. We assume  
that the BH-torus system  of models {\tt NC1} will radiate GWs for 
the entire accretion timescale $t_{\mathrm{acc}}$ (which is the time 
needed for the entire disk to accrete). This is because in these models, 
the $m=1$ overdensity lump and the BH form a long-lasting quasi-binary 
system. Assuming that the accretion rate will remain at the levels it 
attained by the end of the simulations (see Fig.~\ref{fig:mdot}), the 
lowest estimate for the accretion timescale for models {\tt NC1} is 
$t_{\mathrm{acc}} \approx\, 2\times 10^4$. From this timescale, we 
can estimate the number of GW cycles at the peak frequency, which 
yields $\approx 100$ cycles as a lowest estimate for models {\tt NC1}. 
The peak amplitude of the GW strain will then be amplified by 
$\sqrt{N_{\mathrm{cycles}}}\approx 10$, while we assume no such 
amplification for neither models {\tt C1B} (where the PPI is rapidly 
damped after saturation) nor for models {\tt D2} (which are PP-stable). 
We note that these estimates are in very good agreement with the 
corresponding findings reported in~\cite{Kiuchi2011}. The corresponding 
peak amplitudes for the BH-torus systems located at $50\, \mathrm{Mpc}$,
together with the advanced LIGO (aLIGO) sensitivity curve, are plotted 
in Fig.~\ref{fig:heff_adligo}, showing that the GWs emitted by models 
{\tt NC1} could be detectable due to the long-lasting emission of GWs 
due to the persistent $m=1$ mode. Note that the planned factor-of-3 
upgrade of aLIGO~\cite{Hild2011,Adhikari2013,Miller2014} will further 
improve the detectability.

\section{Discussion}
\label{sec:conclusions}

In this paper we have presented the results of an extended set of numerical relativity 
simulations of massive accretion disks around tilted Kerr black holes. We have
considered three different thick accretion disks of varying mass and specific angular
momentum magnitude and distribution, along with different black 
hole configurations with two spin magnitudes $a$ and four initial tilt angles $\beta_0$. On the
one hand the motivation for this work has been to extend the investigations 
of~\cite{Fragile2005,Fragile2007a} in the test-fluid approximation to full general relativity,  
analyzing the effects of the self-gravity of the disk on the BH-torus dynamics. On the other hand, 
our work has also served to enlarge the parameter space of the existing numerical relativity
simulations~\cite{Korobkin2011,Kiuchi2011} by accounting for {\it tilted} BH-torus systems for
the first time. 
 
For the models with disk-to-BH mass ratios of $0.044-0.16$ we have studied,
we have found that the assumption of using a fixed background spacetime is
not complete as we have observed significant 
precession and nutation of the tilted black hole as a result of the disk
evolution that cannot be accounted for in fixed spacetime
simulations.
For some of our models the precession
rate attains a fairly constant value  by the end of the evolution.
The BH nutation rate is also seen to be drastically modulated for those models
that develop the PPI, showing that the development of non-axisymmetric
modes in the disk exerts a torque on the central BH.

Our simulations have revealed that the development of the PPI
seems to be universal for the range of initial spin magnitudes 
and tilt angles investigated. Models {\tt C1B} and {\tt NC1}, both 
PP-unstable when the central BH is non-rotating or untilted, remain PP-unstable 
when the BH is rotating or tilted. The reverse also seems 
to be true: model {\tt D2}, PP-stable for an untilted, non-rotating BH
remains PP-stable for all spins and initial tilt angles considered.
As we mentioned before, model {\tt D2}, having the smallest mass ratio in our study, 
is the most likely outcome of BH-NS mergers based on astrophysical considerations.
Thus, our findings are relevant to gauge 
the practical role of the PPI in thick post-merger accretion disks around 
black holes. 
In agreement with the numerical relativity simulations of~\cite{Kiuchi2011}
the growth of the $m=1$ PPI mode manifests itself in the formation
of a counter-rotating  overdensity lump (or ``planet") that forms a ``quasi-binary" 
with the central BH during its existence. This causes the BH to start
moving in a spiral trajectory for as long as the ``planet'' exists.
For models {\tt C1B} the ``planet'' disperses quickly upon the saturation 
of the PPI, while in models {\tt NC1} the overdensity structure persists much
longer, with the $m=1$ mode amplitude remaining at its level of
saturation. When the PPI develops around a tilted Kerr BH, the binary motion
of BH and ``planet" is not restricted to the $xy$-plane but also shows 
some vertical motion. For models {\tt C1B} in particular, the rapid destruction 
of the non-axisymmetric structure causes a mild kick to the BH-torus system in the
vertical direction which is also imprinted in the time evolution of the linear 
momentum radiated away by gravitational waves in the same direction. 

The evolution of the disk around the tilted BH can cause a significant
twisting (differential precession) and warping (spatially varying tilt) in the disk. 
We have monitored the evolution of the twist $\sigma(r)$ and tilt $\nu(r)$ of 
radial shells during the simulations. For models {\tt C1B} ({\tt NC1}) we have 
found a phase of rapid (mild) realignment of the total angular momentum vector 
of the disk, $J_{\mathrm{Disk}}$, and the BH spin, $J_{\mathrm{BH}}$, during 
the growth of the PPI. We attribute this alignment to the development of the $m=1$ 
structure in the disk. Our simulations have also confirmed the presence of 
significant differential twisting of the disks due to Lense-Thirring precession. 
For all models, the cumulative twist is higher for higher initial BH spins, as expected, 
and the outer regions of the disks precess as a solid body. The evolution of the tilt 
profiles of models {\tt D2} is similar to what~\cite{Fragile2005} observed, with the
development of a noticeable peak in the innermost region of the disk.
For models {\tt C1B} the PPI causes a drastic change in the tilt profile upon saturation. 
Correspondingly, models {\tt NC1} develop strong tilt oscillations in the regions close 
to the BH after the PPI saturates and the $m=1$ structure persists till the end of the 
simulations. By interpreting the long-lived $m=1$ ``planet'' as an eccentric distribution 
of matter, we suggest that the so-called Kozai-Lidov mechanism, in which eccentricity 
and inclination are exchanged in an oscillatory manner, might explain the tilt angle 
oscillations we observe, in analogy with the recent findings of~\cite{Martin2014}.
In none of our models we observe the Bardeen-Petterson effect, as $\nu(r)\neq 0$ in 
the inner region of the disks.

The PPI is thought to activate the outward transport of angular momentum. This is indeed 
the case in our simulations, as the evolution of the angular momentum profiles shows. 
For model {\tt C1Ba0b0}, the angular momentum profile contours start growing in a 
wave-like manner up until the sudden saturation of the PPI. After saturation there is no 
more transport of angular momentum and the evolution resembles that of the PP-stable 
model {\tt D2a01b0}. For model {\tt NC1a0b0} the persistence of the $m=1$ structure in 
the disk continuously transports angular momentum outwards for the entire evolution of 
the disk. 

As~\cite{Kiuchi2011} showed, the development of the PPI and the
corresponding overdensity $m=1$ lump in the disk cause the long-term
radiation of gravitational waves, predominantly in the $l=m=2$
multipole mode, as expected from the radiation emitted by a binary system.
Our simulations have confirmed these results, showing that models {\tt C1B} and 
{\tt NC1} do indeed radiate mainly in that particular mode. The rapid destruction of the 
$m=1$ structure in models {\tt C1B} causes the amplitude of the emitted GW signal
to drop significantly after the PPI saturates, while the peak 
amplitude is closely correlated with the time of saturation. Models
{\tt NC1} on the other hand, emit at an amplitude similar to that attained at 
saturation, as the $m=1$ structure survives until the end of
the simulations. We have also calculated the effective strains of the GW signals. 
While models {\tt C1B} show different spectra for different initial
spin magnitudes and tilt angles, there seems to be no such trend 
for models {\tt NC1}. For those two set of models, the GW emission is predominantly 
in the $l=m=2$ mode while in models {\tt D2} there is also significant power
in the $l=3$, $m=1$ mode. In addition, the strain spectra of models {\tt D2} 
show a clear trend to higher values with increasing initial spin and tilt angle.
Finally, we have briefly touched on the issue of the possible modulation of the GW
signal caused by the BH precession. For the fastest  precessing models of our sample, 
models {\tt C1Ba01}, we have shown that we would need about 80 orbits for a complete
BH precession cycle in order to see such an effect, assuming the precession rate 
remains constant beyond the 20 orbit mark we could afford in our simulations.
We plan on exploring such a long-term simulation in the future in order
to obtain the complex GW signal from the kind of precessing BH-torus systems 
we have explored in this work. 

\begin{acknowledgments}
  We would like to thank the anonymous referees for their very
  helpful comments and suggestions to improve the manuscript.
  It is a pleasure to thank Chris Fragile for his careful reading of the manuscript and for providing 
  useful comments. VM would like to thank the developers of the Einstein Toolkit for their invaluable support during
  the duration of the project. He would also like to thank Ewald M{\"u}ller and the MPA for hospitality 
  and support during his two research visits. FG thanks Luciano Rezzolla for useful discussions on the 
  gravitational wave signals. FG gratefully acknowledges financial support from the ``NewCompStar'' 
  COST Action MP1304 during his visit to the University of Valencia. This work was supported  
  by the Spanish Ministry of Economy and Competitiveness (MINECO) through grants 
  AYA2010-21097-C03-01 and AYA2013-40979-P, by the Generalitat Valenciana  
  (PROMETEOII-2014-069), by the Deutsche Forschungsgemeinschaft (DFG) through its 
  Transregional Center SFB/TR7 ``Gravitational Wave Astronomy'', and by the Helmholtz 
  International Center for FAIR within the framework of the LOEWE program launched by the 
  State of Hesse. The simulations were performed on the Hydra HPC system at the 
  Rechenzentrum Garching.
\end{acknowledgments}

\appendix

\section{The BSSN evolution equations}
\label{sec:BSSN}

The BSSN formulation introduces a set of new
auxiliary variables for the evolution, defined in terms of the ADM
quantities: the conformal factor
\begin{equation}
  \phi \equiv \text{ln} \left(\frac{1}{12} \text{det} \, \gamma_{ij}\right)\,,
\end{equation}
the conformal spatial metric
\begin{equation}
  \tilde{\gamma}_{ij} \equiv e^{- 4 \,\phi} \, \gamma_{ij}\,,
\end{equation}
the trace of the extrinsic curvature
\begin{equation}
  K \equiv \gamma^{ij} \, K_{ij}\,,
\end{equation}
the conformal trace-free extrinsic curvature
\begin{equation}
  \tilde{A}_{ij} \equiv e^{- 4 \,\phi} \left(K_{ij} -\frac{1}{3}  \gamma_{ij} \, K\right)\,,
\end{equation}
and the conformal connections
\begin{equation}
  \tilde{\Gamma}^{i} \equiv \tilde{\gamma}^{jk} \, \tilde{\Gamma}^{i}_{jk}\,.
\end{equation}
Using these auxiliary variables together with appropriate gauge
choices leads to a strongly hyperbolic system~\cite{Beyer2004} that allows for
long-term stable numerical evolutions.

The evolution equations used in \verb:McLachlan: are the following:
\begin{eqnarray}
  \partial_0 \alpha &=& -\alpha^2 \, f(\alpha,\phi,x^{\mu})(K-K_0(x^{\mu}))\,,
  \\
  \p0 K &=& -\gamma^{ij}\tilde{D}_{i} \, \tilde{D}_{j} \, \alpha + \alpha(\tilde{A}^{ij} \, \tilde{A}_{ij} + \frac{1}{3} K^2) 
  \nonumber \\
  && + 4 \pi (E + \gamma^{ij} \, S_{ij})\,,
  \\
  \p0 \beta^i &=& \alpha^2 \, G(\alpha,\phi,x^{\mu}) \,B^i \,,
  \\
  \p0 B^i &=& e^{- 4 \,\phi} \, H(\alpha,\phi,x^{\mu}) \, \p0 \tilde{\Gamma}^i - \eta^i(B^i,\alpha,x^{\mu}) \,,
  \\
  \p0 \phi &=& - \frac{1}{6}\left(\alpha \,K - \partial_k \beta^k \right) \,,
  \\
  \p0 \tilde{\gamma}_{ij} &=& -2 \, \alpha \, \tilde{A}_{ij} + 2 \, \tilde{\gamma}_{k(i}\partial_{j)} \beta^k - \frac{2}{3} \, \tilde{\gamma}_{ij} \, \partial_k \beta^k \,,
  \\
  \p0 \tilde{A}_{ij} &=& e^{- 4 \,\phi} \left( \alpha \, \tilde{R}_{ij} + \alpha \, R^{\phi}_{ij} - \tilde{D}_i  \tilde{D}_j \alpha \right)^{\text{TF}} 
  \nonumber \\
  && + \alpha \, K \, \tilde{A}_{ij} -2 \, \alpha \, \tilde{A}_{ik} \, \tilde{A}^k_{\:j} + 2 \, \tilde{A}_{k(i} \partial_{j)} \beta^k 
  \nonumber \\
  &&  -\frac{2}{3} \tilde{A}_{ij} \, \partial_k \beta^k - 8 \, \pi \, \alpha e^{- 4 \,\phi} \, S^{\text{TF}}_{ij} \,,
  \\
  \p0 \tilde{\Gamma}^i &=& -2 \, \tilde{A}^{ij} \, \partial_j \alpha 
  \nonumber \\
  &&  + 2\, \alpha \left(\tilde{\Gamma}^i_{kl} \, \tilde{A}^{kl} + 6\, \tilde{A}^{ij} \, \partial_j \phi - \frac{2}{3} \, \tilde{\gamma}^{ij} \, K_{,j}\right) 
  \nonumber \\
  &&  - \tilde{\Gamma}^j \, \partial_j \beta^i + \frac{2}{3} \, \tilde{\Gamma}^i \, \partial_j \beta^j + \frac{1}{3} \, \tilde{\gamma}^{ik} \, \beta^j_{,jk} + \tilde{\gamma}^{jk} \, \beta^i_{,jk} 
  \nonumber \\
  &&  -16 \, \pi \, \alpha \, \tilde{\gamma}^{ik} \, S_k \,,
\end{eqnarray}
where we have used the following shorthand notation, $\p0
= \partial_t - \beta^j \, \partial_j$.

The coupling to the stress-energy tensor is done via the usual projections with the spatial metric $\gamma_{ij}$ and normal vector $n_{\mu}$,
\begin{eqnarray}
  E &\equiv& n_{\mu}\, n_{\nu} \, T^{\mu \nu} \\
     &=& \frac{1}{\alpha^2} \left(T_{00} -  \beta^i  T_{0i} + \beta^i  \beta^j T^{ij} \right),
  \\
  S_{ij} &\equiv& \gamma_{i \mu} \, \gamma_{j \nu} \, T^{\mu \nu}\,,
  \\
  S &\equiv& S^i_i = \gamma^{ij} \, S_{ij}\,,
  \\
  S_i &\equiv& - \gamma_{i \mu} \, n_{\nu} \, T^{\mu \nu} = - \frac{1}{\alpha} \left(T_{0i} - \beta^j \, T_{ij} \right)\,.
\end{eqnarray}

This system constitutes the so-called $\phi$-variant of the BSSN
formulation (see~\cite{Pollney2011b} for other possible variations).

\verb:McLachlan: uses the 1+log slicing for the
evolution of the lapse~\cite{Bona1995}:
\begin{eqnarray}
  f(\alpha,\phi,x^{\mu}) &\equiv& \frac{2}{\alpha}\,,
  \\
  K_0(x^{\mu}) &\equiv& 0\,,
\end{eqnarray}
and the following $\Gamma$-driver shift condition~\cite{Alcubierre2003} for the
evolution of the shift:
\begin{eqnarray}
  G(\alpha,\phi,x^{\mu}) &\equiv& \frac{3}{4 \, \alpha^2}\,,
  \\
  H(\alpha,\phi,x^{\mu}) &\equiv& e^{- 4 \,\phi}\,.
  \\
   \eta^i(B^i,\alpha,x^{\mu}) &\equiv& \eta B^i \, q(r)\,,
\end{eqnarray}
where $q(r)$ is a function that attenuates the $\Gamma$-driver 
depending on the radius and $\eta$ is a damping parameter.
We evolve the damping parameter $\eta$ using the formulation 
of~\cite{Alic2010} adapted to a single puncture. This approach 
considers a dynamically evolved damping parameter as well as 
a radial falloff in $\eta$. Assuming a constant gauge
function $\eta$ causes the shift to diverge at mesh refinement
boundaries at already very early times.

\section{GRHD evolution}
\label{sec:grhydro}

In the so-called Valencia formulation that is evolved in \verb:GRHydro:, the evolved variables,
called the \emph{conserved variables}, are defined in terms of the {primitive variables} 
via the relations
\begin{eqnarray}
  && D = \sqrt{\gamma} \, \rho \, W\,, \\
  && S_i = \sqrt{\gamma} \rho \,h \,W^2 \,v_i\,, \\
  && \tau = \sqrt{\gamma} \,(\rho \,h \,W^2 - p) - D\,,
\end{eqnarray}
where $\rho$ is the rest-mass density, $h$ is the specific enthalpy, $h=1+\epsilon+p/\rho$, 
$\epsilon$ is the specific internal energy, $p$ is the isotropic pressure, and $v^i$ is the three-velocity.
In addition, $W$ is the Lorentz factor and $\gamma$ is the determinant of the spatial
metric. The primitive variables are calculated from the conserved quantities
at each time step of the evolution using a root-finding procedure~\cite{Baiotti07}. 

The corresponding first-order flux-conservative hyperbolic evolution system is:
\begin{equation}
  \frac{\partial\boldsymbol{U}}{\partial t} + \frac{\partial\boldsymbol{F}^i}{\partial x^i} = \boldsymbol{S}\,,
\end{equation}
with the vector of conserved quantities
\begin{equation}
  \boldsymbol{U} = \left[D,S_i,\tau\right]\,,
\end{equation}
the vector of fluxes
\begin{equation}
  \boldsymbol{F}^i = \left[D \,\hat{v}^i,S_j \,\hat{v}^i + \delta^i_j \,p,\tau \,\hat{v}^i + p \,v^i \right]\,,
\end{equation}
with $ \hat{v}^i = \alpha v^i-\beta^i$,
and the vector of sources
\begin{equation}
  \boldsymbol{S} = \begin{bmatrix} 0 \\ \\
      \frac{\alpha}{2} S^{kl} \partial_i \gamma_{kl} 
      + S_k \partial_i \beta^k 
      - \left( \tau+ D \right) \partial_i \alpha  \\
      \\
    \alpha  S^{kl}K_{kl} - S^j\partial_j \alpha
  \end{bmatrix}\,.
\end{equation}
The stress-energy tensor $T^{\mu \nu}$ used in~\Eref{eqs:einstein}
is that of a perfect fluid, given by
\begin{equation}\label{eq:str-ene}
  T^{\mu \nu} = \rho \: h \: u^{\mu} \: u^{\nu} + p \: g^{\mu \nu}\,.
\end{equation}
The evolution system is closed with an equation of state (EOS) which
relates the pressure to the other primitive quantities. In our simulations
the torus is initially described by a polytropic EOS
\begin{equation}\label{eq:poly_eos}
p=K\rho^\Gamma\,,
\end{equation}
where $K$ is the polytropic constant and $\Gamma$ the adiabatic exponent.
During the evolution we allow for non-isotropic changes in the fluid flow such as 
shocks and for this reason we use an ideal gas EOS, 
\begin{equation}\label{eq:ig_eos}
p=(\Gamma-1)\rho\epsilon\,,
\end{equation}
where $\Gamma$ is the same adiabatic index in \Eref{eq:poly_eos}.
As indicated in Table~\ref{table:models}, we use $\Gamma=4/3$ for the evolution
of all our models. 
A Gamma-law EOS with $\Gamma=4/3$ is commonly 
adopted to describe a single-component perfect gas in the relativistic 
regime. Our work focuses in particular on the comparison with previous 
work on tilted accretion disks by~\cite{Fragile2005}, as well as the
works of~\cite{Korobkin2011} and~\cite{Kiuchi2011} where this value
of $\Gamma$ was also adopted.
For thick disks formed by the merger of two NS this EOS may be a poor
approximation. We are planning on extending our work to more realistic
EOSs (including the effects of composition and pressure from trapped
neutrinos) such as the one described in~\cite{Paschalidis2011}.

\section{Accuracy and convergence}
\label{sec:convergence}

\begin{figure}[t]
  \centering
  \includegraphics[scale=1.0]{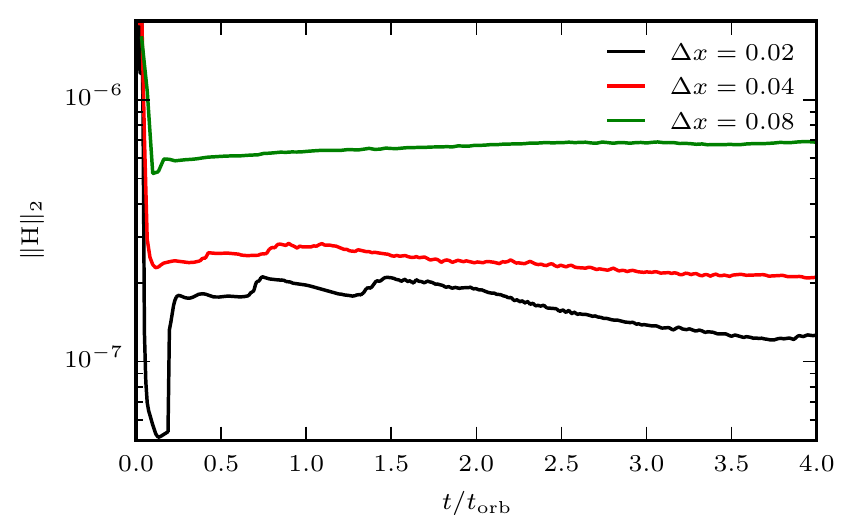}
  \\
  \vspace{-0.57cm}
  \includegraphics[scale=1.0]{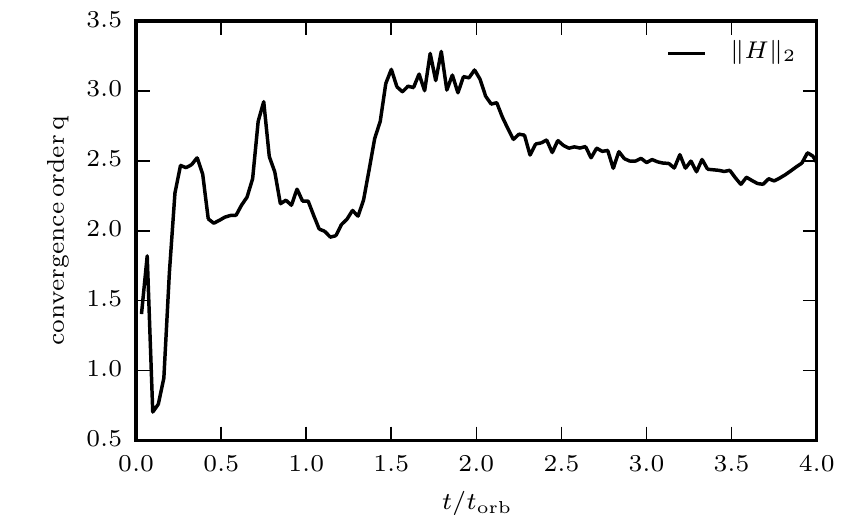}
  \caption{Convergence properties: {\it Top panel:} dependence on grid
    resolution for the $L2$-norm of the Hamiltonian constraint. {\it
      Bottom panel:} corresponding order of convergence. Model {\tt C1Ba01b30} has been used for
    this figure.}
  \label{fig:resolution-comparison}
\end{figure}

This appendix provides details of the accuracy and convergence properties of our 
simulations.
In Fig.~\ref{fig:resolution-comparison} we plot the time evolution
of the $L2$-norm of the Hamiltonian constraint (top panel) and the
convergence order of this quantity (bottom panel) for the first four
orbits of model {\tt C1Ba01b30}. The plot shows the results obtained
by using both the canonical (finest grid resolution $\Delta x=0.02$)
and two lower ($\Delta x=0.04$ and $\Delta x=0.08$) resolutions. We
follow~\cite{Galeazzi2013} and define the convergence order $q$ as
\begin{equation}
  q = \frac{1}{\log(f)} \log \left( \frac{\|\| \mathrm{H}\| _2^{\mathrm{low}} - \| \mathrm{H}\| _2^{\mathrm{med}} \|}{\|\| \mathrm{H}\| _2^{\mathrm{med}} - \| \mathrm{H}\| _2^{\mathrm{high}} \|} \right) \,,
\end{equation}
where $f$ is the refinement factor $f=\Delta x_{\rm{low}}/\Delta
x_{\rm{med}} = \Delta x_{\rm{med}}/\Delta x_{\rm{high}} = 2$.  From
Fig.~\ref{fig:resolution-comparison} we see that the order of
convergence shows some variability between a minimum of $\sim 1$ and a
maximum of about 3. The spacetime evolution is fourth order, while the 
evolution of the hydrodynamics is only second order accurate and
reduces to first order in the presence of shocks. We observe higher
than second order convergence in the $L2$-norm of the Hamiltonian 
constraint. This could be ascribed to the spacetime evolution which, due 
to the small disk-to-BH mass ratio and to the inconsistent ID setup,
could dominate the error budget in our simulations. Furthermore,
it is difficult to assess the order of convergence of our simulations at 
later times once the PPI
sets in, since the truncation error of the different resolutions 
could excite the exponential growth of the non-axisymmetric modes 
at different times causing the error to diverge.

\begin{figure}[t]
  \centering
  \includegraphics[scale=1.0]{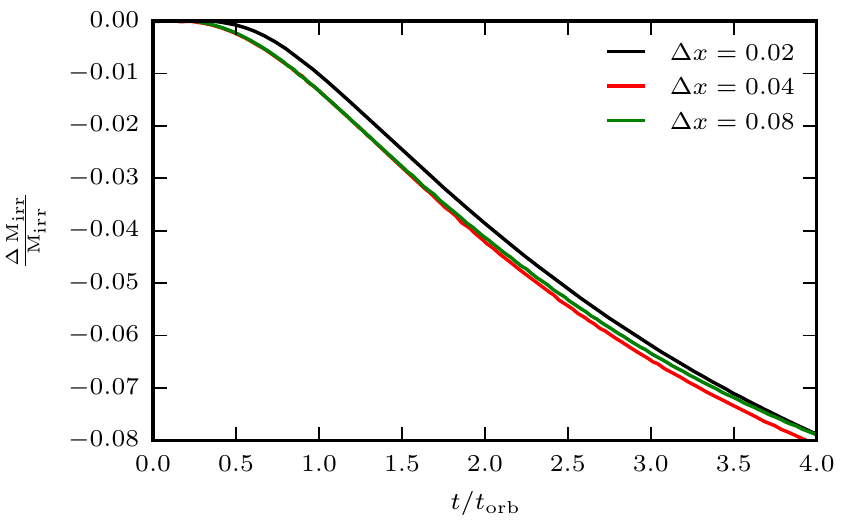}
  \caption{Dependence of the irreducible mass of the BH on grid
    resolution for model {\tt C1Ba01b30} plotted for the first 4 orbits.}
  \label{fig:mirr-resolution-comparison}
\end{figure}

The fractional change in the irreducible mass of the BH for model {\tt C1Ba01b30} 
is plotted in Fig.~\ref{fig:mirr-resolution-comparison}. The most striking feature
of this figure is the absence of an initial drop of the irreducible
mass due to the inconsistent choice of ID, contrary to the behavior
found in the simulations of~\cite{Korobkin2011}. The initial drop
found by~\cite{Korobkin2011} (which is about 8-10\%) happened in the
first orbit and was due to their metric blending technique in the
ID setup. After that significant initial drop, the
irreducible mass in~\cite{Korobkin2011} increases again to remain at around
$97.5$\% of its initial value during the evolution. Such a precise
control of the irreducible mass conservation may be related to the use
of a (quasi) spherical grid. In contrast, in our setup without
AH excision, there is no oscillatory behaviour in the
evolution of the irreducible mass at the early stages of the
evolution, as the ID in quasi-isotropic coordinates
dynamically adapts to the puncture gauge. Nevertheless, the
irreducible mass is seen to start decreasing in a smooth fashion from
$0.5 t_{\mathrm{orb}}$ onwards.  By the 4 orbital periods mark shown
in the figure, the irreducible mass has decreased to a
minimum of about $92$\% of its initial value, to subsequently increase
as the simulation proceeds as matter and angular momentum accrete onto
the BH. Moreover, Fig.~\ref{fig:mirr-resolution-comparison} shows that
this (unphysical) drop in the irreducible mass does not converge away
with resolution. 

\begin{figure}[t]
  \centering
  \includegraphics[scale=1.0]{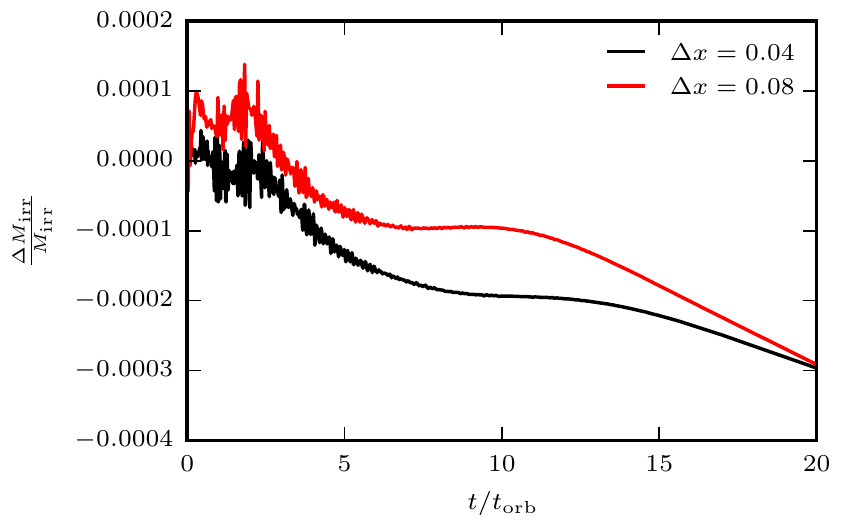}
  \caption{Evolution of the fractional error in the irreducible mass
  of a vacuum Kerr BH 
  with the same mass as that of model {\tt C1B} and a Kerr parameter of
  $a=0.3$. Results are shown for two different resolutions.}
  \label{fig:mirr-vac-resolution-comparison}
\end{figure}

In order to investigate this issue, we have performed two additional {\it vacuum} 
simulations of a BH with the same mass of that of the {\tt C1B} models and with an initial Kerr 
parameter of $a=0.3$.  In Fig.~\ref{fig:mirr-vac-resolution-comparison} we plot the 
corresponding fractional error in the irreducible mass for two (finest grid) resolutions, 
$\Delta x=0.04$ and $0.08$, for a total time of 20 orbits.
Similar to our results for the BH-torus setup, we observe no convergence
in the unphysical drop of the irreducible mass during the first $\approx 10$
orbits, while the loss of irreducible mass has a smaller slope for the higher
resolution run in the later stages of the evolution, as expected. We note,
however, that the magnitude of the fractional error in the vacuum tests is much
smaller (about two orders of magnitude) than that observed in our BH-torus runs.

\begin{figure}[t]
  \centering
  \includegraphics[scale=1.0]{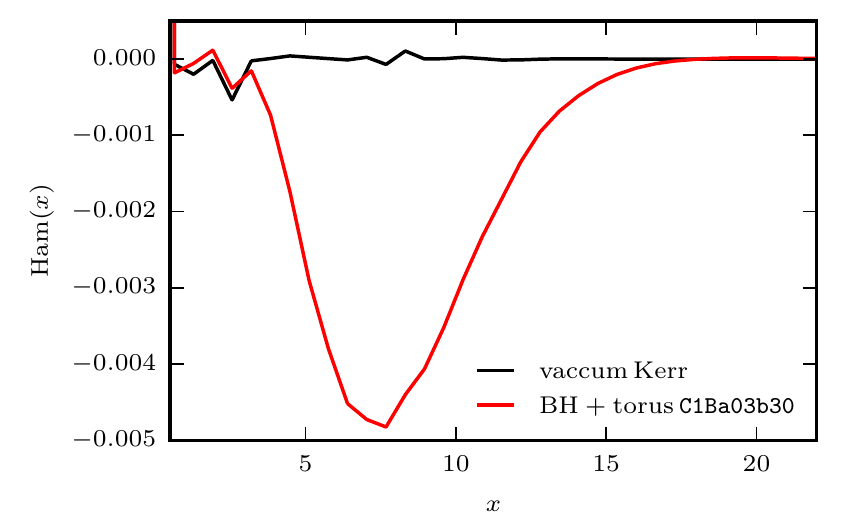}
  \caption{Initial radial profile of the Hamiltonian constraint for a vacuum Kerr 
  BH and for model {\tt C1Ba03b30} along the $x$-axis from the AH outwards.}
  \label{fig:ham-x}
\end{figure}

Unphysical loss of irreducible mass for constraint-violating ID has been reported before
in the literature (see,
e.g.~\cite{Mundim2011,Reifenberger2012,Okawa2014}). It has been 
interpreted as if the effect of the constraint violations were equivalent to the presence of a negative
mass in the system. The absorption of the violations by the BH causes the area of
the AH, and therefore also the irreducible mass, to decrease. As stated in Section~\ref{sec:ID}, 
our ID is manifestly constraint-violating and, furthermore, the  constraint violations do {\it not} 
converge away with resolution. In Fig.~\ref{fig:ham-x} we plot the initial profile of the Hamiltonian 
constraint for the vacuum BH run with $\Delta x = 0.04$ and for the corresponding BH-torus 
{\tt C1Ba03b30} model. We clearly see a large constraint violation (compared to the vacuum 
case) in the region where the disk initially resides. By interpreting the loss of irreducible mass 
as caused by the absorption of constraint violations, the lack of convergence in the irreducible 
mass drop in our BH-torus systems can be explained. As we have also seen in the vacuum tests, 
this transient feature connected to the initial constraint violations (much smaller in this case) is 
very long lived. In  our BH-torus models the loss of irreducible mass only flattens out  once enough 
matter has been accreted. Such large fractional errors are not reported in simulations where 
the BH-torus system forms after the evolution of a binary BH-NS or NS-NS merger, which 
highlights the undesired effect of employing constraint-violating ID.

\bibliographystyle{apsrev4-1-noeprint} \bibliography{references}

\end{document}